\documentclass{aa}  

\DeclareUnicodeCharacter{FEFF}{}
\DeclareUnicodeCharacter{2212}{-}
\usepackage[]{natbib}
\usepackage[T1]{fontenc}
\usepackage{ae,aecompl}
\usepackage[dvipsnames]{xcolor}
\usepackage{psfrag,color}
\usepackage[]{inputenc}
\usepackage{graphicx}	
\usepackage{placeins}
\usepackage{amsmath}	
\usepackage{pifont}
\usepackage{amssymb}	
\usepackage{longtable}
\usepackage{subfigure}
\usepackage{comment}
\usepackage{multirow}
\usepackage{adjustbox}
\usepackage{ulem}       
\usepackage{newtxmath} 
\usepackage{newtxtext}
\usepackage{pdflscape}
\usepackage{supertabular}
\usepackage{rotating}
\usepackage{footnote}
\usepackage{times}
\usepackage{chemmacros}
\usepackage{xspace}

\usepackage[colorlinks=true, urlcolor=blue, linkcolor=blue, citecolor=blue]{hyperref}

\definecolor{lime}{HTML}{A6CE39}
\DeclareRobustCommand{\orcidicon}{
	\begin{tikzpicture}
	\draw[lime, fill=lime] (0,0) 
	circle [radius=0.16] 
	node[white] {{\fontfamily{qag}\selectfont \tiny ID}};
	\draw[white, fill=white] (-0.0625,0.095) 
	circle [radius=0.007];
	\end{tikzpicture}
	\hspace{-2mm}
}

\foreach \x in {A, ..., Z}{\expandafter\xdef\csname orcid\x\endcsname{\noexpand\href{https://orcid.org/\csname orcidauthor\x\endcsname}
			{\noexpand\orcidicon}}
}

\usepackage[]{ulem}  
\newcommand\redout{\bgroup\markoverwith
{\textcolor{red}{\rule[0.5ex]{2pt}{0.8pt}}}\ULon}

\newcommand{\oii}{[O\,\textsc{ii}]}
\newcommand{\oiii}{[O\,\textsc{iii}]}
\newcommand{\nii}{[N\,\textsc{ii}]}
\newcommand{\neiii}{[Ne\,\textsc{iii}]}

\newcommand{\feiii}{[Fe\,\textsc{iii}]}

\newcommand{\cliii}{[Cl\,\textsc{iii}]}
\newcommand{\siii}{[S\,\textsc{iii}]}
\newcommand{\sii}{[S\,\textsc{ii}]}
\newcommand{\ariv}{[Ar\,\textsc{iv}]}
\newcommand{\hi}{H\,\textsc{i}}
\newcommand{\hii}{H\,\textsc{ii}}
\newcommand{\hei}{He\,\textsc{i}}

\newcommand{\ariii}{[Ar\,\textsc{iii}]}

\newcommand{\nel}{$n_{\rm e}$} 
\newcommand{\tel}{$T_{\rm e}$}

\begin{document}

   \title{Alpha-element abundance patterns in star-forming regions of the local Universe}

   \subtitle{}

   \author{C. Esteban\inst{1,2}{\orcidA{}},
          J. E. M\'endez-Delgado\inst{3}{\orcidB{}},
          J. Garc\'{\i}a-Rojas\inst{1,2}{\orcidC{}}, 
          K. Z. Arellano-C\'ordova\inst{4}{\orcidD{}},
          F. F. Rosales-Ortega\inst{5}{\orcidE{}},
          M. Orte-Garc\'{\i}a\inst{1,2}{\orcidF{}},
          E. Reyes-Rodr\'{\i}guez\inst{2,6}{\orcidG{}},
          L. Carigi\inst{3}{\orcidH{}},
          and A. Amayo\inst{3}{{\orcidI{}}}
     }

   \institute{Instituto de Astrof\'isica de Canarias, E-38205           La Laguna,             
             Tenerife, Spain \email{cel@iac.es} 
         \and
             Departamento de Astrof\'isica, Universidad de La Laguna, E-38206 La Laguna, Tenerife, Spain 
        \and
             Instituto de Astronom\'{\i}a, Universidad Nacional Aut\'onoma de M\'exico, Ap. 70-264, 04510 CDMX, Mexico
        \and
             Institute for Astronomy, University of Edinburgh, Royal Observatory, Edinburgh, EH9 3HJ, United Kingdom
        \and
             Instituto Nacional de Astrof\'isica, \'Optica y Electr\'onica (INAOE-CONAHCyT), Luis E. Erro 1, 72840, Tonantzintla, Puebla, Mexico
        \and
            Isaac Newton Group of Telescopes, Apto 321, E-38700 Santa Cruz de La Palma, Canary Islands, Spain
    }

   \date{Received ; accepted }

\authorrunning {Esteban et al.}
\titlerunning {Alpha-element abundance patterns in star-forming regions of the local Universe}
   \date{\today}

 
  \abstract
   {}
   {We make a reassessment of the distribution of the alpha-element abundance ratios Ne/O, S/O and Ar/O with respect to metallicity in a sample of about 1000 spectra of Galactic and extragalactic {\hii} regions and star-forming galaxies (SFGs) of the local Universe. We also analyse and compare different ionisation correction factor (ICF) schemes for each element in order to obtain the most confident determination of total abundances of Ne, S and Ar.}
   {We use the DEep Spectra of Ionised REgions Database (DESIRED) Extended project (DESIRED-E) that comprises about 1000 spectra of {\hii} regions and SFGs with direct determinations of electron temperature ({\tel}). We homogeneously determine the physical conditions and chemical abundances for all the sample objects. We compare the Ne/O, S/O and Ar/O ratios obtained using three different ICF schemes for each element. We also compare the abundance patterns with the predictions of a chemical evolution model of the Milky Way and stellar Ne and S abundance determinations.}
   {After a careful analysis we conclude that one of the tested ICF schemes provides a better match to the observed behaviour of Ne/O, S/O and Ar/O ratios. We find that the distribution of Ne/O ratios in {\hii} regions shows a large dispersion and no clear trend with O/H, indicating that the different ICF(Ne) schemes are not able to provide correct Ne/O ratios for most of these objects. This is not the case for SFGs that show similar linear relations with slightly positive slopes for the distributions of log(Ne/O) with respect to 12+log(O/H) or 12+log(Ne/H). The origin of this abundance pattern may be a combination of a metallicity-dependent dust depletion of O and ICF effects. The log(S/O) versus 12+log(O/H) distribution is consistent with a constant value especially for HII regions and when we consider both types of objects (SFGs + {\hii} regions). However, the log(S/O) vs. 12+log(S/H) distribution shows a quite tight linear fit with a positive slope. This relation seems to flatten at 12+log(S/H) $\lesssim$ 6.0. We consider that the observed behaviour of S/O with S/H is compatible with some contribution of S produced by SNe Ia. Finally, the behaviour of log(Ar/O) vs. 12+log(O/H) is very similar for {\hii} regions and SFGs and seems to be independent of the ionisation degree and the type of ICF(Ar) used, whether based on only the ({\ariii} lines or on the combination of {\ariii} and {\ariv} lines. The linear fit to log(Ar/O) vs. 12+log(O/H) indicates a slight decrease of log(Ar/O) as 12+log(O/H) increases. However, the log(Ar/O) vs. 12+log(Ar/H) relation shows an inverse behaviour, with a small positive slope that could indicate a small contribution of Ar produced by SNe Ia.
}
   {}

   \keywords{ISM: abundances -- \hii~regions -- Galaxies: abundances --  Nucleosynthesis}

   \maketitle
%

\section{Introduction}
\label{sec:intro}

The analysis of emission line spectra of star-forming regions is the main source of information on the chemical composition of the Universe and its evolution over cosmic time. Using standard tools for analysing ionised nebulae in the optical, ultraviolet and infrared spectral ranges, we can determine chemical abundances of an important fraction of the most abundant elements such as He, C, N, O, Ne, S, Cl, Ar, Fe, and Ni. Among these, only O shows bright optical emission lines from all of its possible ionisation states most common in star-forming regions. Therefore, for O its total abundance is usually simply the sum of the ionic abundances of its observed ionisation states, just O$^+$ and O$^{2+}$. In addition to the ease of determining its abundance from nebular spectra, O is the third most abundant chemical element in the Universe, which explains why it is used as a proxy for metallicity in the ionised gas-phase of the interstellar medium (ISM). For the remaining elements apart from O, we need to use an ionisation correction factor (ICF) to estimate their total abundances \citep[e.g.][]{Peimbert:69,Peimbert:77,Stasinska:78b,Mathis:85,Izotov:06,PerezMontero:07,Dors:13,Dors:16,Amayo:21}. These ICFs allow us to estimate the contribution of unseen ionisation states to the total abundance of a given element. 

O is produced in hydrostatic nucleosynthesis in the interior of massive stars ($>$ 8 M$_\odot$) after the triple-alpha process, where helium is fused into C and other heavier elements during the final stages of the star evolution. O is further produced and finally released into the ISM during the stellar wind phase and explosion of the dying star as core-collapse supernovae (CCSNe) on short time-scales \citep[e.g.][]{Woosley:02,Chiappini:03,Limongi:18,Kobayashi:20}. The so-called alpha-elements are those lighter than Fe whose mass number of their most abundant isotope is a multiple of four, that is, they contain an integer number of alpha particles or He nuclei.  Ne, S, and Ar are alpha-elements and, in principle, they should have the same nucleosynthetic origin as O and, therefore, present a similar behaviour in their cosmic evolution \citep[e.g.][]{Izotov:06,Croxall:16,Esteban:20,Rogers:22,ArellanoCordova:24}. Because of that, the abundance ratio of any of these alpha-elements with respect to O should remain constant. However, some nucleosynthesis and chemical evolution models predict some contribution of S and Ar from type I supernovae (SNIa) \citep[e.g.][]{Iwamoto:99, Johnson:19, Kobayashi:20}. Attending to those models, a certain deviation --a slight increase-- in the S/O and Ar/O ratios with respect to O/H --but also with respect to S/H and Ar/H-- would be expected. Whether or not the abundance ratio of alpha-elements with respect to O is constant is not yet a well-established issue from an observational point of view. The analysis of emission-line spectra of ionised nebulae is the main source of information on the abundances of Ne, S and Ar in the cosmos. In particular, for Ar it is the only possibility, since there are no lines of this element in stellar spectra. The situation for Ne is not so unfavorable, although it is only possible to obtain Ne abundances in the spectra of O and B-type stars.

In the local Universe, most studies devoted to chemical abundances in the ionised gas contained in star-forming regions find that Ne/O, S/O and Ar/O ratios do indeed remain basically constant with the metallicity of the gas, considering the rather large statistical dispersion and observational errors \citep[e.g.][]{Kennicutt:03,Izotov:06,ArellanoCordova:20,ArellanoCordova:24,Berg:20,Rogers:22}. But there is no shortage of works that argue the presence of some trends, at least for some alpha-elements. In the case of S/O, \citet{Diaz:22} find that it decreases significantly as metallicity increases but only when the star-forming galaxies (SFGs) are removed from their sample. In the case of Ar, \citet{Arnaboldi:22} report an increase of the Ar/O ratio as the Ar/H increases from the spectra of a sample of planetary nebulae in the M31 galaxy, a behaviour that they interpret as the effect of Ar production by SNe Ia. On the other hand, \citet{Izotov:06}, \citet{PerezMontero:07}, \citet{Kojima:21} or \citet{MirandaPerez:23} find the opposite trend in  Ar/O but only when limited or particular samples of objects are considered. Although, in principle, the situation should be clearer in the case of Ne --the models predict that its nucleosynthesis mechanisms are essentially the same as those for O--  there is also no lack of indications that the Ne/O ratio may not be constant. \citet{Izotov:06} observe a slight increase of Ne/O of about 0.1 dex over the entire metallicity range --12+log(O/H) from 7.1 to 8.6--  of their sample of SFGs. They interpret this feature as due to moderate depletion of O onto dust grains in their most metal-rich objects. On the other hand, although the results obtained by \citet{Croxall:16} for {\hii} regions are consistent with a basically constant value, they also find a population of objects with a significant offset to low Ne/O ratios. The calculation of the total abundance of Ne has the problem that its ionisation corrections are less well constrained than in the case of other elements, so some of the reported trends may depend on the ionisation degree of the objects and therefore be ultimately an effect of the ICF used \citep[e.g][]{Kennicutt:03, Berg:20, ArellanoCordova:24}. 

In recent years, and especially with the advent of the {\it James Webb Space Telescope} (JWST), metallicity determinations from the analysis of emission line spectra are beginning to be obtained for high-{\it z} SFGs
\citep[e.g.][]{ArellanoCordova:22,Schaerer:22, Brinchmann:23,Curti:23,Isobe:23,Nakajima:23}. Even in some objects, the detection of the faint auroral {\oiii} $\lambda$4363 line has permitted to determine the electron temperature, {\tel}, and to apply the so-called direct method \citep{Dinerstein:90,Peimbert:17} to obtain more precise determinations of the O abundance \citep[e.g.][]{ArellanoCordova:22,Schaerer:22,Curti:23,Rhoads:23,Trump:23,Laseter:24}. The total abundances of Ne, S and Ar have been determined for several SFGs at $z$ = 4 $-$ 10, offering the opportunity to study the evolution of the Ne/O, S/O and Ar/O abundance ratios  over cosmic time \citep[e.g.][]{ArellanoCordova:22, Isobe:23, Marques-Chaves:24, Stanton:24}. In the case of Ne, \citet{ArellanoCordova:22,ArellanoCordova:24}, \citet{Isobe:23} and \citet{Stanton:24} find that the Ne/O ratio --comparing its behaviour with that of the local Universe-- does not appear to evolve with redshift. Although, on the other hand, \citet{Stanton:24} find evidence of evolution of the Ar/O ratio with redshift.

To properly analyse the growing amount of data on the chemical content of high-$z$ SFGs, it is of utmost  importance to have high-quality chemical abundance data for a representative sample of objects in the local Universe to compare with. For example, \citet{Izotov:06} presented Ne/O, S/O and Ar/O abundance ratios for a sample of about 400 SFGs. \citet{PerezMontero:07} recalculated abundances for 633 SFGs and 220 {\hii} regions. More recent studies, as that by \citet{Berg:20}, present O, N, Ne, S and Ar abundance data for 190 individual {\hii} regions in four spiral galaxies. Another recent large study is that by \citet{Diaz:22}, who present O and S abundances for 256 {\hii} regions and 95 SFGs. 

The main aim of this work is to make a reassessment of the analysis of Ne/O, S/O and Ar/O abundance ratios by making use of the largest possible sample of high quality spectra from local {\hii} regions and SFGs. For this dataset, we recalculate precise abundances, using the direct method and applying the methodology developed by our group in a homogeneous manner. The characteristics of this sample, collected in the DEep Spectra of ionised REgions Database (DESIRED) project \citep{MendezDelgado:23b} are described in Sect.~\ref{sec:description_sample}. Another aim of this work is to compare the total abundances of Ne, S, and Ar obtained using different ICF schemes in order to select those that show a more independent behaviour with respect to the ionisation degree. In this way, we will try to get a closer look at the actual behaviour of the Ne/O, S/O and Ar/O ratios with respect to metallicity in the local Universe.

\section{Description of the sample of {\hii} regions and star-forming galaxies}
\label{sec:description_sample}

The present work is part of the DESIRED project \citep{MendezDelgado:23b} dedicated to the homogeneous analysis of optical spectra of ionised nebulae. The nucleus of DESIRED is comprised by a set of almost 200 deep intermediate- and high-spectral resolution spectra of Galactic and extragalactic \hii\ regions, SFGs, Galactic planetary nebulae (PNe) and ring nebulae (RNe) around evolved massive stars --as well as a small number of Herbig-Haro objects and protoplanetary discs of the Orion Nebula-- observed by our research group over the last 20 years. In addition to the former  DESIRED objects, we decided to include additional high-quality spectra of \hii\ regions, SFGs and PNe from the literature having direct determination of {\tel}. We call this larger sample DESIRED Extended (DESIRED-E) and it currently contains 2133 spectra as of July 25, 2024. We use this spectra dataset in the present study. 

\begin{figure}[ht!]
\centering    
\includegraphics[width=\hsize ]{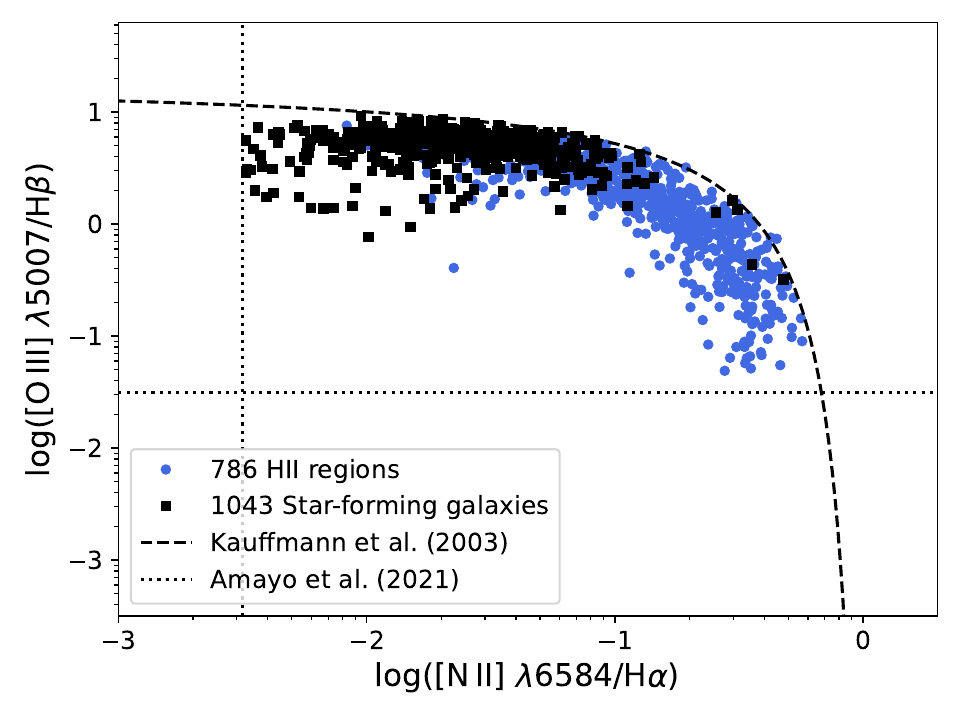}
\caption{BPT diagram of the sample of spectra of Galactic and extragalactic \hii\ regions and  SFGs compiled in DESIRED-E. The dashed line represents the empirical relation by \citet{Kauffmann:03} that distinguishes between star-forming regions and active galactic nuclei (AGNs). The dotted vertical and horizontal lines represent additional filters applied by \citet{Amayo:21} to  photoionisation models to construct their set of ICFs.}
\label{fig:BPT_diagram}
\end{figure}

A first description of DESIRED-E and the format of the datafiles can be found in \citet{MendezDelgado:24}. For each spectra we compile the extinction-corrected intensity ratios --along with their corresponding uncertainties-- of all the emission lines reported in the reference papers. In all the cases, we only consider those line ratios which quoted observational errors were less than 40\%. In this paper, we focus our attention on the study of the abundance ratios of alpha-elements, limiting our abundance calculations to O, Ne, S and Ar in Galactic and extragalactic \hii\ regions (designated simply as `{\hii} regions') and SFGs of DESIRED-E. The extragalactic \hii\ regions comprise individual star-forming nebulae in spiral or irregular galaxies and SFGs correspond to dwarf starburst galaxies whose general spectroscopic properties are indistinguishable from giant extragalactic \hii\ regions \citep[e.g.][]{Sargent:70,Melnick:85,Telles:97}. 

All {\hii} regions and SFGs of DESIRED-E have at least one detection of the following auroral/nebular intensity ratios of collisionally excited lines (CELs): \oiii~$\lambda4363/\lambda 5007$, \nii~$\lambda5755/\lambda6584$, and \siii~$\lambda 6312/\lambda9069$. With these line ratios we can derive \tel\ and they are insensitive to values of the electron density, {\nel}, of the order or lower than 10$^4 \text{ cm}^{-3}$ \citep{FroeseFischer:04,Tayal:11}. As it is well-known, a good determination of \tel\ is necessary to obtain reliable values of ionic abundances from the intensity of CELs. All DESIRED-E objects have direct determinations of the total O abundance --the proxy of metallicity in ionised nebulae studies-- and they cover a range of 12+log(O/H) values from 6.9 to 8.9 approximately. In Fig.~\ref{fig:BPT_diagram} we show the position of the {\hii} regions and SFGs compiled in DESIRED-E in the Baldwin-Phillips-Terlevich (BPT) diagram \citep{Baldwin:81}. Objects located above or at the right of the \citet{Kauffmann:03} line have not been considered in the calculations. Those discarded objects have hard ionising sources that can be associated to AGNs and/or can be affected by shock excitation as in some RNe around Wolf-Rayet stars \citep[e.g.][]{Esteban:16}. We apply additional filters (see Fig.~\ref{fig:BPT_diagram}) to exclude objects that, due to their position on the BPT diagram, are not considered in the ionisation models used by \citet{Amayo:21} to construct their set of ionisation correction factors (ICFs). This is one of the sets of ICFs that we use to determine the total abundances of Ne, S and Ar, as we will see in Sec.~\ref{sec:total_abundances}. The application of the additional filters of \citet{Amayo:21} only involves the rejection of around 2\% of the objects. The total number of DESIRED-E spectra classified as {\hii} regions or SFGs that fulfill our filters in the BPT diagram is 1829, 43.0\% corresponding to {\hii} regions and 57.0\% to SFGs. From this initial sample, we select 1386 objects for which we can determine the total abundance of one, two or three of the alpha-elements --Ne, S and Ar-- that can be detected in the nebular optical spectra.  

In Table D.2 we list the 1386 objects included in the present study, indicating their name, whether they correspond to {\hii} regions or SFGs --both types correspond exactly to 50\% of the objects by a happy coincidence-- and the reference of their published spectra.

\section{Physical conditions and ionic abundances}
\label{sec:physical_chemical}

We make use of \textit{PyNeb 1.1.18} \citep{Luridiana:15, Morisset:20}\footnote{\textit{PyNeb} code is publicly available on GitHub: \href{https://github.com/Morisset/PyNeb\_devel}{https://github.com/Morisset/PyNeb\_devel}.} with the atomic dataset presented in Table~D.1 and the \hi~ effective recombination coefficients from \citet{Storey:95} to determine the physical conditions and the ionic abundances of the ionised nebulae. \textit{PyNeb} is written in Python programming language and is fully vectorized. To calculate the emissivity of collisional excitation lines, \textit{PyNeb} calculates the relative population of the atomic levels of the ions of interest by solving the statistical equilibrium equations. In the case of recombination lines, it interpolates the available emissivity tables.

\subsection{Physical conditions}
\label{subsec:physical_conditions}

We use the \textit{getCrossTemDen} routine of \textit{PyNeb} 
with line intensity ratios sensitive to {\nel} and {\tel}. We cross-correlate the density-sensitive diagnostics \sii~$\lambda 6731/\lambda6716$, \oii~$\lambda 3726/\lambda3729$, \cliii~$\lambda 5538/\lambda5518$, \feiii~$\lambda 4658/\lambda4702$, and \ariv $\lambda 4740/\lambda4711$ with the temperature-sensitive ones \nii~$\lambda 5755/\lambda 6584$, \oiii~$\lambda 4363/\lambda 5007$, \ariii~$\lambda 5192/\lambda 7135$, and \siii~$\lambda 6312/\lambda 9069$ using a Monte Carlo experiment generating 100 random values to propagate uncertainties in line intensities. This number was chosen as a compromise between calculation time and convergence of the result. With this procedure, we  obtain a set of {\nel} and {\tel} values and their associated uncertainties for each diagnostic and individual point of the experiment. The average {\nel}, weighted by the inverse square of the error of the 100 individual points, is adopted as the representative value of the diagnostic. This procedure permits to take into account the small temperature dependence of density diagnostics under typical nebular conditions, which is not considered in many works of the literature. 

Once we obtain the density for each diagnostic, we apply the criteria proposed by \citet{MendezDelgado:23b} to estimate an average density representative of each nebulae. If {\nel}(\sii~$\lambda 6731/\lambda6716) < 100\text{ cm}^{-3}$, we adopt {\nel} = 100$\pm$100 cm$^{-3}$. If $100 \text{ cm}^{-3}\leq$ \nel(\sii~$\lambda 6731/\lambda6716)$ $< 1000\text{ cm}^{-3}$, we adopt the average between {\nel}(\sii~$\lambda 6731/\lambda6716$) and {\nel}(\oii~$\lambda 3726/\lambda3729$). If {\nel}(\sii~$\lambda 6731/\lambda6716)\geq 1000\text{ cm}^{-3}$, we adopt the average of {\nel}(\sii~$\lambda 6731/\lambda6716$), {\nel}(\oii~$\lambda 3726/\lambda3729$), {\nel}(\cliii~$\lambda 5538/\lambda5518$), {\nel}(\feiii~$\lambda 4658/\lambda4702$), and {\nel}(\ariv~$\lambda 4740/\lambda4711$). In cases where a value of \nel\ is not reported or could not be calculated in the source references, we adopt {\nel} = 100$\pm$100 cm$^{-3}$.

We estimate the temperatures {\tel}(\nii~$\lambda 5755/\lambda 6584$), {\tel}(\oiii~$\lambda 4363/\lambda 5007$) and {\tel}(\siii~$\lambda 6312/\lambda 9069$) using the \textit{getTemDen} routine of \textit{PyNeb} and the aforementioned average \nel\ of each object. We also use a Monte Carlo experiment of 100 random values to propagate uncertainties in \nel\ and line intensities ratios in order to estimate the error of each \tel\ diagnostic. To ensure a good determination of \tel\ for each object, we exclude all determinations of auroral lines with errors greater than 40\% and  verify that the flux of the pairs of nebular lines coming from the same upper atomic level used (e.g., \oiii~$\lambda \lambda 5007, 4959$, \siii~$\lambda \lambda 9531, 9069$, \nii~$\lambda \lambda 6584, 6548$) fit with their theoretical predictions, regardless of the physical conditions of the gas \citep{Storey:00}. We discard any diagnostic when the observed nebular line intensity ratios differ by more than 20\% from the theoretical ones. Departures from the theoretical case are rather common in the case of \siii~$\lambda \lambda 9531, 9069$, due to the contamination of telluric absorption bands \citep[see][for a discussion about this issue]{MendezDelgado:24}. There is a small number of objects where only one of the lines of the aforementioned pairs of nebular lines is reported. In these cases, we consider the single nebular line observed in the calculation, assuming that the {\tel} derived is valid. In this paper, we only adopt \tel(\siii) to determine ionic abundances in the absence of \tel(\nii) and \tel(\oiii). We show our derived physical conditions in Table D.3.

\subsection{Ionic abundances}
\label{subsec:ionic_abundances}

We have determined the ionic abundances of O$^{+}$, O$^{2+}$, S$^{+}$, S$^{2+}$, Ne$^{2+}$, Ar$^{2+}$ and Ar$^{3+}$. For O$^{+}$, we use the \oii~$\lambda \lambda 3727, 3729$ doublet, and the sum of the auroral lines \oii~$\lambda \lambda 7319, 7320, 7330, 7331$ when the bluer \oii~doublet lines are not available. In the case of O$^{2+}$, we use the sum of the bright \oiii~$\lambda \lambda 4959, 5007$ nebular lines. For S$^{+}$ and S$^{2+}$ we use the sum of \sii~$\lambda \lambda 6716, 6731$ and \siii~$\lambda \lambda 9069, 9531$ doublets, respectively. There are spectra that do not cover the near-IR range, so in such cases we use the \siii~$\lambda 6312$ line to derive the S$^{2+}$ abundance. For Ne$^{2+}$, we use the \neiii~$\lambda 3868$ line because \neiii~$\lambda 3967$ is usually blend with {\hi}~$\lambda 3970$ in most of the spectra. In the case of Ar$^{2+}$, we use the \ariii~$\lambda 7136$ line alone or in combination with \ariii~$\lambda 7751$ when both lines are available. Finally, we use the \ariv~$\lambda \lambda 4711, 4740$ doublet to derive the Ar$^{3+}$ abundance . The \hei~$\lambda$4713 line may contaminate \ariv~$\lambda 4711$ in the low-resolution spectra of high-ionisation objects, especially in SFGs. As we will see in Sect.~\ref{subsec:argon}, there is no indication of a significant sample of objects with abnormally high Ar/O ratios that might be affected by this problem.

We assume {\tel}(\nii~$\lambda 5755/\lambda 6584$) and the adopted value of \nel\ in the \textit{getIonAbundance} routine of \textit{PyNeb} to calculate the abundance of the low-ionisation ions: O$^{+}$ and S$^{+}$, propagating the uncertainties in \nel, \tel\ and line ratios through 100-point Monte Carlo experiments. When {\tel}(\nii~$\lambda 5755/\lambda 6584$) is not available for a given spectrum, we apply the temperature relations of \citet{Garnett:92} to estimate it from {\tel}(\oiii~$\lambda 4363/\lambda 5007$), or {\tel}(\siii~$\lambda 6312/\lambda 9069$), when the first diagnostic is also absent. The O$^{2+}$, Ne$^{2+}$ and Ar$^{3+}$ abundances are determined using {\tel}(\oiii~$\lambda 4363/\lambda 5007$) and the adopted value of \nel. In the spectra where {\tel}(\oiii~$\lambda 4363/\lambda 5007$) is not available, we use the temperature relations of \citet{Garnett:92} along with the values of  {\tel}(\nii~$\lambda 5755/\lambda 6584$) and/or {\tel}(\siii~$\lambda 6312/\lambda 9069$) when one of these are determined from the spectra. The abundance of the intermediate-ionisation ions, S$^{2+}$ and Ar$^{2+}$, is calculated using he adopted value of \nel\ and {\tel}(\siii~$\lambda 6312/\lambda 9069$) or 
{\tel}(\nii~$\lambda 5755/\lambda 6584$) or {\tel}(\oiii~$\lambda 4363/\lambda 5007$) along with the temperature relations of \citet{Garnett:92} when {\tel}(\siii~$\lambda 6312/\lambda 9069$) is not available. The ionic abundances of O$^{+}$, O$^{2+}$, S$^{+}$, S$^{2+}$, Ne$^{2+}$, Ar$^{2+}$ and Ar$^{3+}$ of the objects included in this study are shown in Table~D.4. 

\section{Total O, Ne, S and Ar abundances}
\label{sec:total_abundances}

The total O abundances have been calculated as the sum of the 
O$^{+}$/H$^{+}$ and O$^{2+}$/H$^{+}$ ratios. The contribution of O$^{3+}$/H$^{+}$ to the total O/H ratio was not considered, as observational studies of metal-poor regions \citep[e.g.][]{Izotov:99a,Berg:21,DominguezGuzman:22} find small values up to the order of 5\% (0.02 dex at maximum) that can be considered negligible compared to the typical errors of the O/H ratio. 

As it is said in Sect.~\ref{sec:intro}, we have to consider ICFs to determine the total abundances of Ne, S, and Ar.  Following a philosophy similar to that of previous works in {\hii} regions and SFGs \citep[e.g.][]{ArellanoCordova:20,ArellanoCordova:24}, we have compared the elemental abundances and their ratios obtained using different correction schemes to achieve the most model-independent and robust total abundance determinations possible. For each element, we compare and analyse the results using three ICF schemes. Two of them: \citet{Amayo:21} and \citet{Izotov:06} are common to Ne, S and Ar, and the third scheme is different for each element, \citet{Dors:13} for Ne, \citet{Dors:16} for S and \citet{PerezMontero:07} for Ar. In Fig.~\ref{fig:comparion_ICFs}, in the Appendix, we include the differences between the Ne, S and Ar abundances (in logarithmic form) obtained for the spectra of the DESIRED-E sample using the different ICF schemes. The shape of the curves helps to interpret the distribution of abundance ratios that will be discussed below. Moreover, The range of values covered by the observational points in these figures --which depends on the element, the emission lines used and the degree of ionisation-- gives us an idea of the uncertainty introduced by the use of an ICF on the value of the total abundance of a given element and object.

\subsection{Neon}
\label{subsec:neon}

\begin{figure}[ht!]
\centering    
\includegraphics[width=\hsize ]{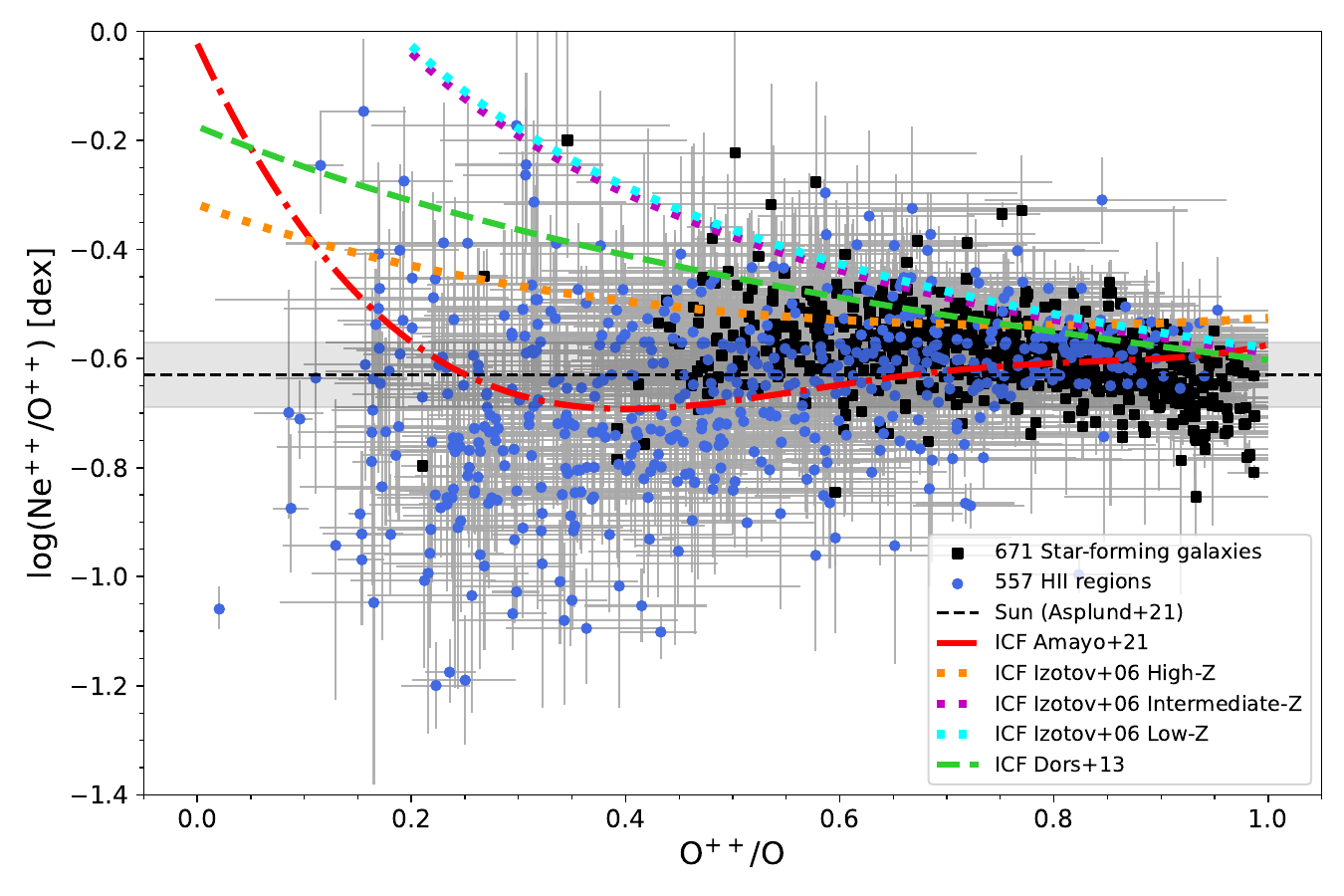}
\includegraphics[width=\hsize ]{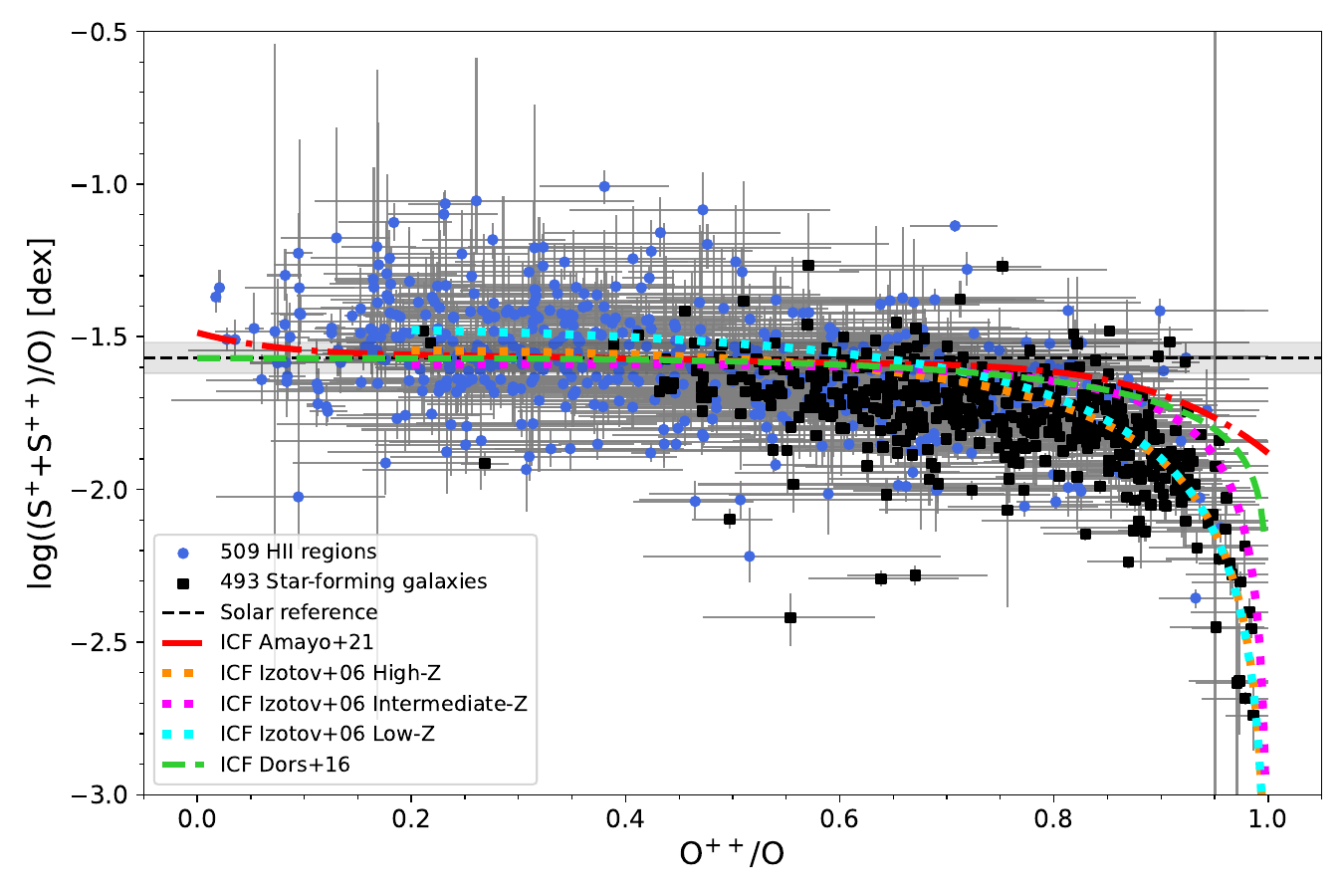}
\caption{Log(Ne$^{2+}$/O$^{2+}$) (top) and log((S$^{+}$+S$^{2+}$)/O) (bottom) as a function of the ionisation degree, O$^{2+}$/O, for DESIRED-E sample of \hii\ regions (blue circles) and SFGs (black squares). The black dashed lines and the grey bands show the solar log(Ne/O) and log(S/O) values and their associated uncertainty, respectively, from \citet{Asplund:21}. The different curves represent the ICF schemes used in this study: \citet[][red dashed-dotted lines]{Amayo:21}, \citet[][dotted lines with colours corresponding to three metallicity ranges]{Izotov:06}, \citet[][green dashed line in top panel]{Dors:13} and \citet[][green dashed line in bottom panel]{Dors:16}.} 
\label{fig:ionic_ratios_NeS}
\end{figure}

\begin{figure*}[ht!]
\centering    
\includegraphics[scale=0.38]{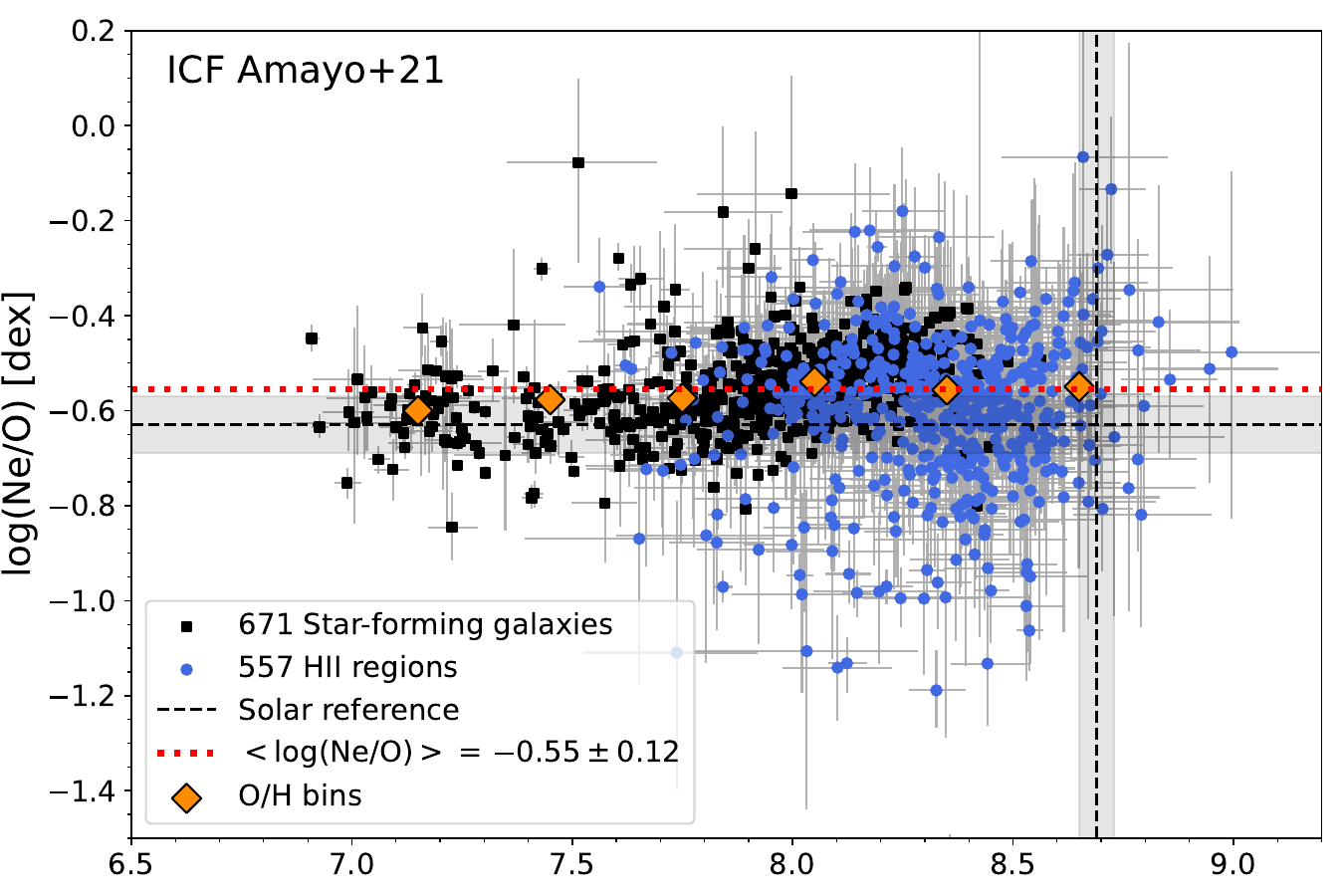}
\includegraphics[scale=0.38]{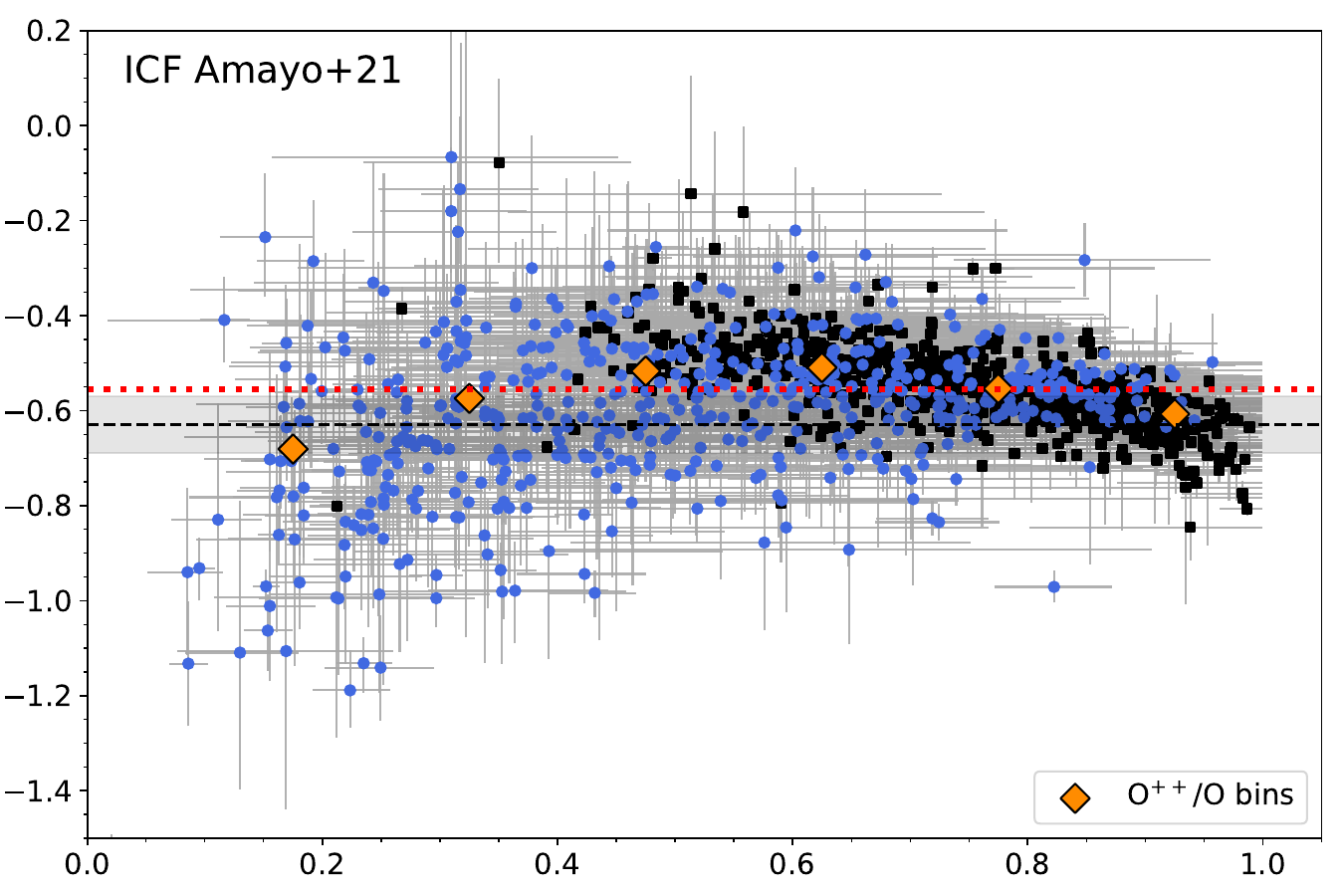}
\\
\includegraphics[scale=0.38]{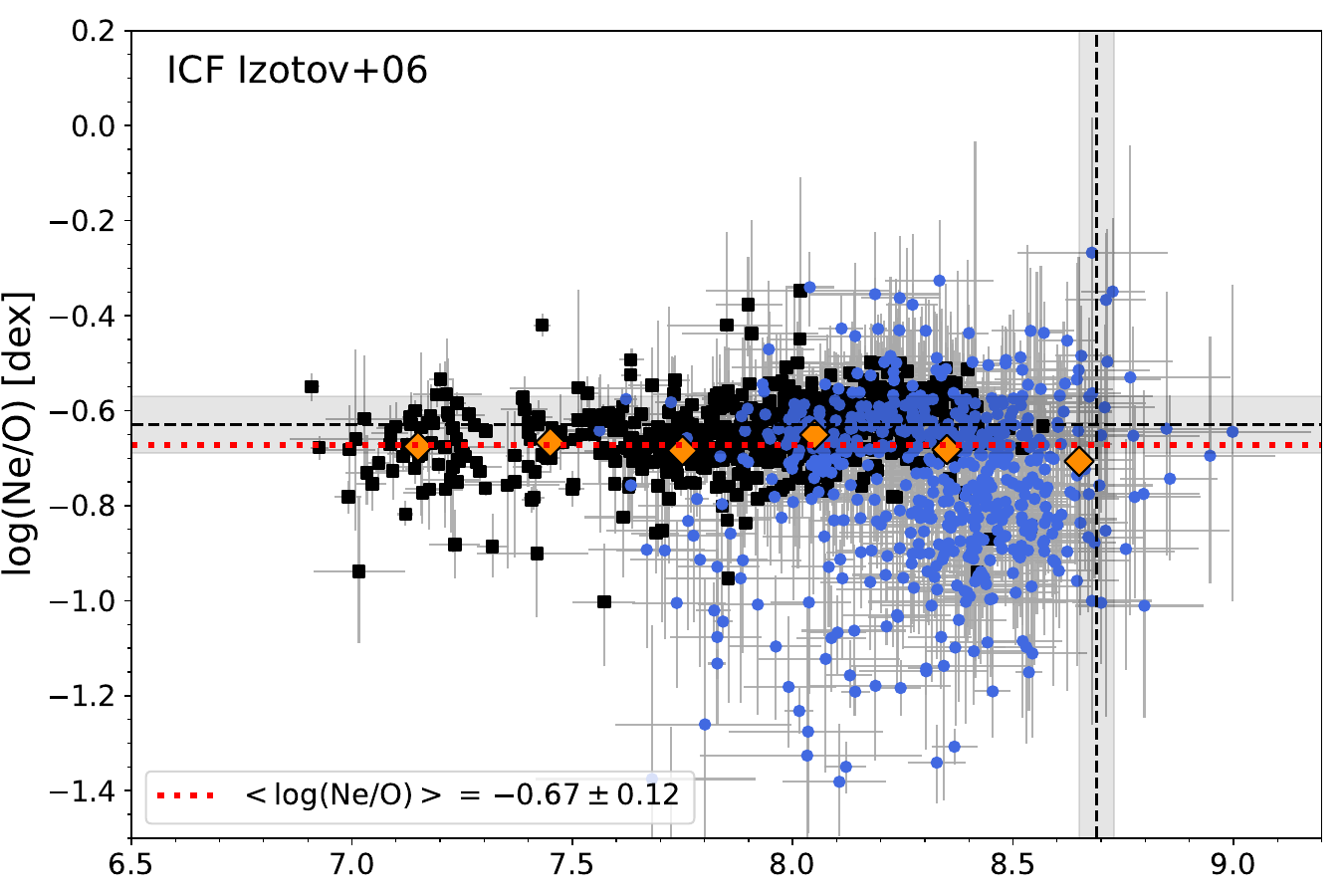}
\includegraphics[scale=0.38]{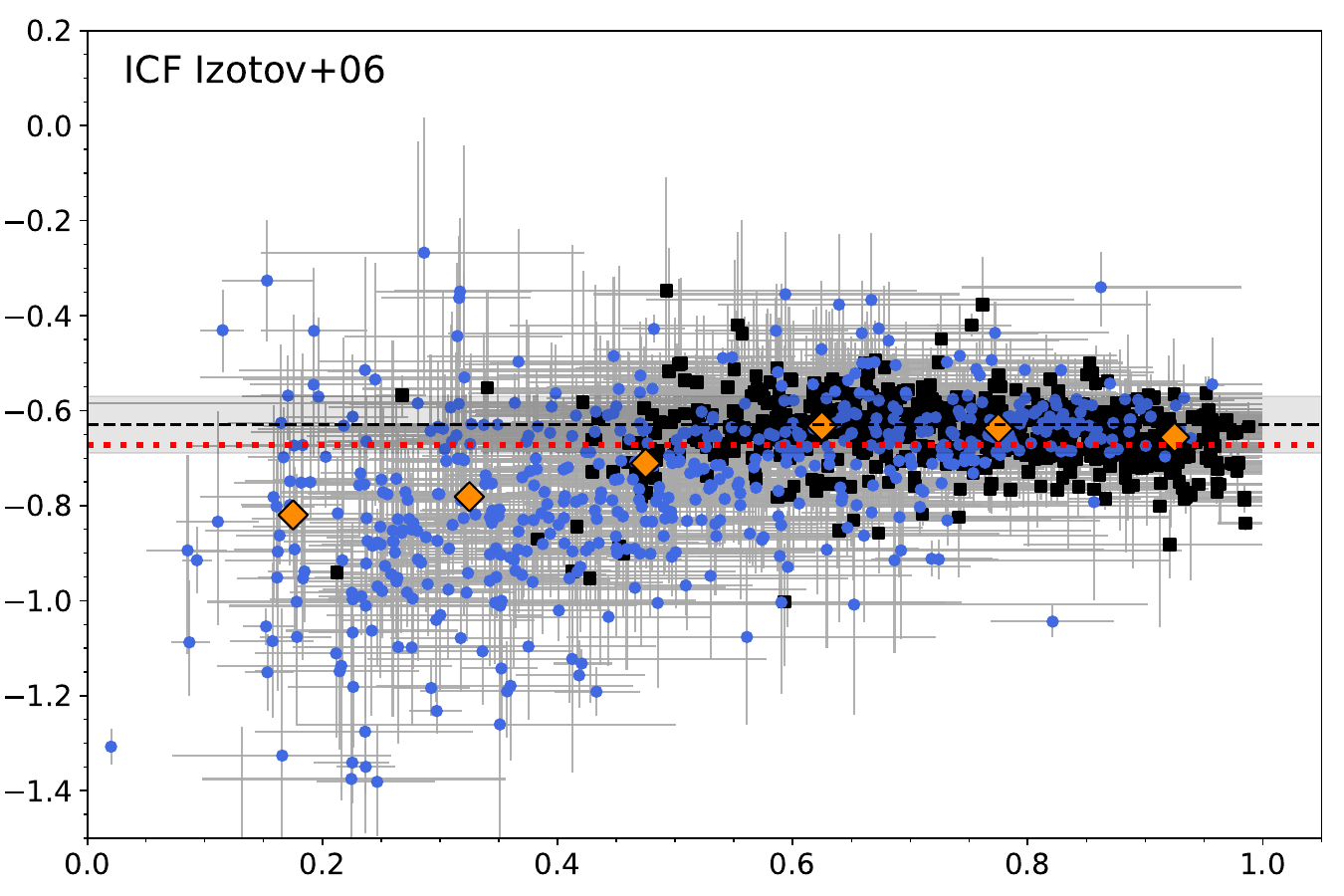}
\\
\includegraphics[scale=0.38]{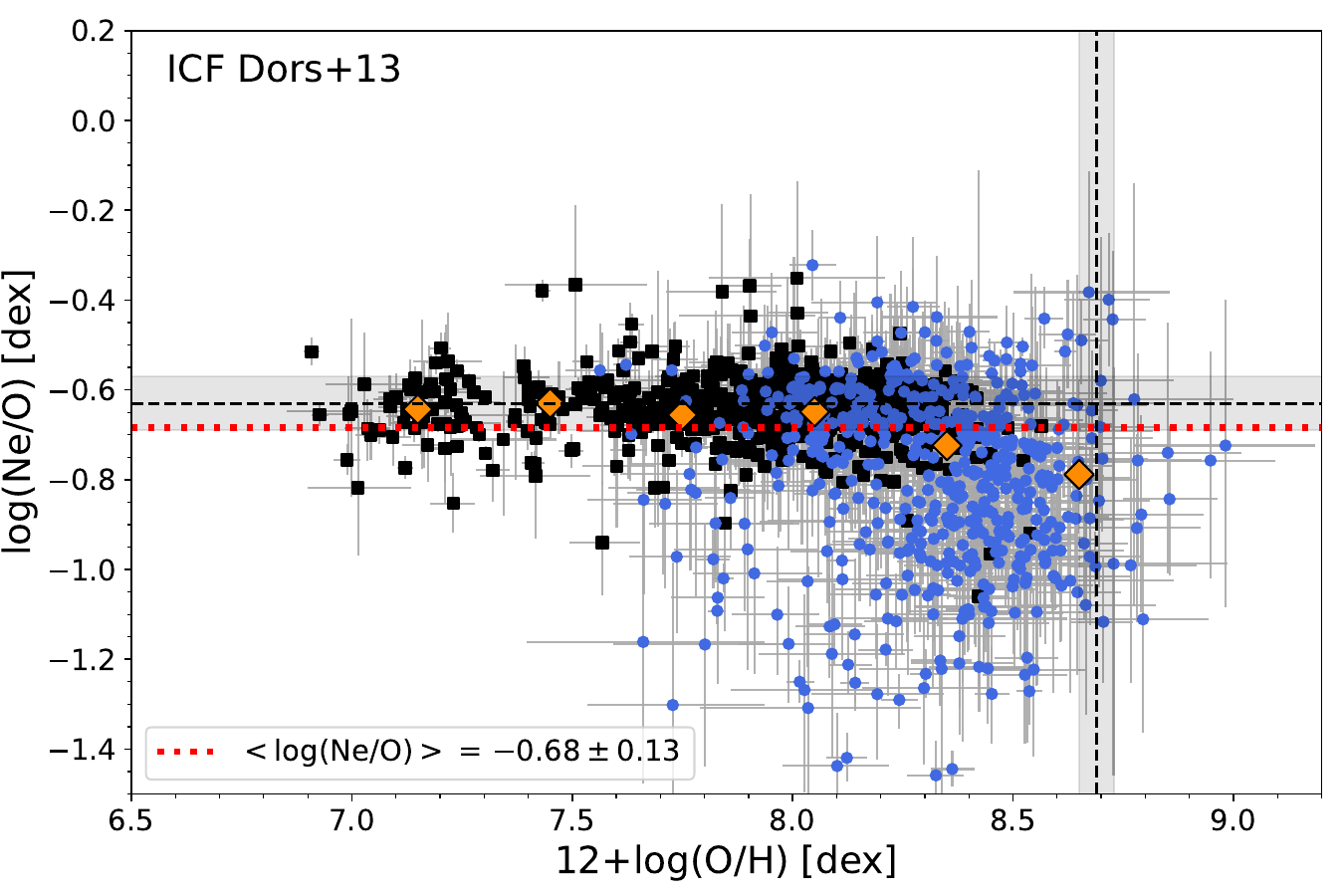}
\includegraphics[scale=0.38]{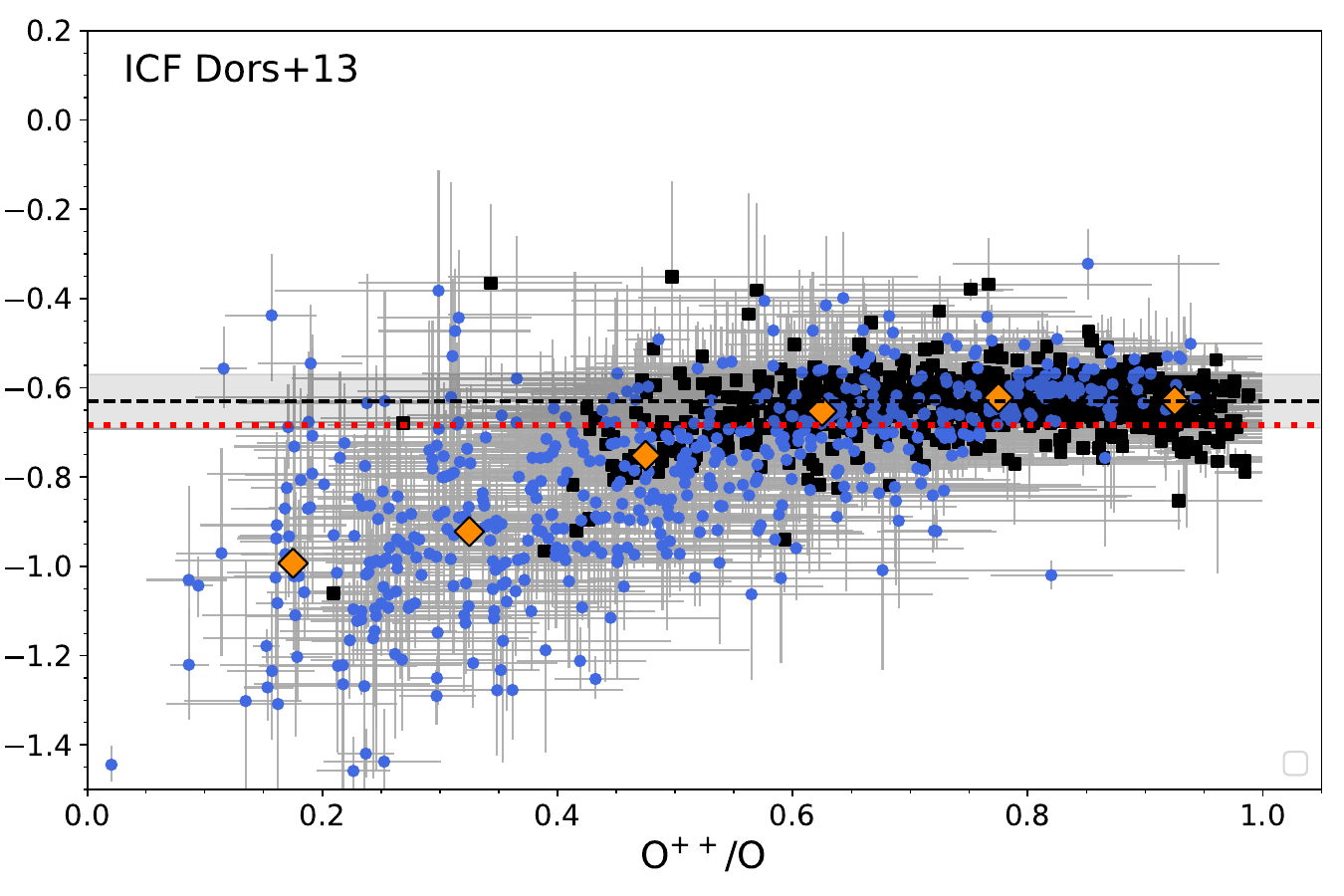}
\caption{log(Ne/O) as a function of 12+log(O/H) (left) and the ionisation degree, O$^{2+}$/O (right), for the DESIRED-E sample of \hii\ regions (blue circles) and SFGs (black squares). Panels in the same row represent log(Ne/O) calculated using the ICF(Ne) scheme of 
\citet[][top panels]{Amayo:21}, \citet[][middle panels]{Izotov:06} and \citet[][bottom panels]{Dors:13}. The dotted red line represents the mean value of the log(Ne/O) obtained using each ICF(Ne). The orange diamonds indicate the mean log(Ne/O) values considering bins in 12+log(O/H) or O$^{2+}$/O. The black dashed lines and the grey bands show the solar 12+log(O/H) and log(Ne/O) and their associated uncertainties, respectively, from \citet{Asplund:21}.} 
\label{fig:ICFNe_analysis}
\end{figure*}

Ne$^{2+}$ is the only ionisation state of Ne that presents emission lines in the optical range of the spectrum. UV photons with energies between 41.0 and 63.5 eV can produce Ne$^{2+}$, therefore, we expect an important fraction of unobservable Ne$^+$ to be present in typical star-forming ionised nebulae. To obtain the Ne/O ratio we apply the relation: 
\begin{equation}
\label{eq:ICFNe}
\frac{\text{Ne}}{\text{O}} = \frac{\text{Ne}^{2+}}{\text{O}^{2+}} \times \text{ICF(Ne)}. 
\end{equation}

As indicated previously, we use three different ICF(Ne)\footnote{Rigorously, in the form we define our ICFs, they should be expressed as ICF(Ne$^{2+}$/O$^{2+}$) for Ne, ICF($\frac{{\rm S}^+ + {\rm S}^{2+}}{\rm O}$) for S, and ICF(Ar$^{2+}$/O$^{2+}$) or ICF($\frac{{\rm Ar}^{2+} + {\rm Ar}^{3+}}{{\rm O}^{2+}}$) for the two different cases used for Ar. To make the text easier to read, we decided to abbreviate them simply as ICF(Ne), ICF(S) and ICF(Ar).} schemes to derive the total Ne abundance. The set by \citet{Amayo:21} is based on a grid of photoionisation models from the Mexican Million Models data base \citep[3MdB][]{Morisset:09} under the ‘BOND’ reference \citep{ValeAsari:16}. This grid was selected by applying several filters to resemble the properties of a large observational sample of extragalactic H II regions and the integrated spectra of SFGs. For their ionisation models, \citet{Amayo:21} adopt stellar spectral energy distributions (SEDs) obtained with the PopStar code \citep{Molla:09}. All the ICFs of \citet{Amayo:21} used in this study depend on the mean ionisation degree of the nebulae --parameterized by the O$^{2+}$/O ratio-- and are, in principle, valid for the whole range of ionisation degrees. The second ICF(Ne) scheme we have used is that by \citet{Izotov:06}, who use the grid of photoionisation models of SFGs presented by \citet{Stasinska:03} and input stellar SED models from Starburst99 \citep{Leitherer:99}. The ICFs by \citet{Izotov:06} depend on the ionisation degree and the metallicity. They propose different equations for 'low' (12+log(O/H) $\leq$ 7.2), 'intermediate' (7.2 $<$ 12+log(O/H) $<$ 8.2) and 'high' (12+log(O/H) $\geq$ 8.2)  metallicity that are used to interpolate the appropriate value of the ICF using the O/H ratio of the object. In the case of the ICF(Ne), the relations given by \citet{Izotov:06} are valid for any value of the O$^{2+}$/O ratio. The last ICF(Ne) scheme considered in this study is that by \citet{Dors:13}, which is based on photoionisation models of extragalactic {\hii} regions and SFGs that use ionising SEDs from  Starburst99 \citep{Leitherer:99}. They provide an analytical expression (their equation 14, applicable for data obtained from optical spectra) parameterized by the O$^{2+}$/O ratio and valid for all the range of ionisation degrees. 

In the top panel of Fig.~\ref{fig:ionic_ratios_NeS}, we plot  log(Ne$^{2+}$/O$^{2+}$) as a function of  O$^{2+}$/O for 557 spectra of {\hii} regions and 671 of SFGs of the DESIRED-E sample (1228 in total). The coloured curves represent the different ICF(Ne) values defined by Eq.~\ref{eq:ICFNe} provided by each of the ICF schemes considered: \citet[][red dashed-dotted lines]{Amayo:21}, \citet[][dotted lines with colours corresponding to the three metallicity ranges]{Izotov:06} and \citet[][green dashed line]{Dors:13}. The horizontal black dashed line indicates the solar log(Ne/O) of $-0.63 \pm 0.06$ --along as its uncertainty represented by the grey band-- as reference \citet{Asplund:21}, this line also represents the classical ICF of \citet{Peimbert:69}, that assumes Ne$^{2+}$/O$^{2+}$ = Ne/O. Fig.~\ref{fig:ionic_ratios_NeS} clearly indicates that the dispersion of the log(Ne$^{2+}$/O$^{2+}$) values of objects classified as {\hii} regions is much larger than of those classified as SFGs. This observational fact was reported by \citet{Kennicutt:03} and \citet{Croxall:16}, among others. Fig.~\ref{fig:ionic_ratios_NeS} also shows that the ICF(Ne) by \citet{Amayo:21} seems to better reproduce the behaviour of log(Ne$^{2+}$/O$^{2+}$) for objects with O$^{2+}$/O $<$ 0.5 (mostly {\hii} regions), but giving fairly higher values of log(Ne/O) --between 0.1 and 0.3 dex higher-- than the ICF(Ne) schemes by \citet{Izotov:06} and \citet{Dors:13} in the whole ionisation range. This can be noted very clearly in the different distribution of the blue and magenta points in the top left panel of Fig.~\ref{fig:comparion_ICFs}. The Ne abundances obtained using the different ICFs can differ by up to 0.3 dex when O$^{2+}$/O $<$ 0.5, although this difference decreases at higher ionisation degrees, reaching values below 0.1 dex when O$^{2+}$/O $>$ 0.8.

In Fig.~\ref{fig:ICFNe_analysis} we show the log(Ne/O) values obtained with the three ICF(Ne) schemes we use to derive the total Ne abundance for our DESIRED-E objects. We represent log(Ne/O) as a function of 12+log(O/H) (left panels) and O$^{2+}$/O (right panels). We can derive the Ne abundances for a total number of 1228 spectra, 45.4\% of them correspond to {\hii} regions and 54.6\% to SFGs. In the figure, we can note that in most panels, the  distribution of the points show systematic trends that move  away from a line of constant value, which is in contradiction with be expected behaviour for two elements with similar nucleosynthetic origin as it is thought to be the case of Ne and O. To better illustrate the behaviour of log(Ne/O), we have represented its mean value for several bins (orange diamonds) in the areas of the figures having a significant number of points (more than 10), for values of 12+log(O/H) from 7.0 to 9.1 (bins 0.30 dex wide) and for O$^{2+}$/O values from 0.1 to 1.0 (bins 0.15 wide). 

The two top panels of Fig.~\ref{fig:ICFNe_analysis} show the log(Ne/O) values obtained with the ICF(Ne) by \citet{Amayo:21}. The upper left panel shows a rather slight increase of log(Ne/O) with respect to the O abundance. In fact, the bin corresponding to the highest 12+log(O/H) values has a mean log(Ne/O) 0.05 dex higher than the bin containing the objects with lowest O abundances. On the other hand, in the upper right panel, the data points are distributed following a curved distribution with a convex shape (looking from positive ordinates), with an amplitude of variation of log(Ne/O) of about 0.17 dex along the whole O$^{2+}$/O range. The ICF(Ne) of \citet{Amayo:21} tends to give oversolar log(Ne/O) values, in fact the mean log(Ne/O) obtained from all the data points is $-$0.55 $\pm$ 0.12, larger than the solar value of $-$0.63 $\pm$ 0.06 \citep{Asplund:21}, although still consistent within the errors. 

In the two middle panels of Fig.~\ref{fig:ICFNe_analysis}, we represent the log(Ne/O) values obtained using the ICF(Ne) scheme by \citet{Izotov:06}. It can be seen that the log(Ne/O) vs. 12+log(O/H) distribution is somewhat flatter in this case, log(Ne/O) tends to decrease in about 0.03 dex from lowest to highest metallicities. However, it is clear that the ICF(Ne) scheme by \citet{Izotov:06} does not correct appropriately for the presence of Ne$^+$ in objects with O$^{2+}$/O < 0.5. The difference between the bins at the extremes is $\sim$0.19 dex. The ICF(Ne) by \citet{Dors:13} shows a similar trend in the log(Ne/O) vs. O$^{2+}$/O panel, but with a more pronounced difference between the bins at the extremes ($\sim$0.37 dex). The bottom left panel of Fig.~\ref{fig:ICFNe_analysis} illustrates that the ICF(Ne) by \citet{Dors:13} produces a decrease of log(Ne/O) as the metallicity increases, with a difference of about 0.15 dex between the bins at the extremes of the 12+log(O/H) range. The ICF(Ne) by \citet{Izotov:06} and \citet{Dors:13} give mean values of the Ne/O ratio of $-$0.67 $\pm$ 0.12 and $-$0.68 $\pm$ 0.13, respectively, below but quite in agreement with the solar value of $-$0.63 $\pm$ 0.06 \citep{Asplund:21}. 

Fig.~\ref{fig:ICFNe_analysis} illustrates that all the ICF(Ne) schemes considered fail to give correct Ne/O ratios for nebulae having O$^{2+}$/O $\leq$ 0.5, that would correspond mainly to intermediate and high-metallicity objects. The vast majority  of the objects of the DESIRED-E sample in that part of the log(Ne/O) vs. O$^{2+}$/O diagram correspond to {\hii} regions. In fact, as this is an expected result according to the behaviour of the ICF(Ne) curves against the observed distribution of the Ne$^{2+}$/O$^{2+}$ ratios of the DESIRED-E objects shown in the upper panel of Fig.~\ref{fig:ionic_ratios_NeS}. The total O and Ne abundances determined using the different ICF schemes considered --as well as the total abundances of S and Ar-- are collected in Table~D5.

\begin{figure*}[ht!]
\centering    
\includegraphics[scale=0.38]{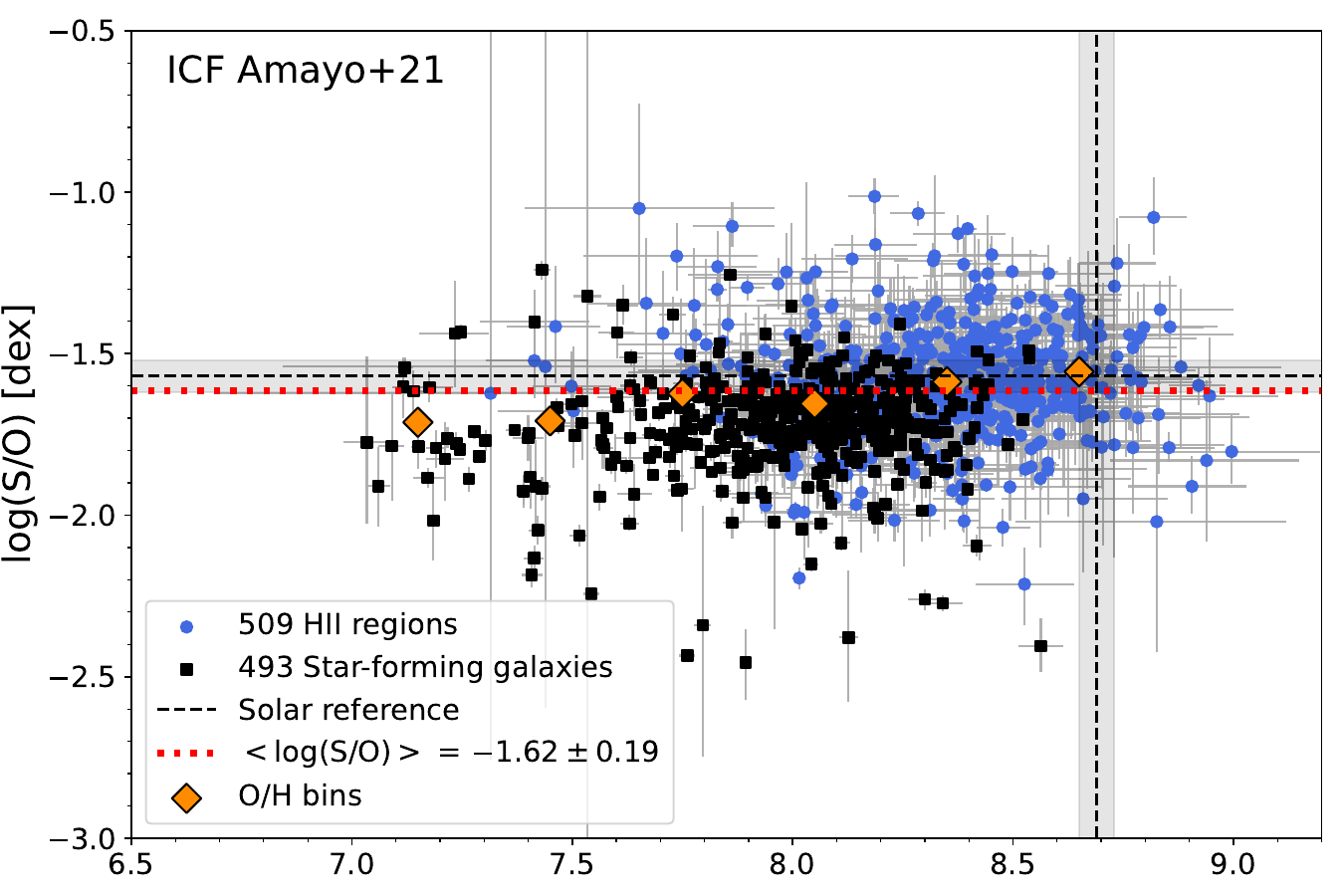}
\includegraphics[scale=0.38]{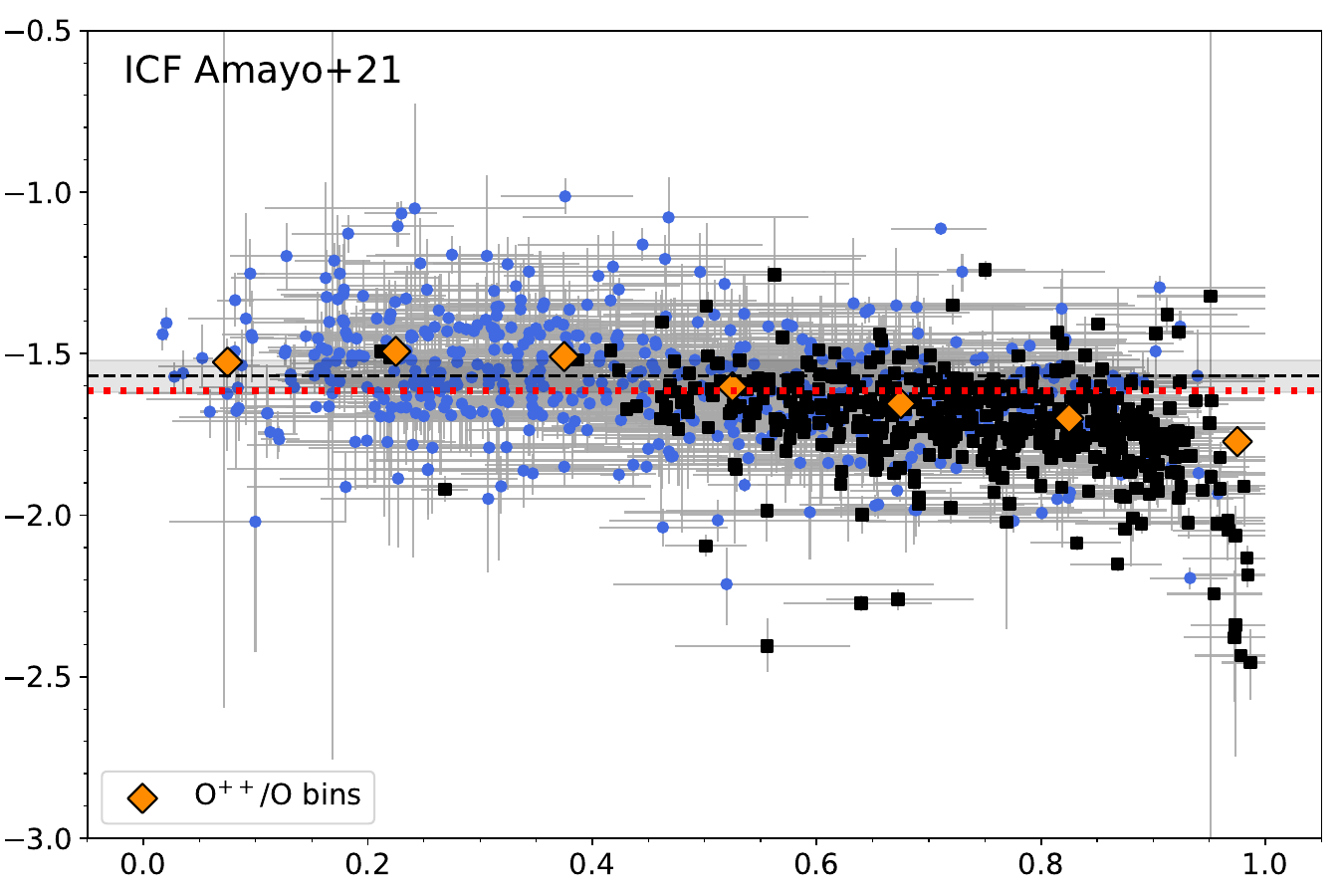}
\\
\includegraphics[scale=0.38]{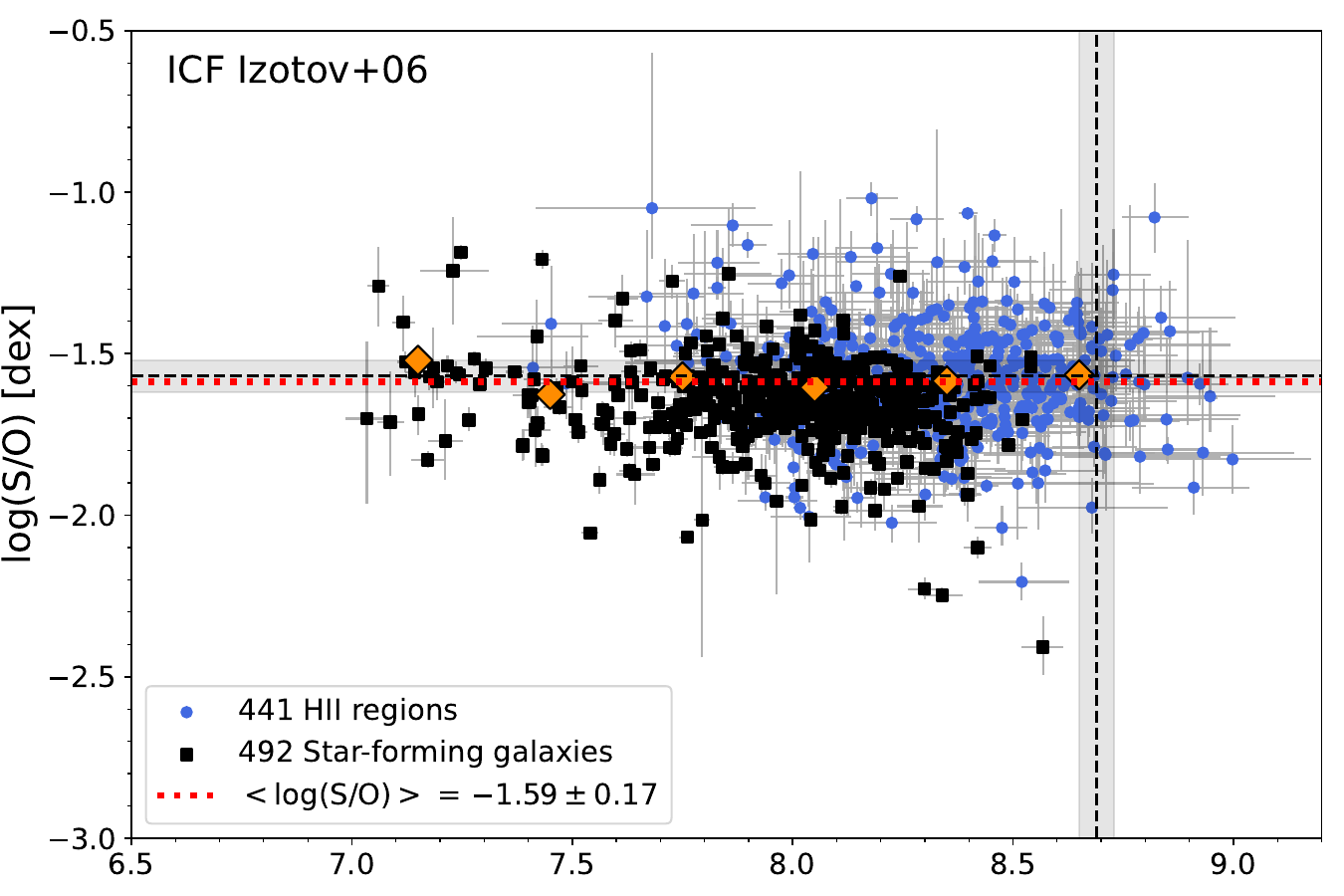}
\includegraphics[scale=0.38]{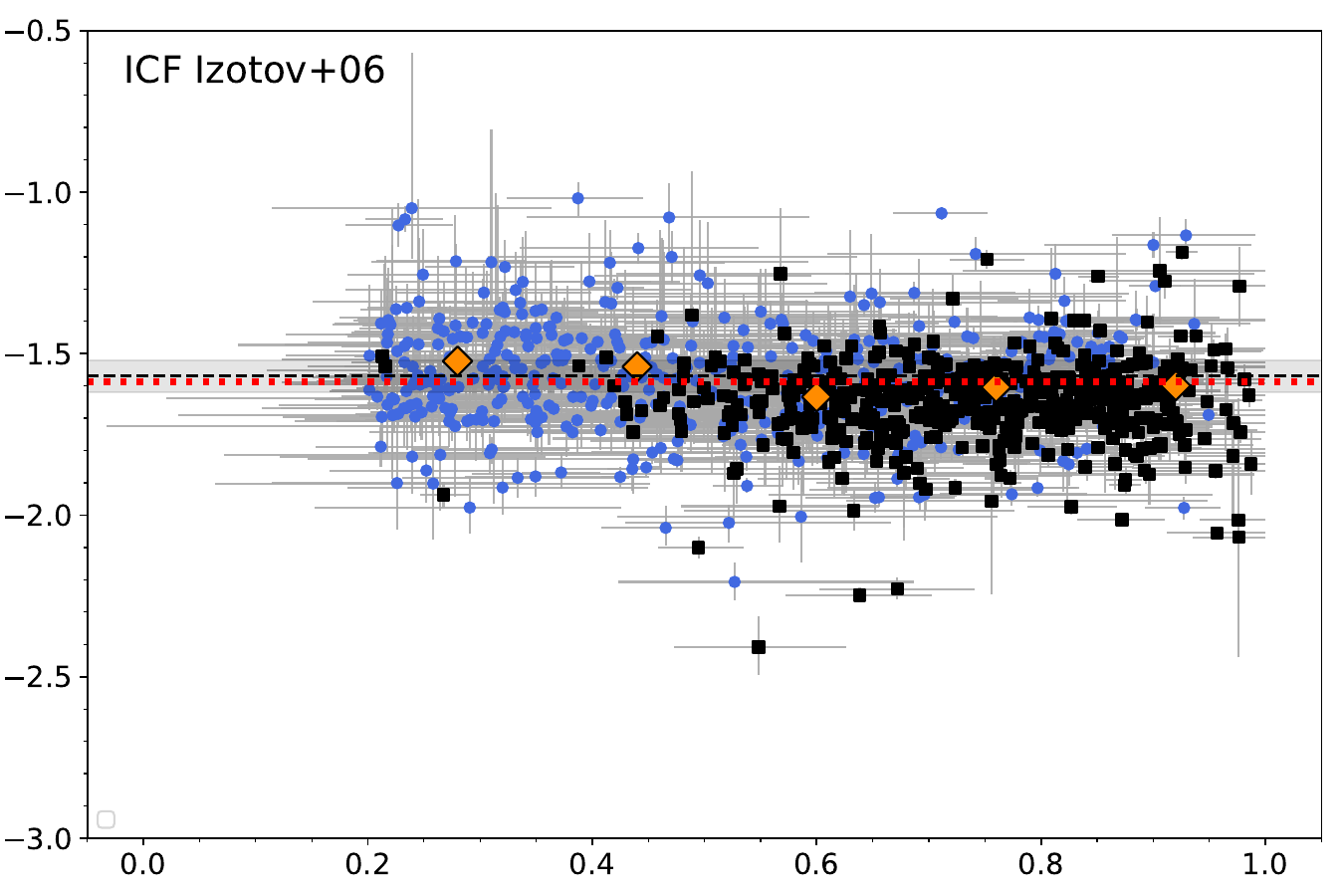}
\\
\includegraphics[scale=0.38]{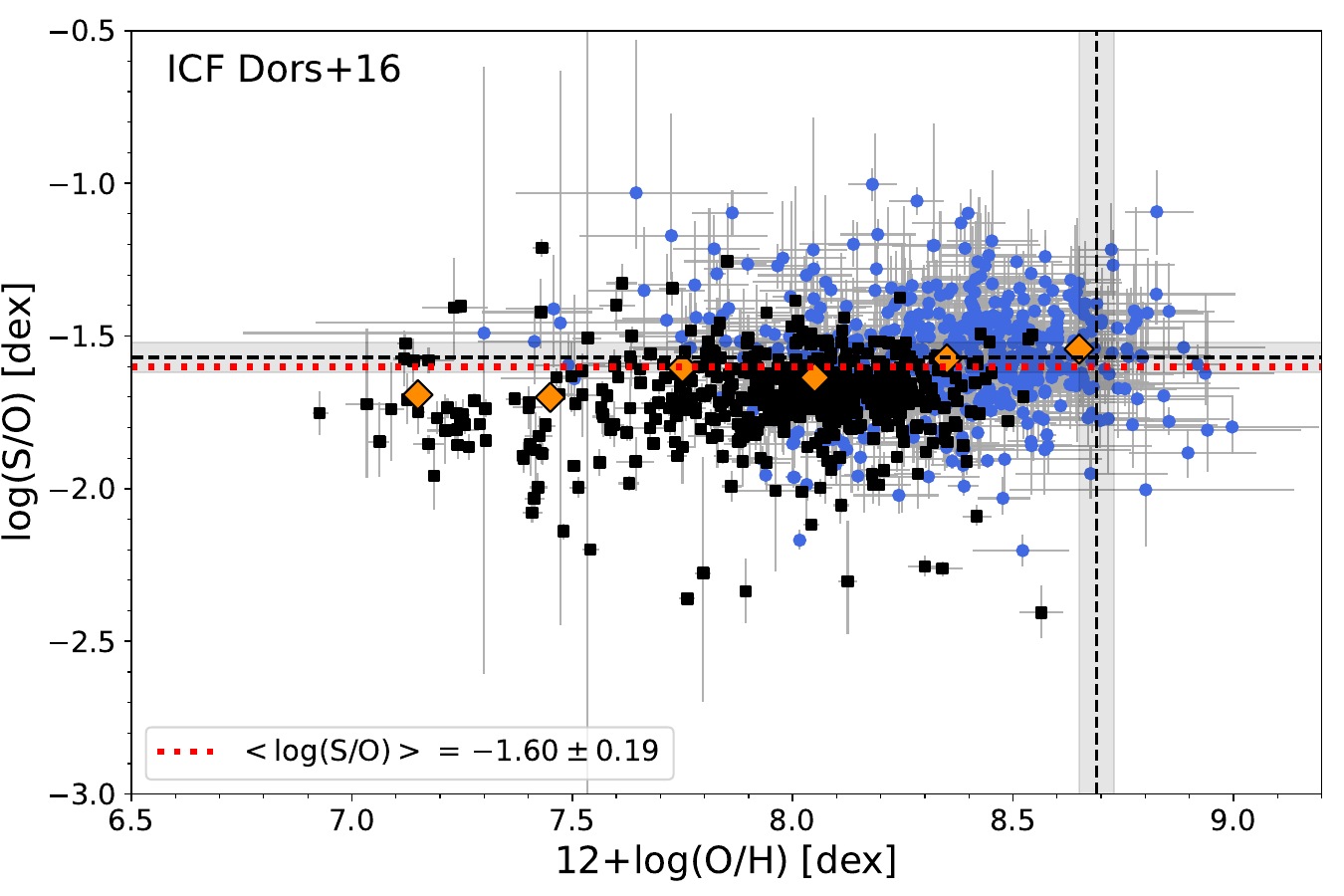}
\includegraphics[scale=0.38]{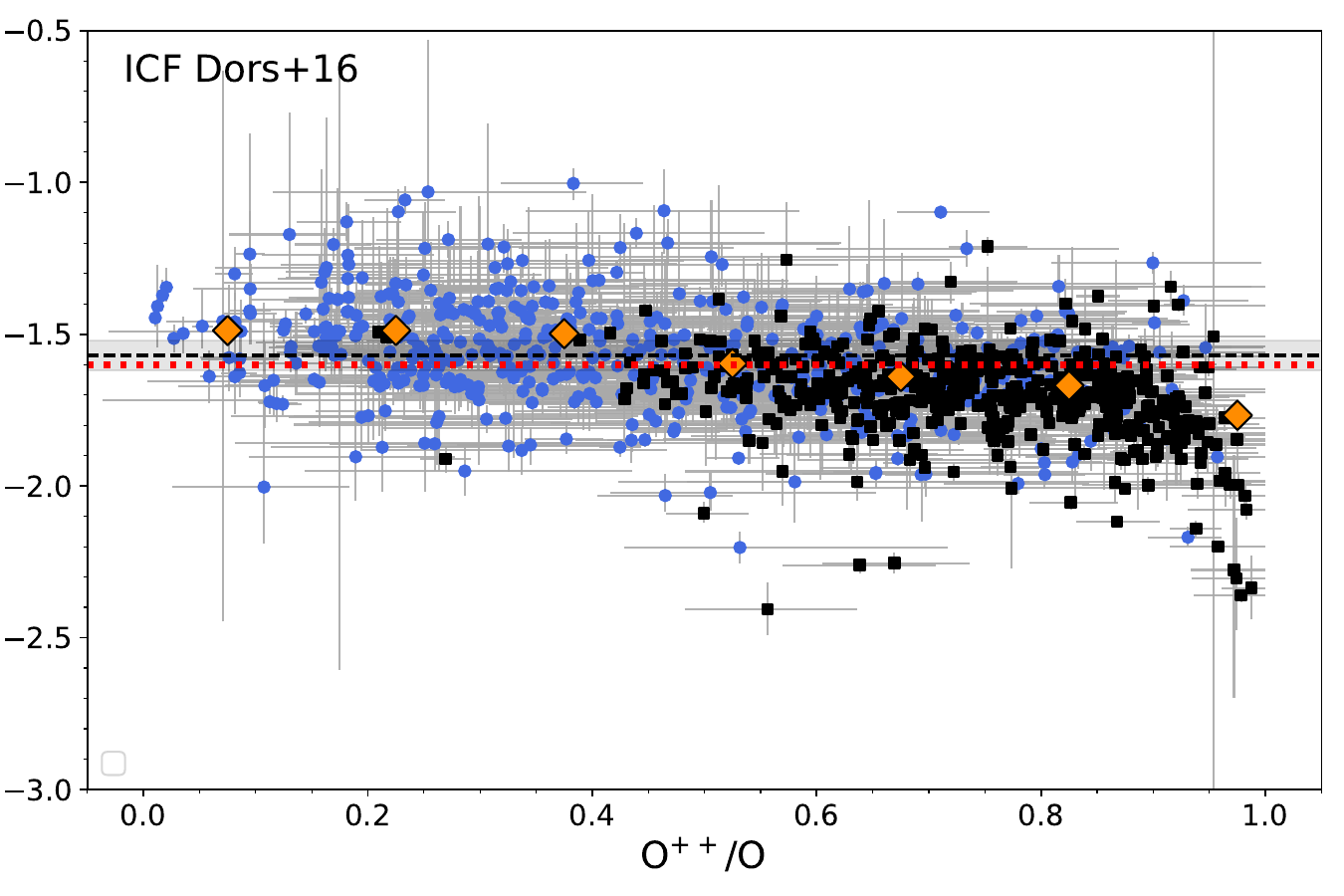}
\caption{Log(S/O) as a function of 12+log(O/H) (left) and the ionisation degree, O$^{2+}$/O (right), for \hii\ regions or SFGs of the DESIRED-E sample having both S$^+$ and S$^{2+}$ abundances. Panels in the same row of the figure represent log(S/O) ratios calculated using the ICF(S) scheme by 
\citet[][top panels]{Amayo:21}, \citet[][middle panels]{Izotov:06} and \citet[][bottom panels]{Dors:16}. The dotted red line represents the mean value of the log(S/O) obtained using each ICF(S). The orange diamonds indicate the mean log(S/O) values considering bins in 12+log(O/H) or O$^{2+}$/O. The black dashed lines and the grey bands show the solar 12+log(O/H) and log(S/O) ratios and their associated uncertainties, respectively, from \citet{Asplund:21}.} 
\label{fig:ICFS_analysis}
\end{figure*}

\subsection{Sulfur}
\label{subsec:sulfur}

S presents bright emission lines of S$^+$ and S$^{2+}$ in the optical spectra of ionised nebulae. S$^+$ can be ionised by photons with energies between 10.4 to 22.3 eV and S$^{2+}$ from 22.3 to 34.8 eV. Therefore, we expect a certain fraction of S$^{3+}$ (34.8 $-$ 47.2 eV) to be present under the typical ionisation conditions of {\hii} regions and SFGs. In many nebular spectra of faint objects only {\sii} lines can be observed, so the only possibility to derive their total S/H is from the S$^+$ abundance. \citet{Amayo:21} obtain an expression for an ICF(S) based only on the intensity of {\sii} lines but they find large differences and uncertainties with respect to the values obtained using {\sii} and {\siii} lines. Therefore, \citet{Amayo:21} do not recommend to use an ICF(S) based on S$^+$ alone. Considering this, we only obtain the S/O ratio for those objects for which we can derive both, S$^+$ and S$^{2+}$ abundances. We apply the relation (see footnote in Sect.~\ref{subsec:neon}):

\begin{equation}
\label{eq:ICFS}
\frac{\text{S}}{\text{O}} = \left(\frac{\text{S}^{+}+\text{S}^{2+}}{\text{O}}\right) \times \text{ICF(S)}. 
\end{equation}

We use three different ICF(S) schemes to derive the S abundance. The first two are the same that we use in the case of Ne, the sets by \citet{Amayo:21} and \citet{Izotov:06}, described in Sect.~\ref{subsec:neon}. The third ICF(S) scheme is the one proposed by \citet{Dors:16}, based --as the ICF(Ne) by \citet{Dors:13} described in Sect.~\ref{subsec:neon}-- on photoionisation models for extragalactic \hii\ regions and SFGs using ionising SEDs from Starburst99. The values of the IFC(S) provided by the three schemes are parameterized by the O$^{2+}$/O ratio. The ICF(S) by \citet{Amayo:21} and \citet{Dors:16} are valid for all the range of ionisation degrees, but the one by \citet{Izotov:06} can only be applied for objects with O$^{2+}$/O $\geq$ 0.2.

In the bottom panel of Fig.~\ref{fig:ionic_ratios_NeS}, we plot  log((S$^{+}+$S$^{2+}$)/O) as a function of the ionisation degree, O$^{2+}$/O, for 509 {\hii} regions and 493 SFGs of the DESIRED-E sample with measurements of both, {\sii} and {\siii} lines. The coloured curves represent the behaviour of the different ICF(S) schemes considered: \citet[][red dashed-dotted lines]{Amayo:21}, \citet[][dotted lines with colours corresponding to the three metallicity ranges]{Izotov:06} and \citet[][green dashed line]{Dors:16}. The horizontal black dashed line shows the value of the solar log(S/O) ratio --along as its uncertainty represented by the grey band-- recommended by \citet{Asplund:21}.  Fig.~\ref{fig:ionic_ratios_NeS} shows that from values of O$^{2+}$/O larger than about 0.5, the contribution of S$^{3+}$ in the nebulae (with an IP of 34.8 eV) becomes important. In the figure, we can see that only the ICF(S) scheme by \citet{Izotov:06} reproduces the observed log((S$^{+}+$S$^{2+}$)/O) values for very high ionisation degree objects (with O$^{2+}$/O $\geq$ 0.9), mostly corresponding to SFGs. 

In Fig.~\ref{fig:ICFS_analysis} we show the log(S/O) values obtained with the three ICF(S) schemes we use to derive the total S abundance for the DESIRED-E spectra. The exact number depends on the ICF(S) scheme used, 1002 with \citet{Amayo:21} or \citet{Dors:16} (50.8\% {\hii} regions and 49.2\% SFGs) and 933 with \citet[][47.2\% {\hii} regions and 52.8\% SFGs]{Izotov:06}. Similarly to  Fig.~\ref{fig:ICFNe_analysis} for log(Ne/O), in Fig.~\ref{fig:ICFS_analysis} we represent log(S/O) as a function of 12+log(O/H) and O$^{2+}$/O. In the right panels of Fig.~\ref{fig:ICFS_analysis}, we can note that the  distribution of the sample points using the ICF(S) by \citet{Amayo:21} and \citet{Dors:16} show a systematic tendency to give lower log(S/O) values as the ionisation degree of the objects increases. The mean log(S/O) corresponding to the different O$^{2+}$/O bins decreases almost monotonically 0.30 and 0.27 dex, respectively for the two aforementioned ICF(S) schemes along the whole range of O$^{2+}$/O ratios. This trend: (a) clearly indicates that the ICF(S) by \citet{Amayo:21} and \citet{Dors:16} do not completely correct for the ionisation dependence the conversion from the sum of the ionic abundances to the final S/H ratio; (b) can also explain, at least partially --due to the rough relation between O/H and O$^{2+}$/O (see Fig.~\ref{fig:Ovsion})--, the increase in log(S/O) as a function of 12+log(O/H) that can be seen in the top and bottom panels on the left columns of Fig.~\ref{fig:ICFS_analysis}. The 12+log(O/H) bins plotted in both panels show a systematic increase of +0.17 dex in log(S/O) as the metallicity increases.

The two panels of the middle row of Fig.~\ref{fig:ICFS_analysis} show the results using the ICF(S) by \citet{Izotov:06}, which clearly provides flatter trends than the other schemes. In fact, the maximum variation between the extreme values of the log(S/O) of the bins represented in both panels is only of the order of 0.1 dex. Another argument to claim that the ICF(S) by \citet{Izotov:06} is more appropriate is the shortage of objects with very low log(S/O) at O$^{2+}$/O $\geq$ 0.95,  which appears when the other two IFC(S) are applied and which is explained by the sharp drop in the curves at those high ionisation degrees that can be seen in Figs.~\ref{fig:ionic_ratios_NeS} and \ref{fig:comparion_ICFs}. The mean log(S/O) obtained with the different ICF(S) schemes are fairly similar: $-$1.62 $\pm$ 0.19 for \citet{Amayo:21}, $-$1.59 $\pm$ 0.17 for \citet{Izotov:06} and $-$1.60 $\pm$ 0.19 for \citet{Dors:16}. All of these values are consistent with the solar value of $-$1.57 $\pm$ 0.05 \citep{Asplund:21}. In Fig.~\ref{fig:comparion_ICFs} we can see that the three ICF(S) used give very similar log(S/H) values when the objects are in the interval 0.1$<$O$^{2+}$/O$<$0.5, in that case the differences are of the order or even smaller than 0.05 dex.

\begin{figure}[ht!]
\centering    
\includegraphics[width=\hsize ]{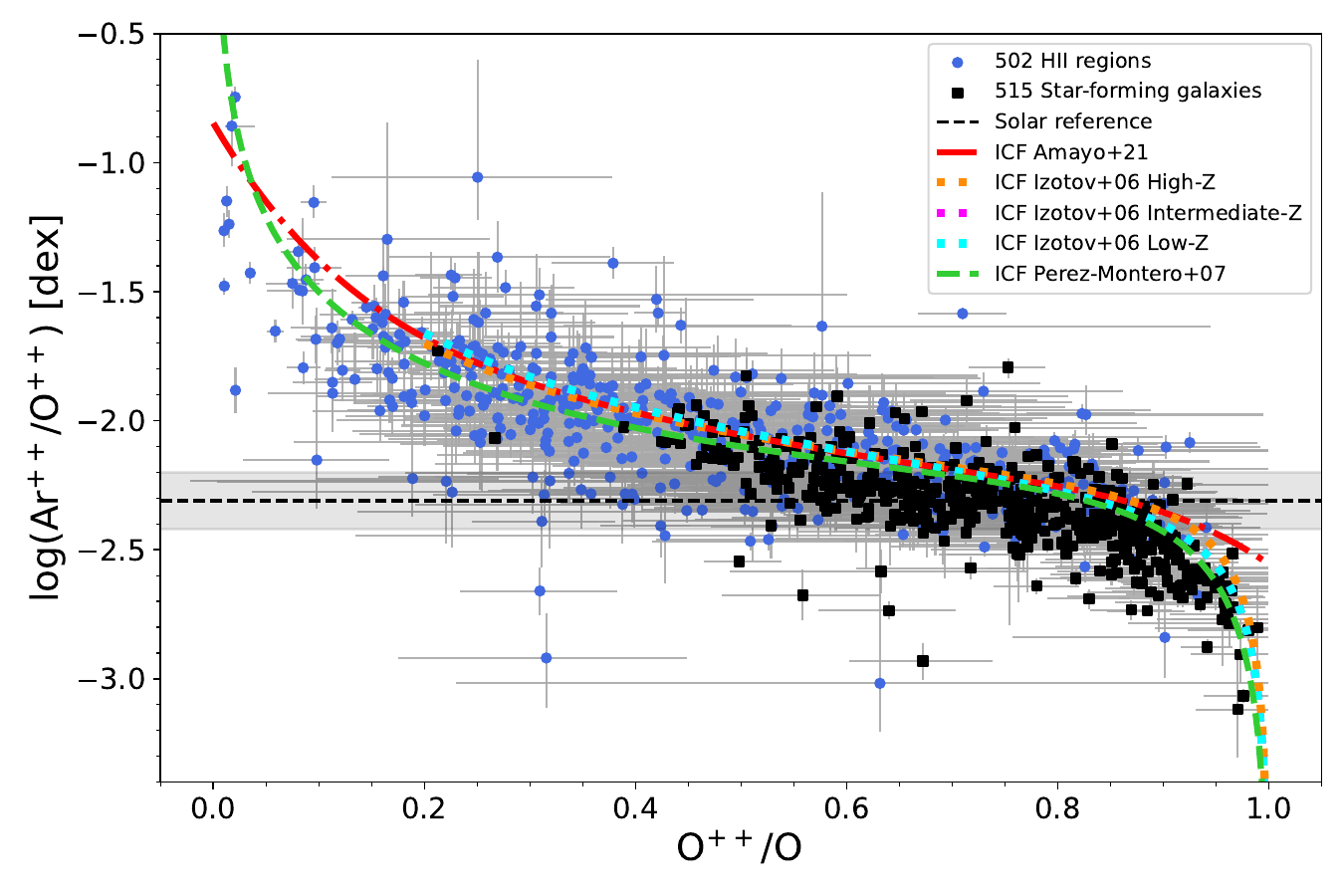}
\includegraphics[width=\hsize ]{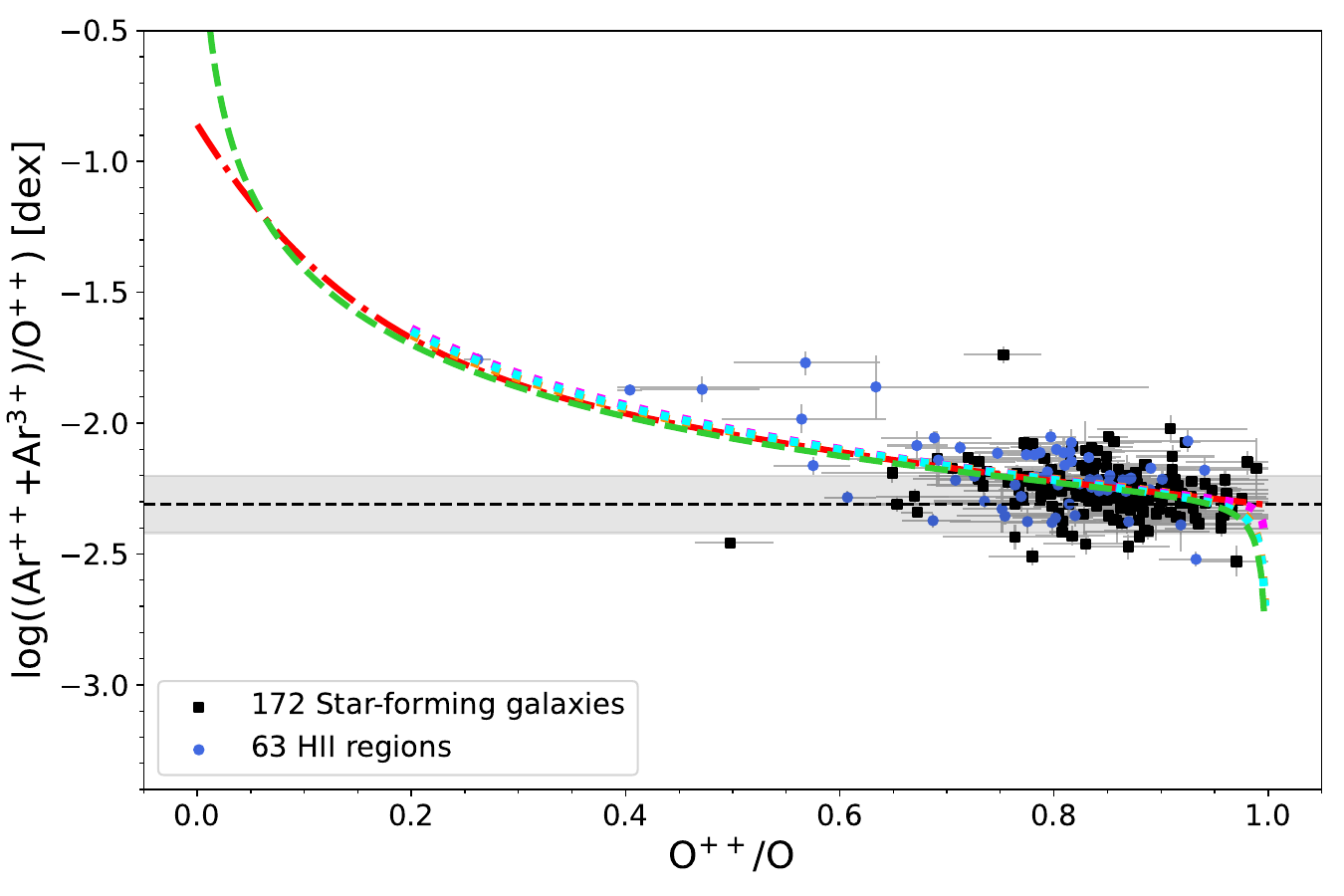}
\caption{Log(Ar$^{2+}$/O$^{2+}$) (top) and log((Ar$^{2+}$+Ar$^{3+}$)/O$^{2+}$) (bottom) as a function of the ionisation degree, O$^{2+}$/O, for DESIRED-E sample of \hii\ regions (blue circles) and SFGs (black squares). The black dashed lines and the grey bands show the solar log(Ar/O) value and their associated uncertainty, respectively, from \citet{Asplund:21}. The different curves represent the ICF(Ar) schemes used in this study: \citet[][red dashed-dotted lines]{Amayo:21}, \citet[][dotted lines with colours corresponding to three metallicity ranges]{Izotov:06} and \citet[][green dashed lines]{PerezMontero:07}.} 
\label{fig:ionic_ratios_Ar}
\end{figure}

\begin{figure*}[ht!]
\centering    
\includegraphics[scale=0.38]{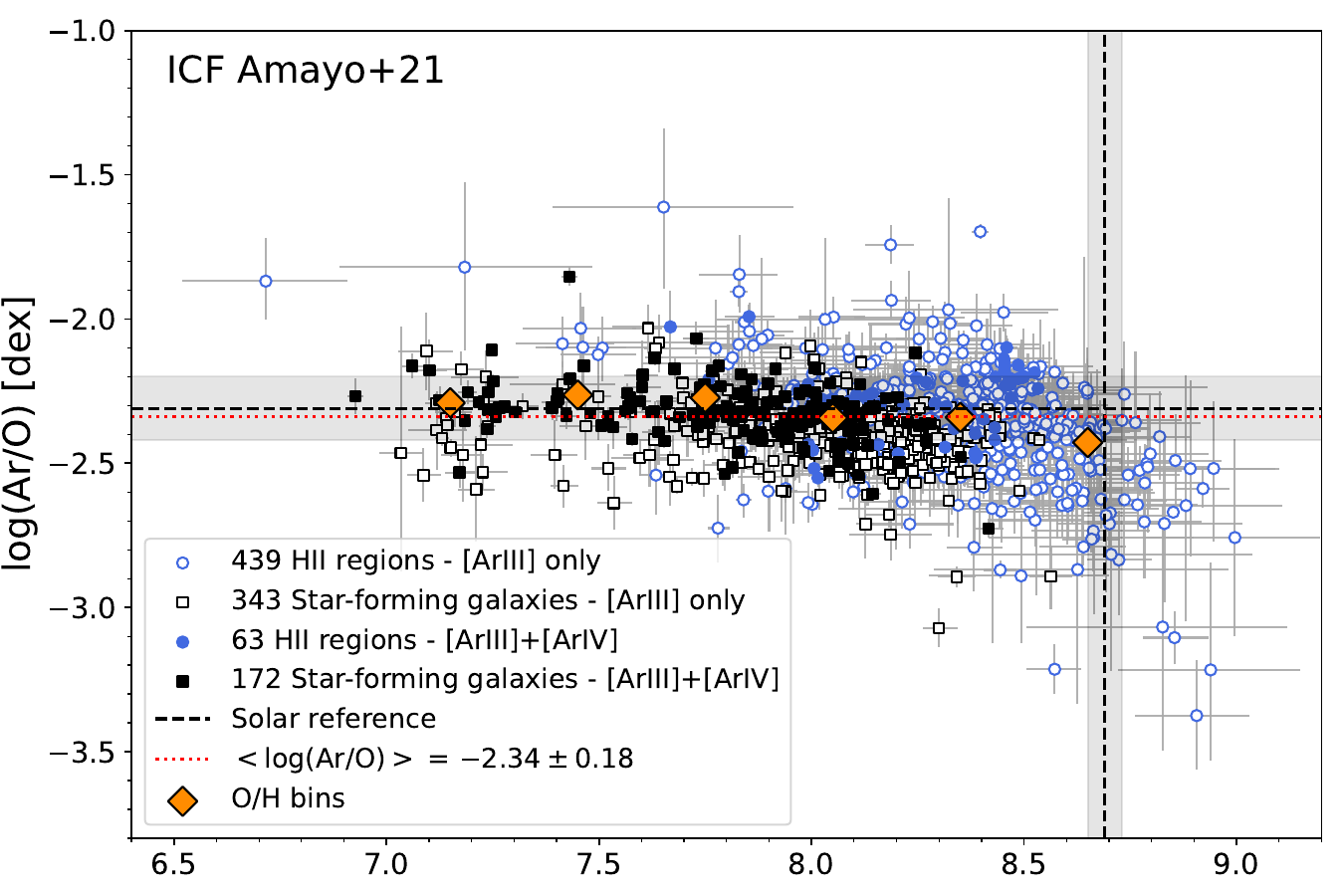}
\includegraphics[scale=0.38]{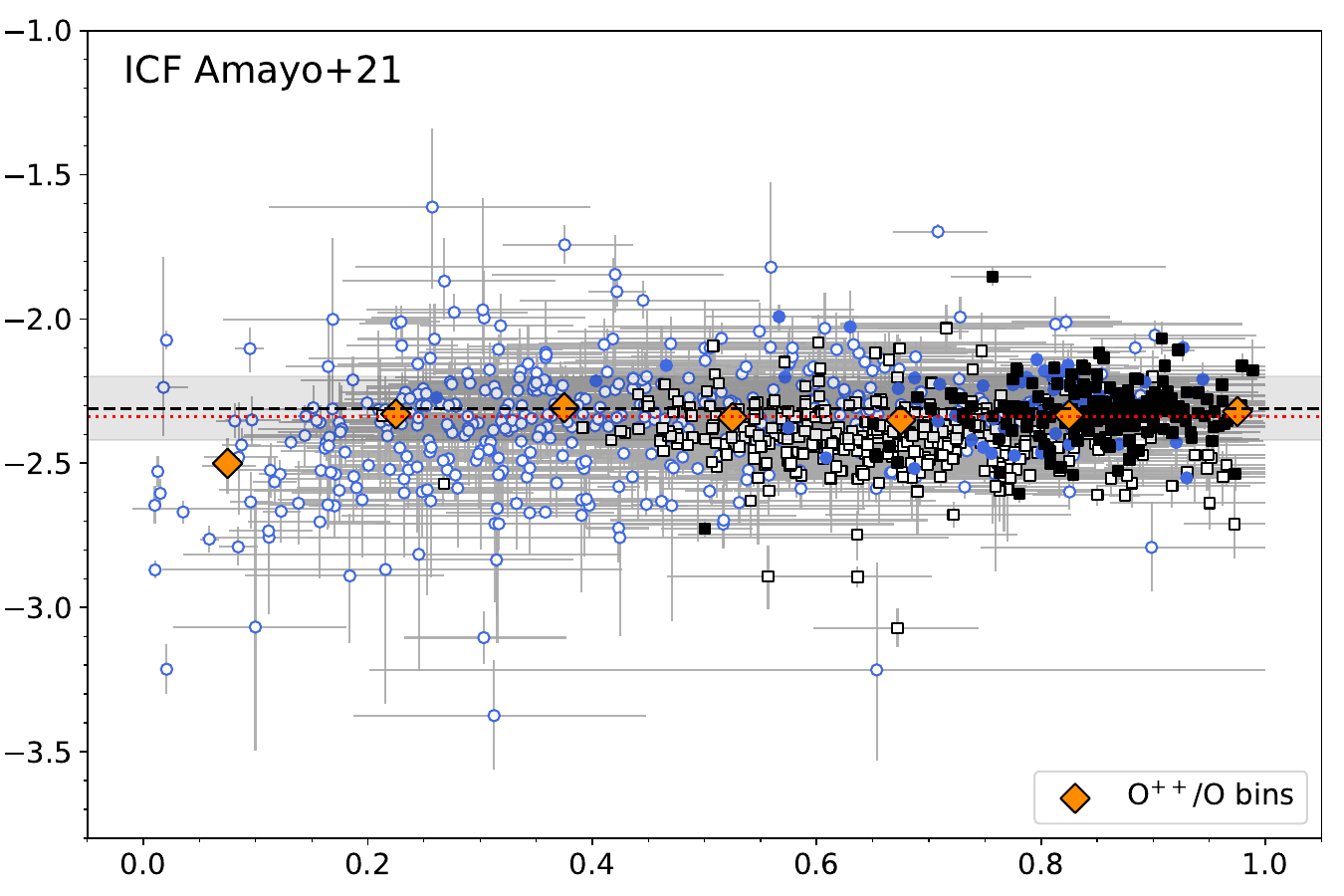}
\\
\includegraphics[scale=0.38]{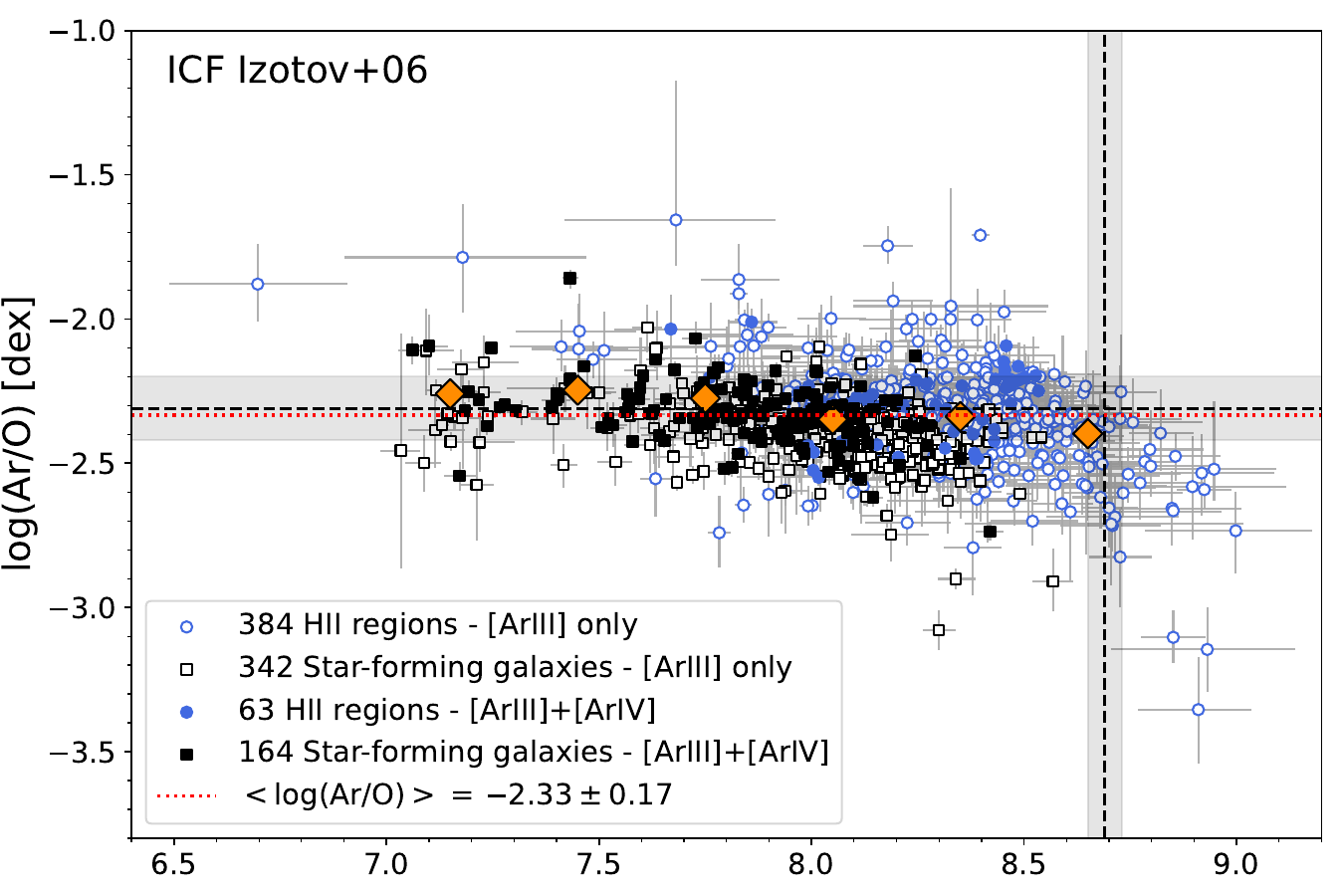}
\includegraphics[scale=0.38]{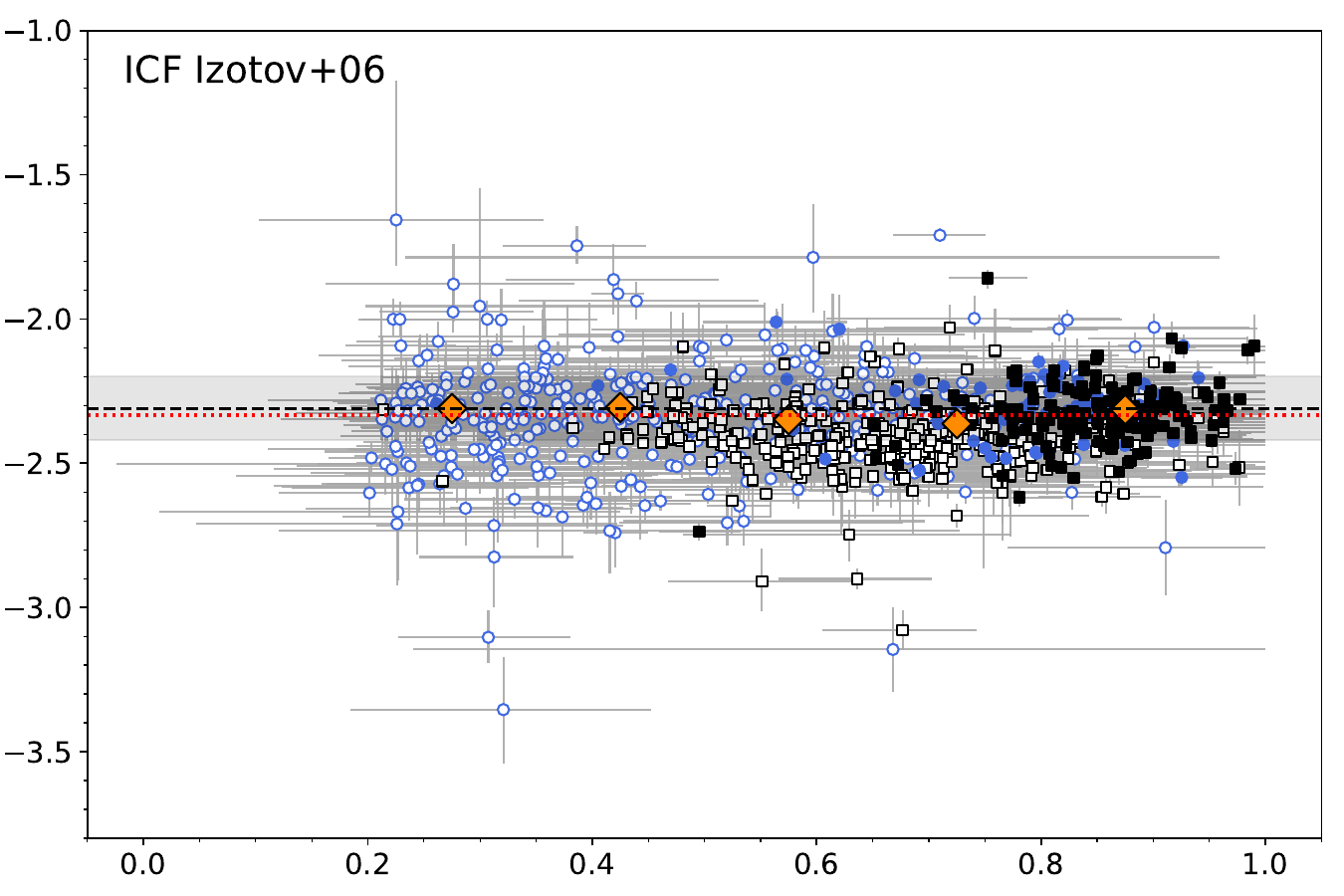}
\\
\includegraphics[scale=0.38]{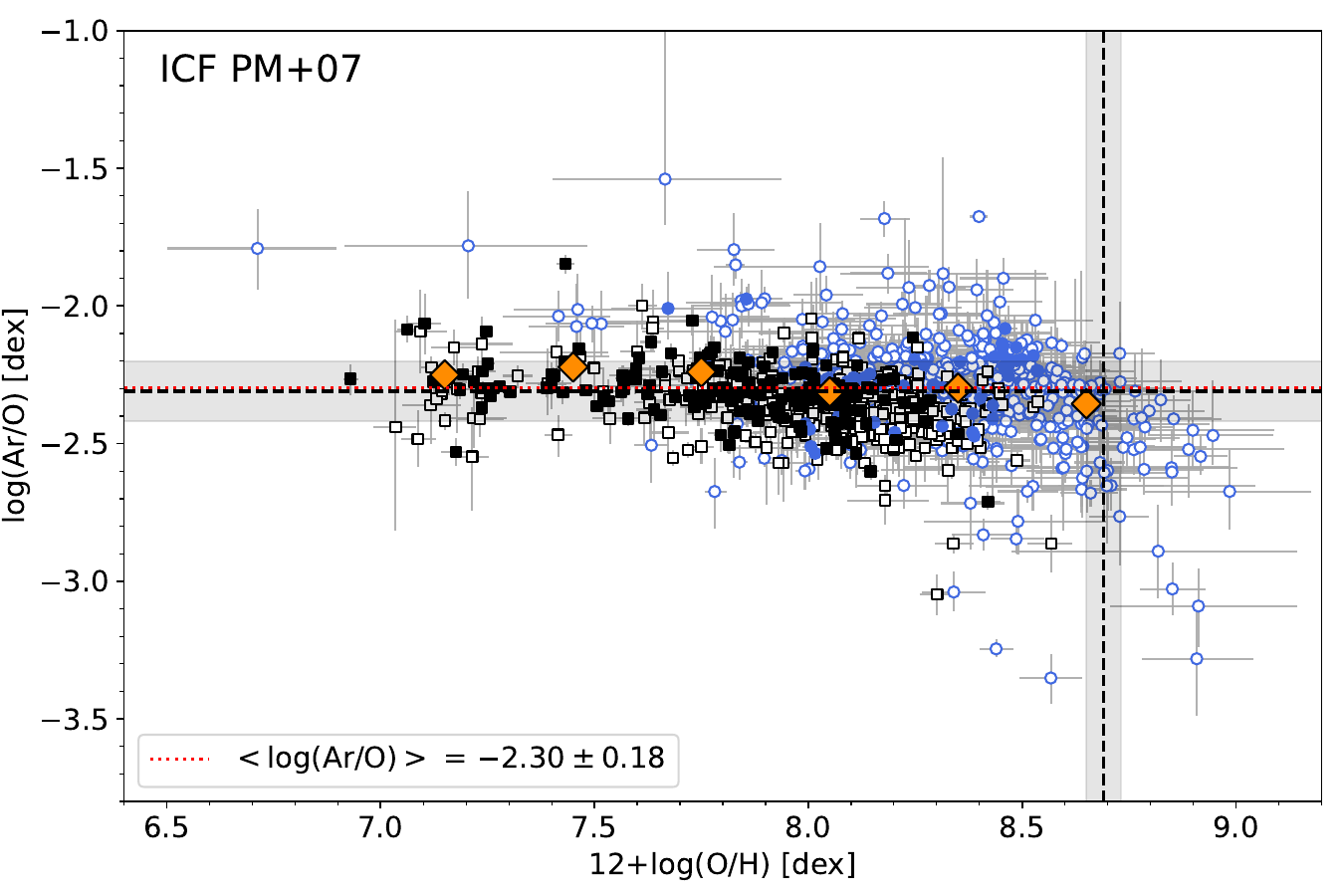}
\includegraphics[scale=0.38]{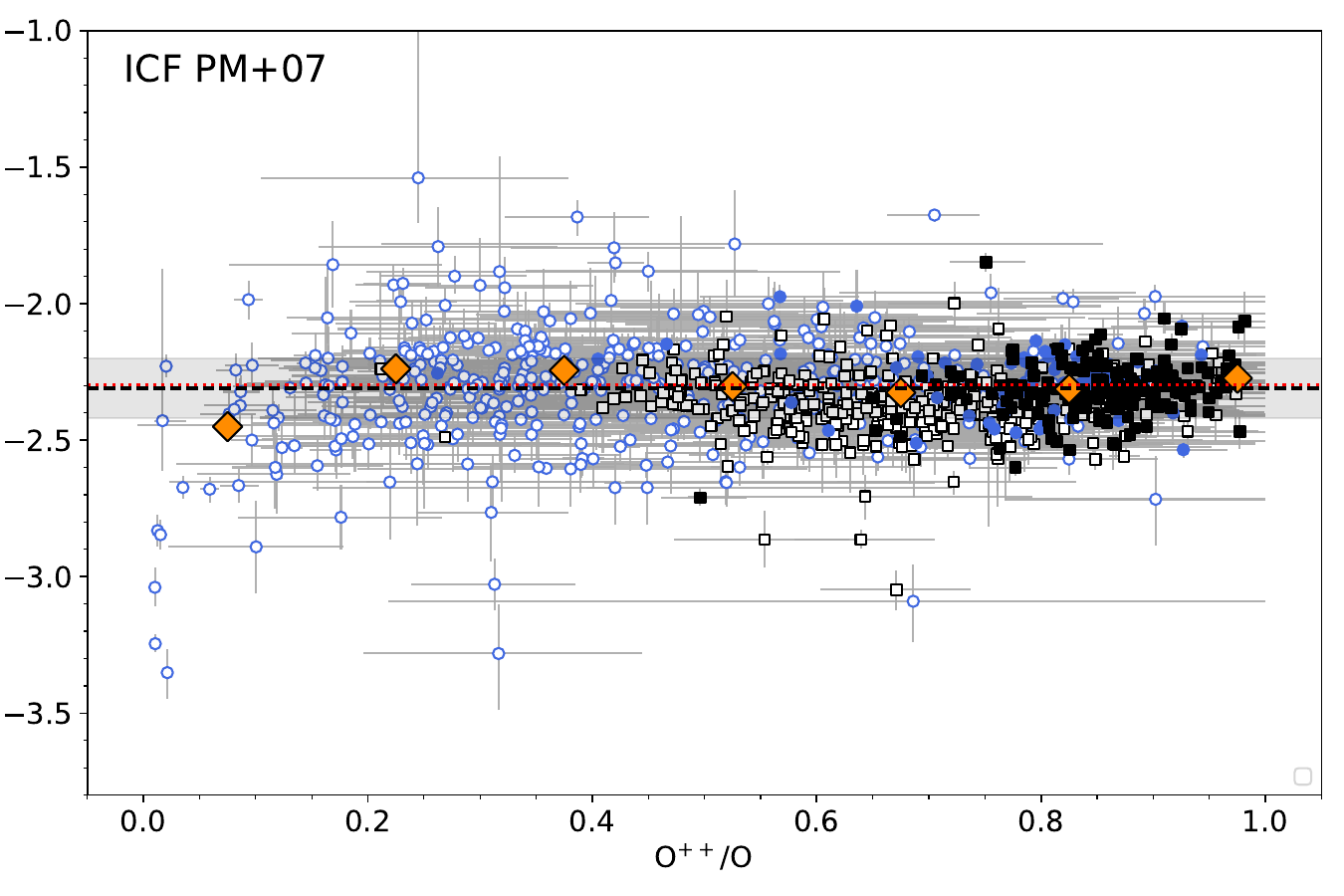}
\caption{Log(Ar/O) as a function of 12+log(O/H) (left) and the ionisation degree, O$^{2+}$/O (right), for the DESIRED-E \hii\ regions (blue circles) or SFGs (black squares) having only Ar$^{2+}$ (empty symbols) or Ar$^{2+}$ and Ar$^{3+}$ abundances (full symbols). Panels in the same row of the figure represent log(Ar/O) calculated using the ICF(Ar) scheme of 
\citet[][top panels]{Amayo:21}, \citet[][middle panels]{Izotov:06} and \citet[][bottom panels]{PerezMontero:07}. The dotted red line represents the mean value of log(Ar/O) obtained using each ICF(Ar). The orange diamonds indicate the mean log(Ar/O) values defined by all the points represented considering bins in 12+log(O/H) or O$^{2+}$/O. The black dashed lines and the grey bands show the solar 12+log(O/H) and log(Ar/O) and their associated uncertainties, respectively, from \citet{Asplund:21}.} 
\label{fig:ICFAr_analysis}
\end{figure*}

\subsection{Argon}
\label{subsec:argon}

Ar presents emission lines of Ar$^{2+}$ and Ar$^{3+}$ in the optical spectra of \hii\ regions and SFGs. Ar$^{2+}$ can be produced by photons with energies between 27.6 and 40.7 eV and Ar$^{3+}$ by those having energies from 40.7 to 59.8 eV. Under the typical ionisation conditions of {\hii} regions and SFGs, we expect a certain fraction of unseen Ar$^{+}$ (15.8 $-$ 27.6 eV) to be present in the ionised gas. The brightest optical Ar lines in {\hii} regions and SFGs are {\ariii} $\lambda\lambda$7136, 7751. From one or the sum of both lines we can derive the Ar$^{2+}$ abundance in 1017 objects of the DESIRED-E sample. Due to the high ionisation potential of Ar$^{3+}$ ion, the {\ariv} $\lambda\lambda$4711, 4740 doublet can only be observed --and their corresponding Ar$^{3+}$/H$^+$ abundance calculated-- in 235 objects, a 23.1\% of the total number of Ar/H determinations. When we have ionic abundances of Ar$^{2+}$ and Ar$^{3+}$ for a given object, the total Ar abundance is determined using the relation (see footnote in Sect.~\ref{subsec:neon}):  

\begin{equation}
\label{eq:ICFAr34}
\frac{\text{Ar}}{\text{O}} = \left(\frac{\text{Ar}^{2+}+\text{Ar}^{3+}}{\text{O}^{2+}}\right) \times \text{ICF(Ar)}.  
\end{equation}

When only Ar$^{2+}$/H$^+$ is available, the used relation is: 

\begin{equation}
\label{eq:ICFAr3}
\frac{\text{Ar}}{\text{O}} = \left(\frac{\text{Ar}^{2+}}{\text{O}^{2+}}\right) \times \text{ICF(Ar)},   
\end{equation}

\noindent obviously, the ICF(Ar) values used in both expressions are in principle different for a same given object. 

As for Ne and S, we consider three different ICF(Ar) schemes to derive the Ar abundance. We consider two versions of each ICF(Ar), one when having Ar$^{2+}$ and Ar$^{3+}$ abundances and the other when only Ar$^{2+}$ is available. We use the sets by \citet{Amayo:21} and \citet{Izotov:06}, described in Sect.~\ref{subsec:neon}, and the one proposed by \citet{PerezMontero:07}, based on photoionisation models made with CLOUDY v06.02 \citep{Ferland:98} for spherical geometry and constant low-density ionised nebulae covering a range of  values of the ionisation parameter and effective temperatures of the ionising sources. They use SEDs of O and B stars calculated with the WM-BASIC v1.11 code \citep{Pauldrach:01}. The values of the IFC(Ar) by \citet{Amayo:21} and \citet{PerezMontero:07} are valid for all the whole range of possible values of the O$^{2+}$/O ratio, but the ones by \citet{Izotov:06} only for O$^{2+}$/O $\geq$ 0.2. 

In Fig.~\ref{fig:ionic_ratios_Ar}, we plot log(Ar$^{2+}$/O$^{2+}$) (top) and log((Ar$^{2+}$+Ar$^{3+}$)/O$^{2+}$) (bottom) as a function of O$^{2+}$/O, for {\hii} regions (blue circles) and SFGs (black squares) of the DESIRED-E sample. The fraction of objects with Ar$^{2+}$ and Ar$^{3+}$ abundances is 33.4\% in SFGs, which is higher than in {\hii} regions (12.5\%). This is because SFGs tend to have higher O$^{2+}$/O ratios than {\hii} regions in galaxies (this is illustrated in Fig.~\ref{fig:Ovsion}). In Fig.~\ref{fig:ionic_ratios_Ar}, the coloured curved lines represent the behaviour of the different ICF(Ar) schemes considered: \citet[][red dashed-dotted lines]{Amayo:21}, \citet[][dotted lines with colours corresponding to the three metallicity ranges]{Izotov:06} and \citet[][green dashed line]{PerezMontero:07}. The horizontal black dashed line shows the value of the solar log(Ar/O) ratio --along as its uncertainty represented by the grey band-- recommended by \citet{Asplund:21}. The top panel of Fig.~\ref{fig:ionic_ratios_Ar} shows that the curves of the different ICF(Ar) schemes reproduce fairly well the behaviour of the observational points except the one by \citet{Amayo:21}, that does not reproduce the position of SFGs with O$^{2+}$/O $\geq$ 0.9. In the case of the bottom panel of Fig.~\ref{fig:ionic_ratios_Ar}, we can see that all the curves reproduce quite well the distribution of the observational points. This is because the  objects showing {\ariv} lines in their spectra have necessarily a high ionisation degree and the contribution of unseen Ar$^{+}$ should be very small. Moreover, the bottom right panel of Fig.~\ref{fig:comparion_ICFs} shows that the Ar abundance obtained with the three ICF(Ar) for the same object differs less than ~0.04 dex, except for those with O$^{2+}$/O $\geq$ 0.95 where the differences can be as large as 0.3 dex for those objects that lack {\ariv} lines.

In Fig.~\ref{fig:ICFAr_analysis} we represent the log(Ar/O) values obtained using the three ICF(Ar) schemes considered to derive the total Ar abundance for 1017 DESIRED-E objects \citep[953 in the plots of the middle row using the ICF(Ar) by][]{Izotov:06}. About half of them are {\hii} regions and the rest SFGs. Similarly to Figs.~\ref{fig:ICFNe_analysis} and \ref{fig:ICFS_analysis}, we represent log(Ar/O) as a function of 12+log(O/H) (left) and O$^{2+}$/O (right).  
In Fig.~\ref{fig:ICFAr_analysis} we distinguish between the objects whose Ar abundance has been determined solely from Ar$^{2+}$/H$^+$ (empty symbols) and those that has been calculated from (Ar$^{2+}$+Ar$^{3+}$)/H$^+$ (full symbols).  In the left panels of Fig.~\ref{fig:ICFAr_analysis} we can note a small tendency for the log(Ar/O) to decrease by $\sim$ 0.15 dex with the different ICF(Ar) schemes. However, the behaviour of  log(Ar/O) with respect to O$^{2+}$/O shows a slightly undulating behaviour in all the panels, with maximum amplitudes between 0.04 and 0.09 dex once we exclude the bin corresponding to the lowest values of O$^{2+}$/O in the ICF(Ar) by \citet{Amayo:21} and \citet{PerezMontero:07}. This undulating behaviour is most probably  introduced by the curves of the ICF(Ar) schemes for the objects based only in Ar$^{2+}$/H$^+$ (see Fig.~\ref{fig:ionic_ratios_Ar}). 

All the ICF(Ar) schemes provide very similar mean values and standard deviations of log(Ar/O): $-$2.34 $\pm$ 0.18 for \citet{Amayo:21}, $-$2.33 $\pm$ 0.17 for \citet{Izotov:06} and $-$2.30 $\pm$ 0.18 for \citet{PerezMontero:07}, entirely consistent with the solar value of $-$2.31 $\pm$ 0.11 recommended by \citet{Asplund:21}. 

\section{The Ne/O, S/O and Ar/O abundance ratios in the local Universe}
\label{sec:abundance_ratios}

In this section we will discuss the results obtained in Sect.~\ref{sec:total_abundances} in order to explore the behaviour of Ne/O, S/O and Ar/O ratios with respect to metallicity in star-forming regions of the local Universe. To do so, we will first select the ICF scheme that we consider works better for each element and focusing the subsequent discussion on the results obtained with that chosen ICF. A second step is to study how the behaviour of these ratios is as a function of metallicity to analyse whether the nucleosynthetic origin of Ne, S and Ar is, in fact, similar to that of O and try to interpret possible departures. We will also explore whether there are differences between the behaviour of these ratios as a function of the type of object: {\hii} regions or SFGs. Finally, we will establish precise values of the Ne/O, S/O and Ar/O ratios representative of the ionised gas-phase of the ISM in the local Universe. 

\begin{figure*}[ht!]
\centering    
\includegraphics[scale=0.38]{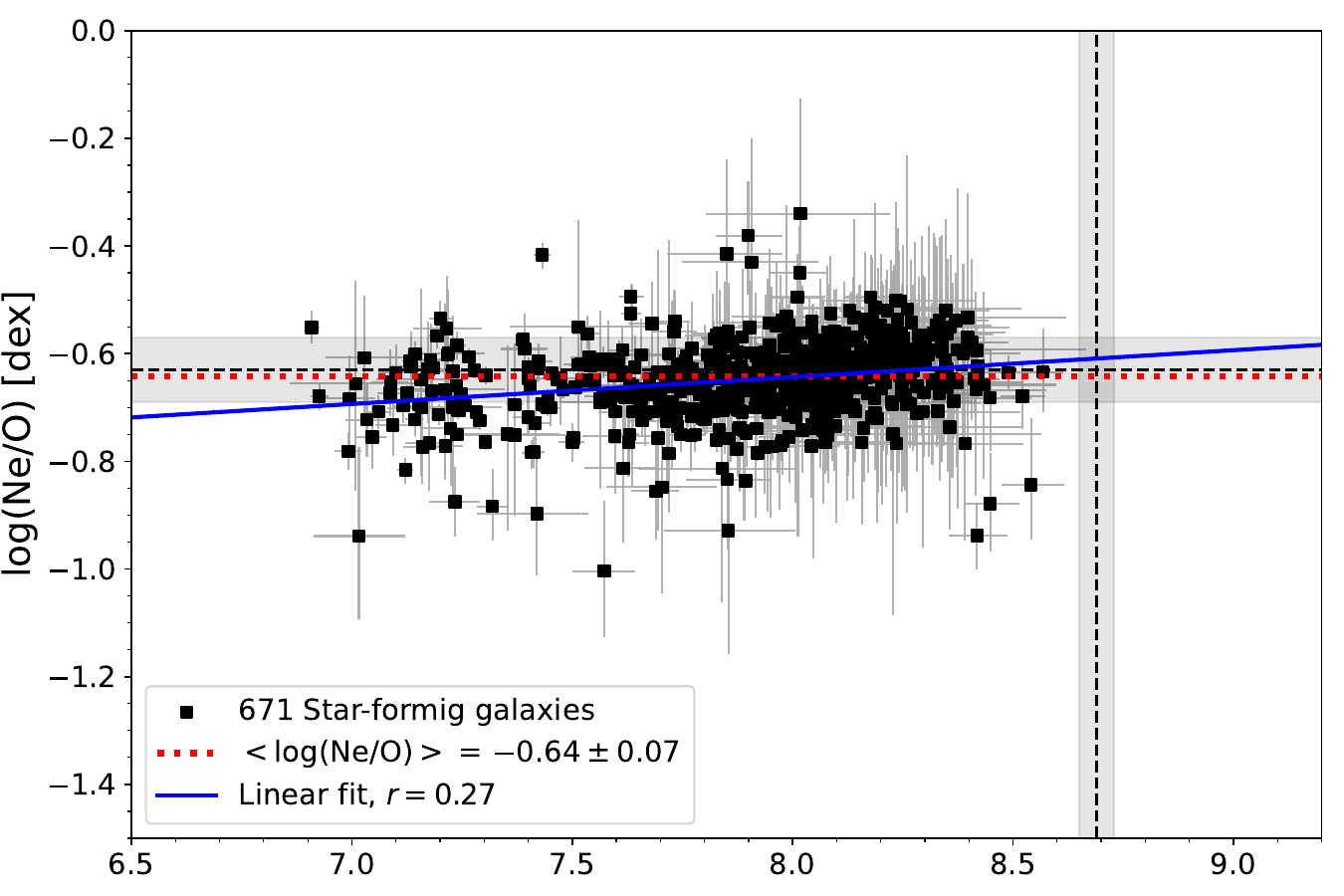}
\includegraphics[scale=0.38]{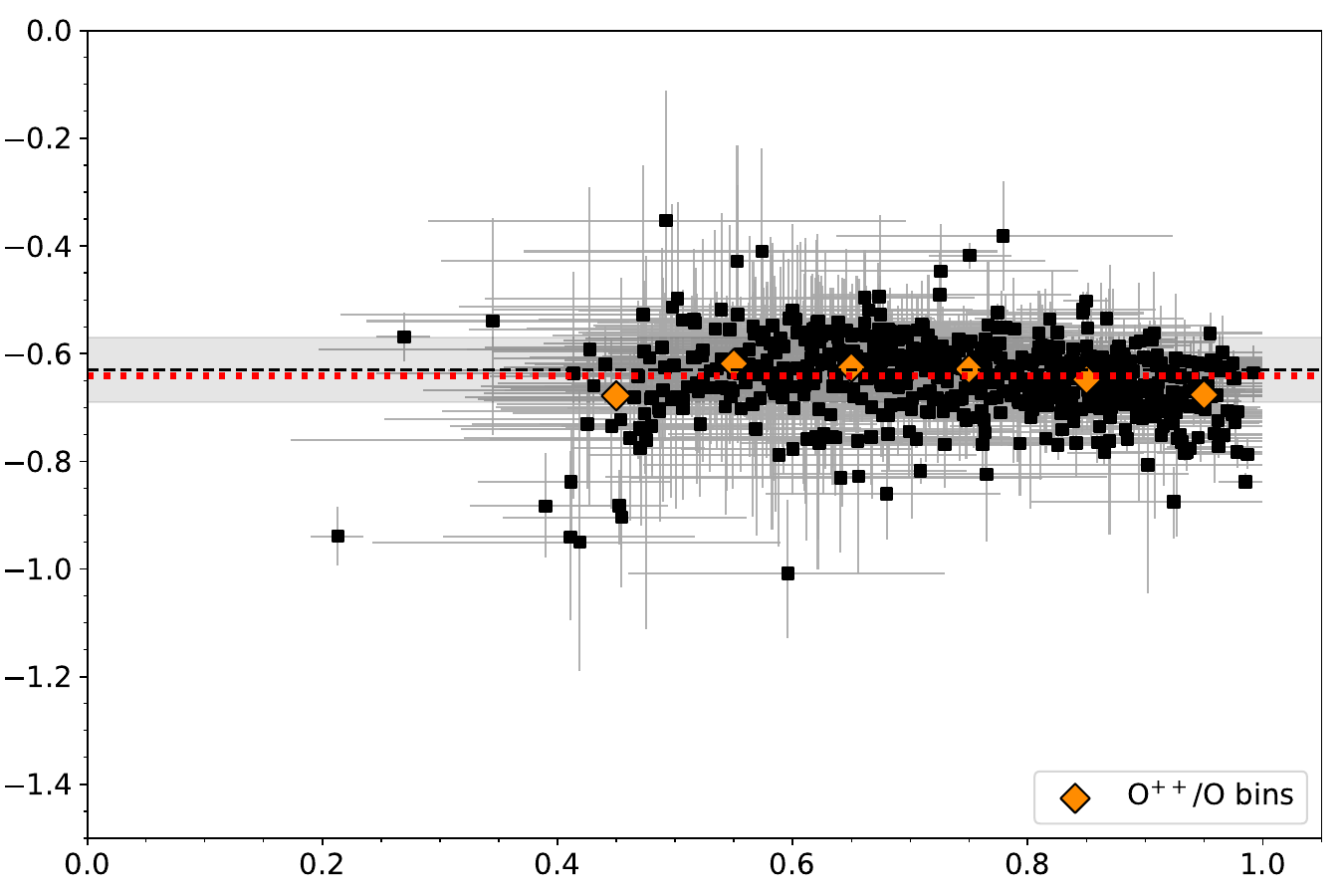}
\\
\includegraphics[scale=0.38]{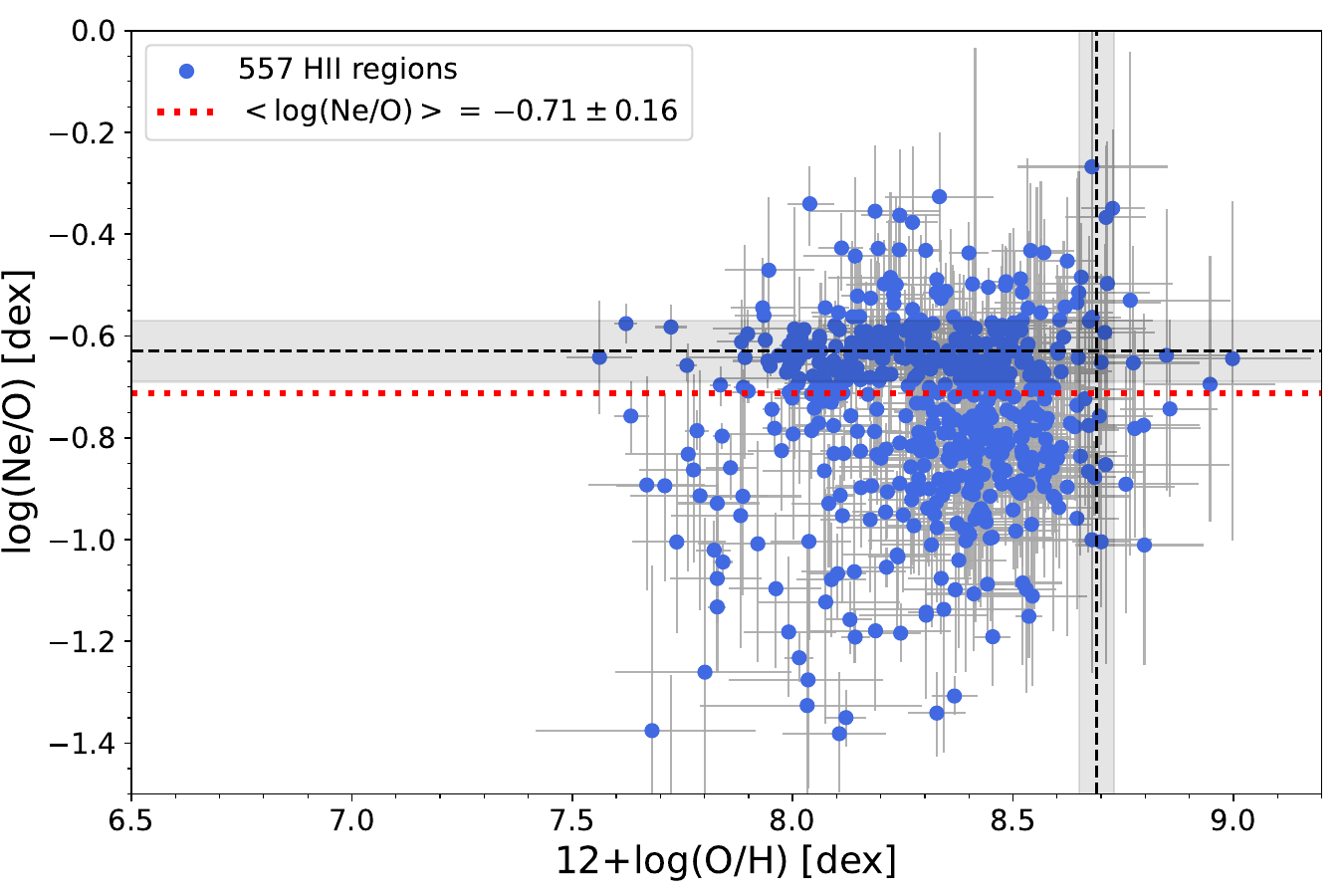}
\includegraphics[scale=0.38]{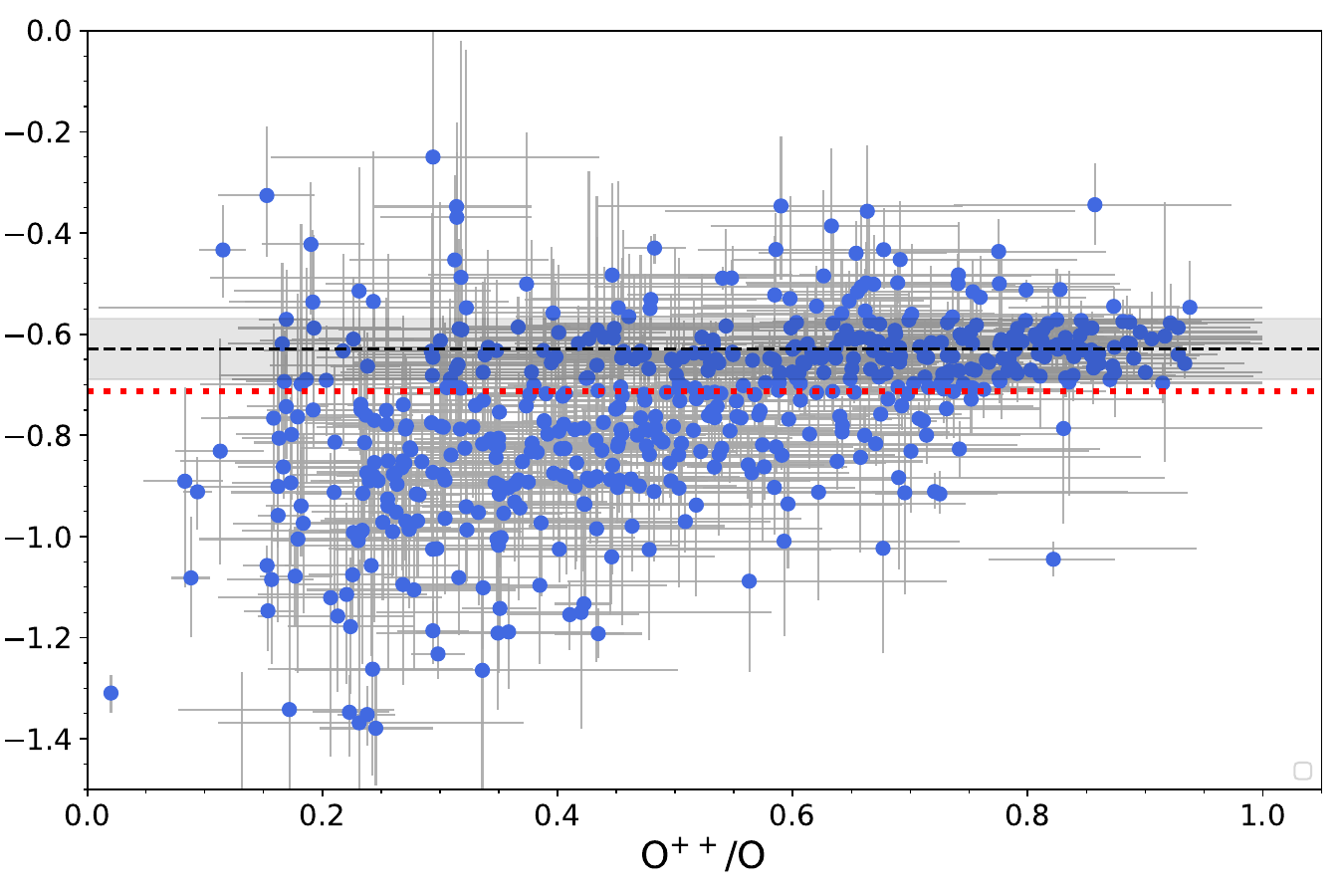}
\caption{log(Ne/O) as a function of 12+log(O/H) (left) and the ionisation degree, O$^{2+}$/O (right), for the DESIRED-E sample. The values of log(Ne/O) are calculated using the ICF(Ne) scheme by \citet{Izotov:06}. The top panels show the points corresponding to SFGs (black squares) and the bottom panels the corresponding to {\hii} regions (blue circles). Orange diamonds in the top right panel indicate the mean log(Ne/O) values considering bins in O$^{2+}$/O. The blue continuous line included in the top left panel represents a linear fit to the data of SFGs. The dotted red line represents the mean value of the log(Ne/O) obtained for each kind of object.  The black dashed lines and the grey bands show the solar 12+log(O/H) and log(Ne/O) and their associated uncertainties, respectively, from \citet{Asplund:21}.} 
\label{fig:NeO_separated}
\end{figure*}

\subsection{Ne/O}
\label{subsec:NeO}

\begin{table*}
    \caption{Mean and median abundance ratios per element X and type of object.}
    \label{table:mean_ratios}
    \begin{tabular}{ccccccc}
        \hline
        \noalign{\smallskip}
        & & & Mean 12+log(O/H) & Mean log(X/O) &  Median log(X/O) & log(X/O)$^{a}_\odot$\\
        X & Type & No. & [dex] & [dex] & [dex] & [dex] \\
        \noalign{\smallskip}
        \hline
        \multirow{2}{*}{Ne} & SFGs & 671 & $8.02 \pm 0.26$ & $-0.64 \pm 0.07$ & $-$0.64 & \multirow{2}{*}{$-0.63 \pm 0.06$} \\
        & {\hii} regions & 557 & $8.37 \pm 0.23$ & $-0.71 \pm 0.16$ & $-$0.71 \\
        \hline
        \multirow{2}{*}{S} & SFGs & 492 & $8.06 \pm 0.24$ & $-1.63 \pm 0.13$ & $-$1.65 & \multirow{2}{*}{$-1.57 \pm 0.05$} \\
        & {\hii} regions & 441 & $8.41 \pm 0.26$ & $-1.54 \pm 0.19$ & $-$1.58 \\
        \hline
        \multirow{2}{*}{Ar} & SFGs & 506 & $8.06 \pm 0.24$ & $-2.37 \pm 0.12$ & $-$2.38 & \multirow{2}{*}{$-2.31 \pm 0.11$} \\
        & {\hii} regions & 447 & $8.39 \pm 0.26$ & $-2.30 \pm 0.20$ & $-$2.33 \\
        \noalign{\smallskip}
        \hline
    \end{tabular}
    \tablebib{ $a$: \citet{Asplund:21}}
\end{table*}

\begin{table*}
    \caption{Linear fits log(X/O) = $m$ $\times$ [12+log(Y/H)] + $n$ per element X and type of object.}
    \label{table:fits}
    \begin{tabular}{cccccccc}
        \hline
        \noalign{\smallskip}
        X & Type & No. & Y & Pearson $r$ & $p$-value & Slope ($m$) & Intercept ($n$) \\
        \noalign{\smallskip}
        \hline
        \multirow{2}{*}{Ne} & \multirow{2}{*}{SFGs} & \multirow{2}{*}{671} & O & 0.27 & 2.23$\times$10$^{-12}$ & $0.051 \pm 0.006$ & $-1.052 \pm 0.049$ \\
        & & & Ne & 0.45 & 3.43$\times $10$^{-34}$ & $0.072 \pm 0.006$ & $-1.167 \pm 0.040$ \\
        \hline
        \multirow{6}{*}{S} & \multirow{2}{*}{SFGs} & \multirow{2}{*}{492} & O & $-$0.17 & 1.69$\times$10$^{-4}$ & $-0.076 \pm 0.024$ & $-1.061 \pm 0.190$ \\
        & & & S & 0.27 & 3.98$\times$10$^{-10}$ & $0.246 \pm 0.021$ & $-3.197 \pm 0.130$ \\  
        & \multirow{2}{*}{{\hii} regions} & \multirow{2}{*}{441} & O & $-$0.10 & 0.027 & $-0.030 \pm 0.034$ & $-1.324 \pm 0.280$ \\
        & & & S & 0.48 & 4.89$\times$10$^{-26}$ & $0.416 \pm 0.024$ & $-4.339 \pm 0.157$ \\
        & \multirow{2}{*}{SFGs+{\hii} regions} & \multirow{2}{*}{933} & O & 0.03 & 0.339 & $0.072 \pm 0.017$ & $-2.206 \pm 0.139$ \\
        & & & S & 0.45 & 4.89$\times$10$^{-26}$ & $0.277 \pm 0.013$ & $-3.403 \pm 0.081$ \\
        \hline
        \multirow{6}{*}{Ar} & \multirow{2}{*}{SFGs} & \multirow{2}{*}{506} & O & $-$0.43 & 3.64$\times$10$^{-24}$ & $-0.201 \pm 0.015$ & $-0.768 \pm 0.122$ \\
        & & & Ar & $-$0.05 & 0.299 & $-0.046 \pm 0.020$ & $-2.098 \pm 0.109$ \\  
        & \multirow{2}{*}{{\hii} regions} & \multirow{2}{*}{447} & O & $-$0.44 & 4.11$\times$10$^{-22}$ & $-0.163 \pm 0.023$ & $-0.987 \pm 0.191$ \\
        & & & Ar & 0.22 & 3.51$\times$10$^{-6}$ & $0.164 \pm 0.023$ & $-3.282 \pm 0.137$ \\
        & \multirow{2}{*}{SFGs+{\hii} regions} & \multirow{2}{*}{953} & O & $-$0.28 & 9.20$\times$10$^{-19}$ & $-0.100 \pm 0.012$ & $-1.541 \pm 0.095$ \\
        & & & Ar & 0.17 & 1.41$\times$10$^{-7}$ & $0.063 \pm 0.012$ & $-2.697 \pm 0.069$ \\        \noalign{\smallskip}
        \hline
    \end{tabular}
\end{table*}

Summarizing the results analysed in Sec.~\ref{subsec:neon} and Figs.~\ref{fig:ionic_ratios_NeS} and \ref{fig:ICFNe_analysis}, the ICF(Ne) by \citet{Izotov:06} shows a log(Ne/O) vs. O$^{2+}$/O ratio relation flatter than the other ICF(Ne) schemes as well as the lowest standard deviation and mean and median values of log(Ne/O) ($-$0.67 $\pm$ 0.12 and $-$0.66, respectively) more consistent with the solar ratio of $-$0.63 $\pm$ 0.06 \citep{Asplund:21}. For these reasons we will use the Ne abundances obtained with the ICF(Ne) by \citet{Izotov:06} for the discussion we will carry out in the present section. However, even if we make this selection, determining total Ne abundances using ICFs has serious limitations. As we pointed out in Sec.~\ref{subsec:neon}, an important fraction of the objects at values of O$^{2+}$/O $<$ 0.5, show very low log(Ne/O) values (see Fig.~\ref{fig:ICFNe_analysis}). In Fig.~\ref{fig:ionic_ratios_NeS} it is evident that the position of those points cannot be correctly reproduced by any of the ICF(Ne) schemes considered. The strong increased scatter of the Ne$^{2+}$/O$^{2+}$ ratio at lower O$^{2+}$/O is not something new, it was formerly reported by \citet{Kennicutt:03}. \citet{Croxall:16}, in their study of {\hii} regions in the spiral galaxy NGC~5457 (M101), confirm a similar scatter, claiming for the existence of a large population of {\hii} regions with significant offset to low values of log(Ne/O). 

In Fig.~\ref{fig:NeO_separated}, we show the distribution of log(Ne/O) as a function of 12+log(O/H) and O$^{2+}$/O separately for {\hii} regions and SFGs, for which Ne abundance has been derived using the ICF(Ne) by \citet{Izotov:06}. The very different behaviour between both types of objects is  striking. Since the distribution of the 557 data points classified as {\hii} regions shows no significant indication of correlation (especially evident in the bottom left panel), the 671 SFGs represented in the top panels of Fig.~\ref{fig:NeO_separated} show a rather flat trend. The log(Ne/O) vs. 12+log(O/H) relation for SFGs seems to follow a weak positive linear correlation with a rather small slope. In Fig.~\ref{fig:NeO_separated} we include a linear fit to the data points of SFGs considering their errors, represented by a continuous line. The mean and median values of log(Ne/O) and 12+log(O/H) for each type of object are given in Table~\ref{table:mean_ratios}. The parameters of the linear fit to the log(Ne/O) vs. 12+log(O/H) and vs. 12+log(Ne/H) distributions of the SFGs are given in Table~\ref{table:fits}. Although the sample Pearson correlation coefficient of the fit to log(Ne/O) vs. 12+log(O/H) is rather low ($r$ = 0.27), it seems to reproduce reasonably well the weak observational trend. In Table~\ref{table:fits} we also include the $p$-value of the fits, that is used to evaluate the plausibility of the null hypothesis, which indicates the absence of correlation between the variables involved in the linear fit\footnote{A value of $p$ $>$ 0.05 indicates that the null hypothesis is plausible, i.e. there is not significant linear fit between the variables. On the other hand, when $p$ $\leq$ 0.05 it can be said that there is a significant correlation between both variables.}. The fit shown in the upper left panel of Fig~\ref{fig:NeO_separated} indicates a small increase of log(Ne/O) of 0.08 dex for the interval of 12+log(O/H) between 7.0 and 8.5. \citet{Izotov:06}, for a sample of 414 SFGs --most of them included in the DESIRED-E sample used in this paper-- found also a slight increase of $\sim$0.1 dex in the same interval of O abundances. Similar trends are found in the log(Ne/O) vs. 12+log(O/H) distributions obtained by \citet{Amayo:21} and \citet{MirandaPerez:23}. \citet{ArellanoCordova:24}, who study 43 local SFGs of the COS Legacy Archive Spectroscopic SurveY (CLASSY), find a group of 5 SFGs ($\sim$10\% of their sample) with 12+log(O/H) of the order or even above the solar value, showing log(Ne/O) $\geq$ $-$0.5, fact that they suggest may reflect some overproduction of Ne. Remarkably, our sample of SFGs do not present objects with such characteristics. 

\begin{figure}[ht!]
\centering    
\includegraphics[scale=0.38]{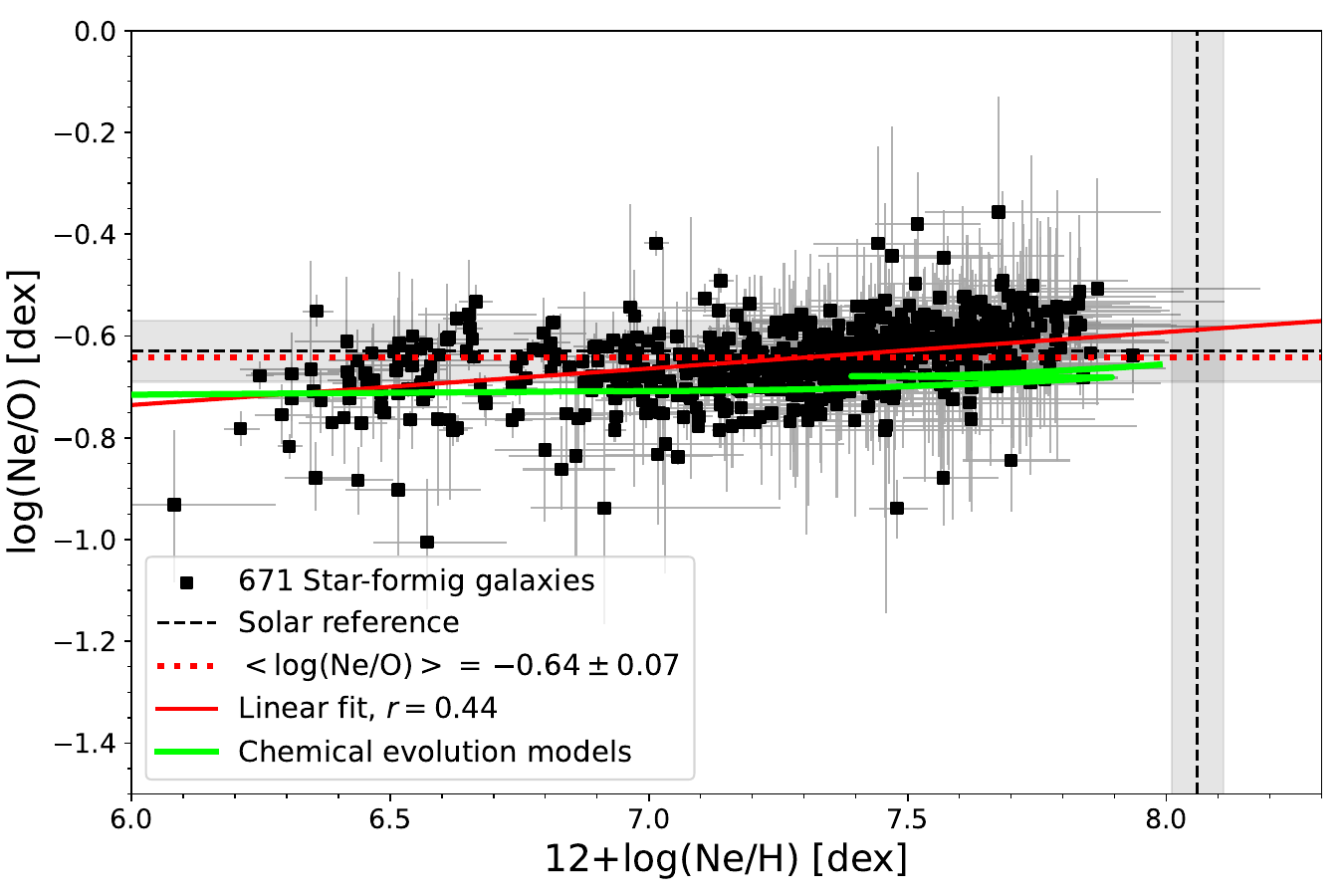}
\caption{log(Ne/O) as a function of 12+log(Ne/H) for SFGs of the DESIRED-E sample, which Ne/O ratios have been calculated using the ICF(Ne) scheme by \citet{Izotov:06}. The red continuous line represents a linear fit to the data. The green continuous line shows the time evolution of Ne and O abundances of the ISM predicted by the CEM of the Milky Way of \citet{Medina-Amayo:23}. The dotted red line represents the mean value of log(Ne/O). The black dashed lines and the grey bands show the solar 12+log(Ne/H) and log(Ne/O) and their associated uncertainties, respectively, from \citet{Asplund:21}.} 
\label{fig:NeOvsNe}
\end{figure}

 In Fig.~\ref{fig:NeOvsNe} we represent log(Ne/O) vs. 12+log(Ne/H) for the DESIRED-E sample of SFGs. The continuous red line shows a least-squares linear fit to the data considering their errors. The sample Pearson correlation coefficient of the fit is $r$ = $-$0.64, a stronger linear correlation to that obtained for the log(Ne/O) vs. 12+log(O/H) distribution shown in the top left panel of Fig.~\ref{fig:NeO_separated}. The fit gives a slope only slightly higher than the one with respect to O/H and an increase of log(Ne/O) of 0.09 dex for the interval of 12+log(Ne/H) between 6.5 and 7.8. As we can see the fits to log(Ne/O) vs. 12+log(O/H) and vs. 12+log(Ne/H) of the SFGs are fairly similar and both clearly incompatible with the null hypothesis (see Table~\ref{table:fits}). 
 
Nucleosynthesis models seem to agree in predicting a basically lockstep evolution of O and Ne \citep[e.g.][]{Iwamoto:99, Johnson:19, Kobayashi:20}. In Fig.~\ref{fig:NeOvsNe} we also include a curve (green line) representing the evolution of Ne and O abundances of the ISM predicted by the chemical evolution model (CEM) of the Milky Way obtained by \citet{Medina-Amayo:23}. This CEM was developed for the solar neighborhood of the Galactic disc (7$-$9 kpc from the Galactic centre) within a two-infall gas formation scenario, similar to that of \citet{Spitoni:21}. During the first 4 Gyr, the thick disc formed from a short accretion of primordial gas. Then, the thin disc began forming from a prolonged second accretion (the loop in the curve), which has continued to the present. The star formation rate and initial mass function (IMF) are based on the formulations of \citet{Kennicutt:98} and \citet{Kroupa:02}, respectively. The model uses stellar yields for low-and intermediate-mass stars by Ventura’s group \citep[e.g.][and references therein]{Ventura:22}, for SNe Ia by \citet{Leung:18}, and for massive stars by \citet{Nomoto:13}. An upper mass limit of the IMF at 37 M$_\odot$ was inferred to match the present-day O abundance of {\hii} regions obtained from CELs at $\sim$8 kpc \citep{MendezDelgado:22}. In Fig.~\ref{fig:CEM} we show the cumulative fraction of $^{16}$O, $^{20}$Ne, $^{32}$S, $^{36}$Ar --the most abundant stable isotopes of each element, that represent between 85 to 95\% of the total amount of them-- produced by different types of stars as a function of 12+log(O/H) predicted by the CEM. We distinguish the fractions produced by low- and intermediate-mass stars in the asymptotic giant branch phase (AGBs, cyan line) or  type Ia supernovae (SNe Ia, dark red line) and massive stars in core collapse supernovae (CCSNe, magenta line). It is important to remember that the CEM used is built to reproduce the abundance patterns observed in the Milky Way. Objects in other galaxies with a different star-formation history may be located in somewhat different places in the abundance ratios diagrams. For example, the infall of unprocessed gas into a star-forming region in a galaxy may decrease the metallicity while keeping the abundance ratio of two alpha-elements constant.

As we can see in Fig.~\ref{fig:NeOvsNe}, this CEM reproduces the log(Ne/O) of about $-$0.70 dex that show the low-metallicty SFGs represented in Fig.~\ref{fig:NeOvsNe}. However, the curve provides a rather slight increase of only +0.06 dex in the whole Ne/H range of the objects, an increase that seems insufficient to reproduce the trend we observe in our log(Ne/O) vs. 12+log(Ne/H) distribution at higher metallicities. In principle, this behaviour may be interpreted as a slight  metallicity-dependent Ne production but this is not predicted by nucleosynthesis models. In fact, Fig.~\ref{fig:CEM} shows that the SNe Ia and AGB contribution to the total amount of $^{16}$O and $^{20}$Ne is very small and that the time evolution of Ne/O should produce a constant ratio. \citet{Izotov:06}, assuming this lockstep evolution of O and Ne, interpret the increase of about 0.1 dex they find in their log(Ne/O) vs. 12+log(O/H) relation --very similar to ours-- as an effect of O depletion onto dust grains as a function of metallicity\footnote{Ne, is a noble gas, with a very low condensation temperature (9 K) and is not expected to be depleted into dust grains in the ISM \citep{Jenkins:09}}. For the Orion Nebula --an example of local {\hii} region which chemical composition is representative of the solar neighborhood-- the fraction of dust trapping O is up to $\sim$ 0.1 dex \citep[e.g.][]{MesaDelgado:09, Peimbert:10}. In a recent paper, \citet{MendezDelgado:24} estimate that at the metallicities of the Magellanic Clouds \citep[Large Magellanic Cloud, LMC: 12+log(O/H) = 8.36, Small Magellanic Cloud, SMC: 8.03;][]{DominguezGuzman:22}, such O depletion should account for up to 0.07 dex and 0.03 dex for LMC and SMC, respectively, becoming negligible at 12+log(O/H) $<$ 8.0. This seems to be roughly consistent with the hypothesis of the dust depletion to explain the observed log(Ne/O) vs. 12+log(O/H) or vs. 12+log(Ne/H) correlations, but also with the offset of about $\sim$0.10 dex between the log(Ne/O) given by the CEM predictions and observations at high metallicities. In principle, the dust depletion hypothesis could be verified if we found a similar behaviour with other noble gases such as Ar \citep[also proposed by][]{Amayo:21}. However, as we will discuss in Sect.~\ref{subsec:ArO} and can be seen in  Fig.~\ref{fig:CEM}, the nucleosynthetic origin of Ar does not seem to be entirely similar to that of O or Ne, so the behaviour of the Ar/O ratio may be modulated by other phenomena apart from a different degree of dust depletion relative to O.
 
 Another possibility to explain the slight increase of log(Ne/O) as a function of metallicity is that it is an artifact produced by the ICF(Ne) itself. As we can see, the mean log(Ne/O) values corresponding to the 0.1 dex wide bins in O$^{2+}$/O shown in the upper right panel of Fig.~\ref{fig:NeO_separated} trace a rather slight tendency to decrease as O$^{2+}$/O increases. This might ultimate contribute in some way to the increase of log(Ne/O) with metallicity we observe. In fact the maximum amplitude of the variation of log(Ne/O) between the O$^{2+}$/O bins represented in Fig.~\ref{fig:NeO_separated} is of about 0.06 dex.

As we pointed out above, in Fig.~\ref{fig:NeO_separated}, the behaviour of {\hii} regions is very different to that shown by SFGs. In the log(Ne/O) vs. 12+log(O/H) diagram, the observational points classified as {\hii} regions are distributed in a quite defined range of metallicity but with a much larger dispersion of log(Ne/O). We can distinguish that although most of the objects are concentrated at log(Ne/O) values close to the solar one, there seems to be a secondary concentration at 12+log(O/H) $\sim$ 8.4 and log(Ne/O) between $-$1.0 and $-$0.7. 
In fact, in Table~\ref{table:mean_ratios} the mean log(Ne/O) of the {\hii} regions is 0.07--0.08 dex lower than the mean of the SFGs or the solar value. 
\citet{Croxall:16} and \citet{Berg:20} find a similar log(Ne/O) vs. 12+log(O/H) distribution, although this was expected given that the observational data on which those studies are based are included in the DESIRED-E sample. The log(Ne/O) vs. O$^{2+}$/O diagram of the bottom row of Fig.~\ref{fig:NeO_separated} shows that the dispersion towards lower values of log(Ne/O) is clearly a function of O$^{2+}$/O, indicating that the ICF(Ne) is not able to correct appropriately for the presence of Ne$^+$ in {\hii} regions data points at O$^{2+}$/O $\leq$ 0.75, which are the vast majority of the objects of this type in the DESIRED-E sample. \citet{Kennicutt:03} suggested that this behaviour is due to the high sensitivity of Ne ionisation models to the input stellar atmosphere fluxes, which in turn are very much dependent on the treatment of opacity and stellar winds. Considering the difficulties of obtaining consistent Ne/O values from Galactic and extragalactic {\hii} regions, we have chosen not to take them into account for the study of the behaviour of this ratio as a function of metallicity and in the data shown in Table 2.

\begin{figure}[ht!]
\centering    
\includegraphics[scale=0.38]{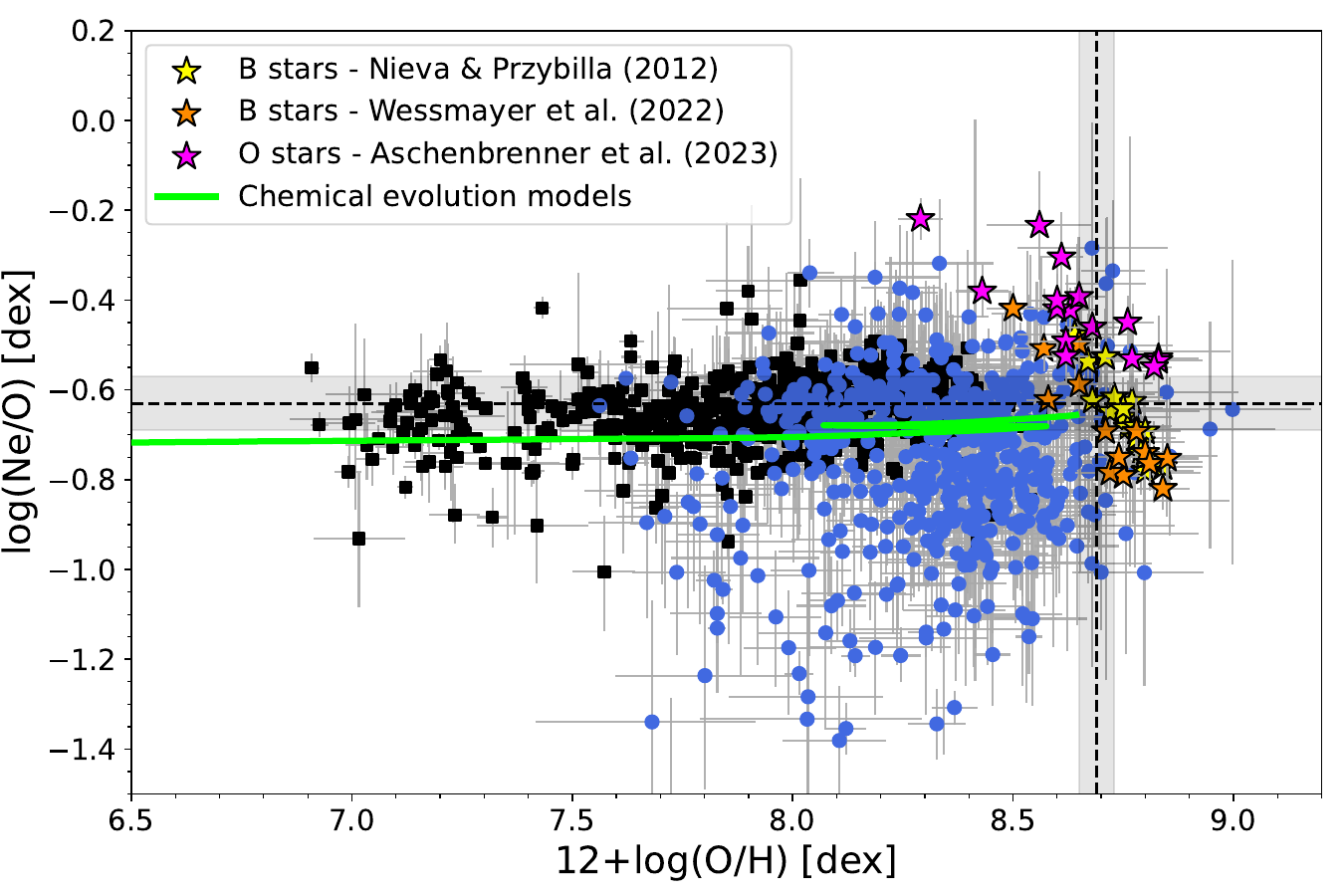}
\caption{log(Ne/O) as a function of 12+log(O/H). Black squares and blue circles represent SFGs and {\hii} regions of the DESIRED-E sample, which Ne/O ratios have been calculated using the ICF(Ne) scheme by \citet{Izotov:06}. The green continuous line shows the time evolution of Ne and O abundances of the ISM predicted by the CEM of the Milky Way of \citet{Medina-Amayo:23}. Yellow and orange stars represent the abundances obtained from quantitative spectroscopic analysis of Galactic B stars by \citet{Nieva:12} and \citet{Wessmayer:2022}, respectively. Magenta stars represent the abundances of Galactic O stars determined by \citet{Aschenbrenner:23}. The black dashed lines and the grey bands show the solar 12+log(O/H) and log(Ne/O) and their associated uncertainties, respectively, from \citet{Asplund:21}.} 
\label{fig:NeO_stars}
\end{figure}

\begin{figure*}[ht!]
\centering    
\includegraphics[scale=0.38]{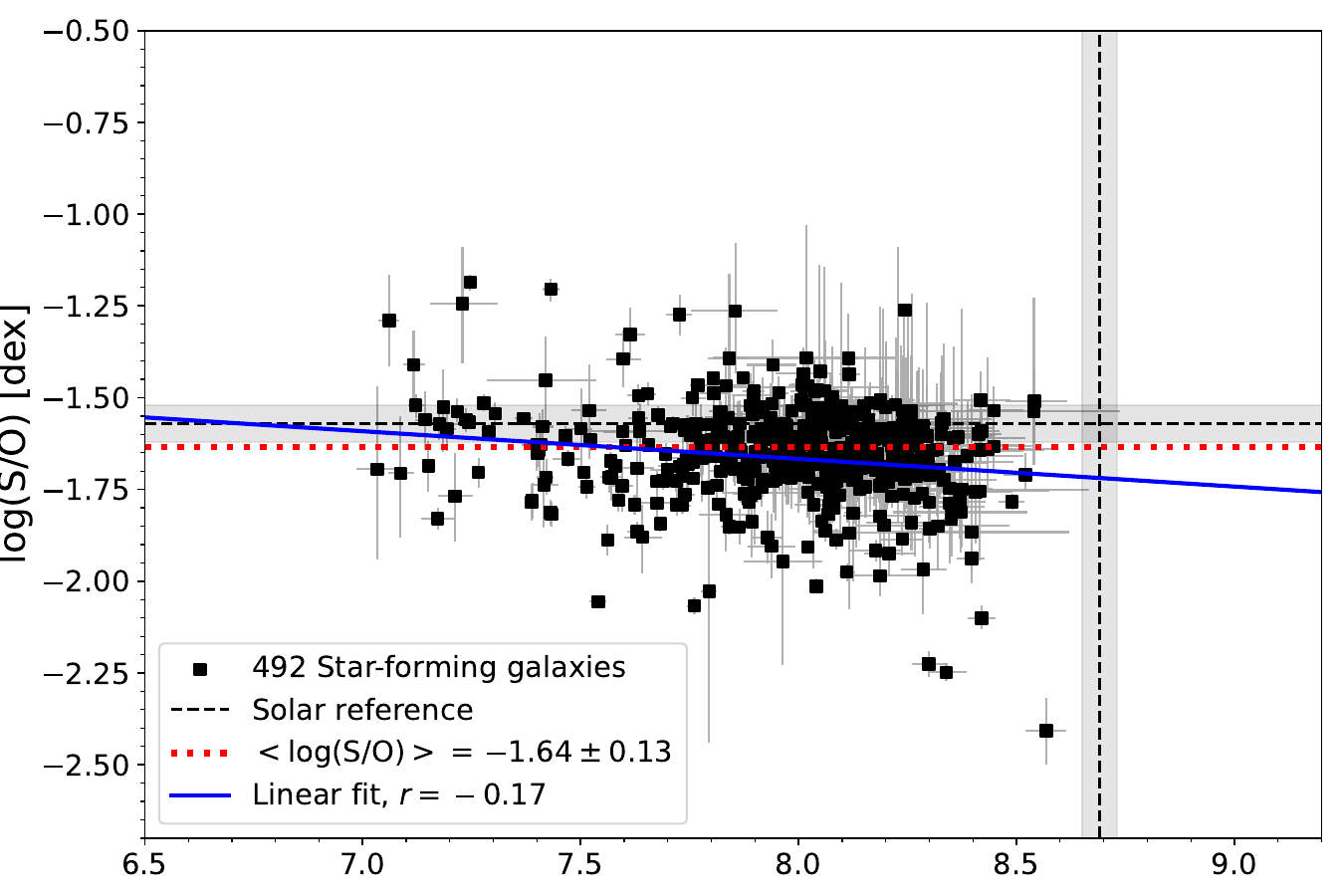}
\includegraphics[scale=0.38]{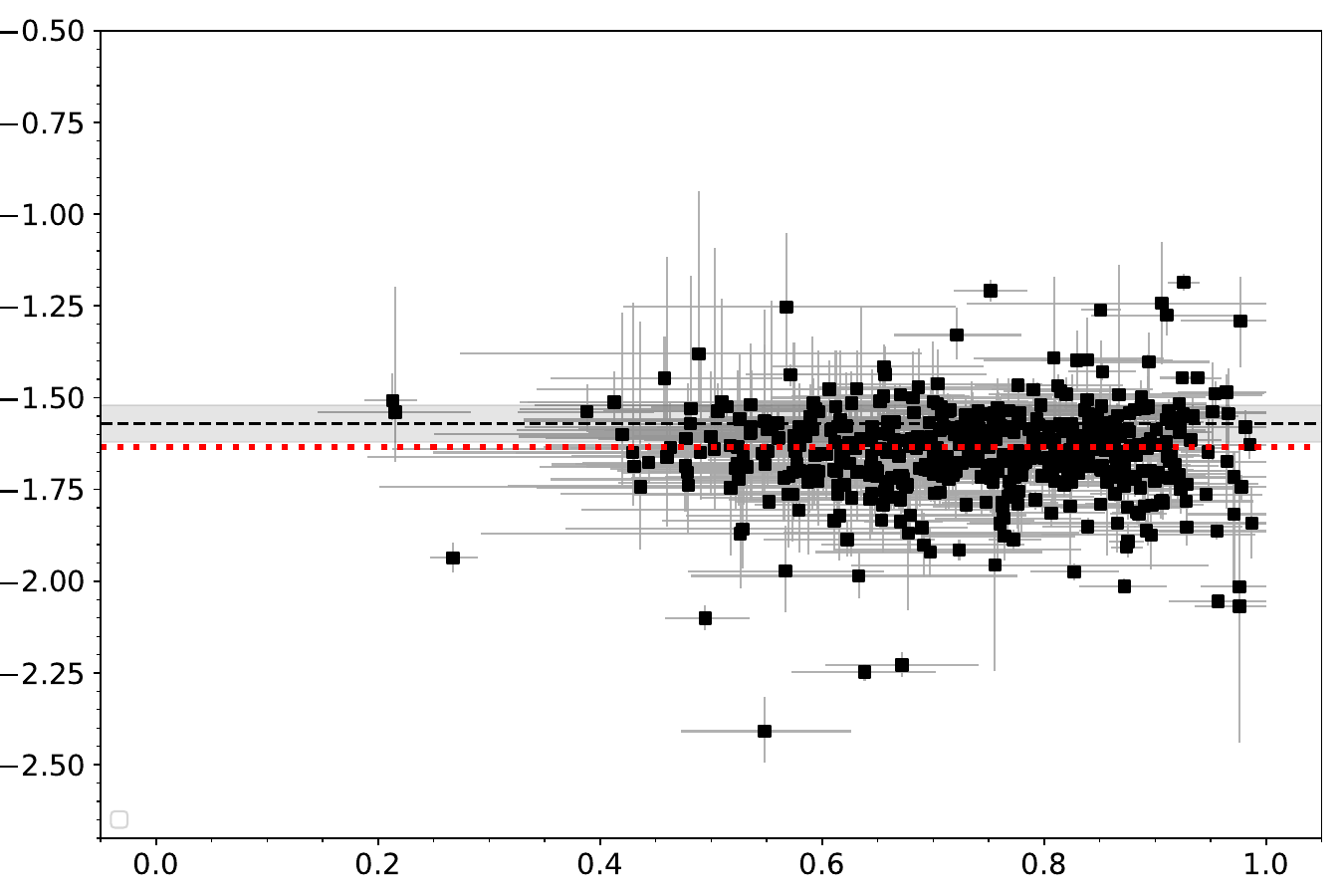}
\\
\includegraphics[scale=0.38]{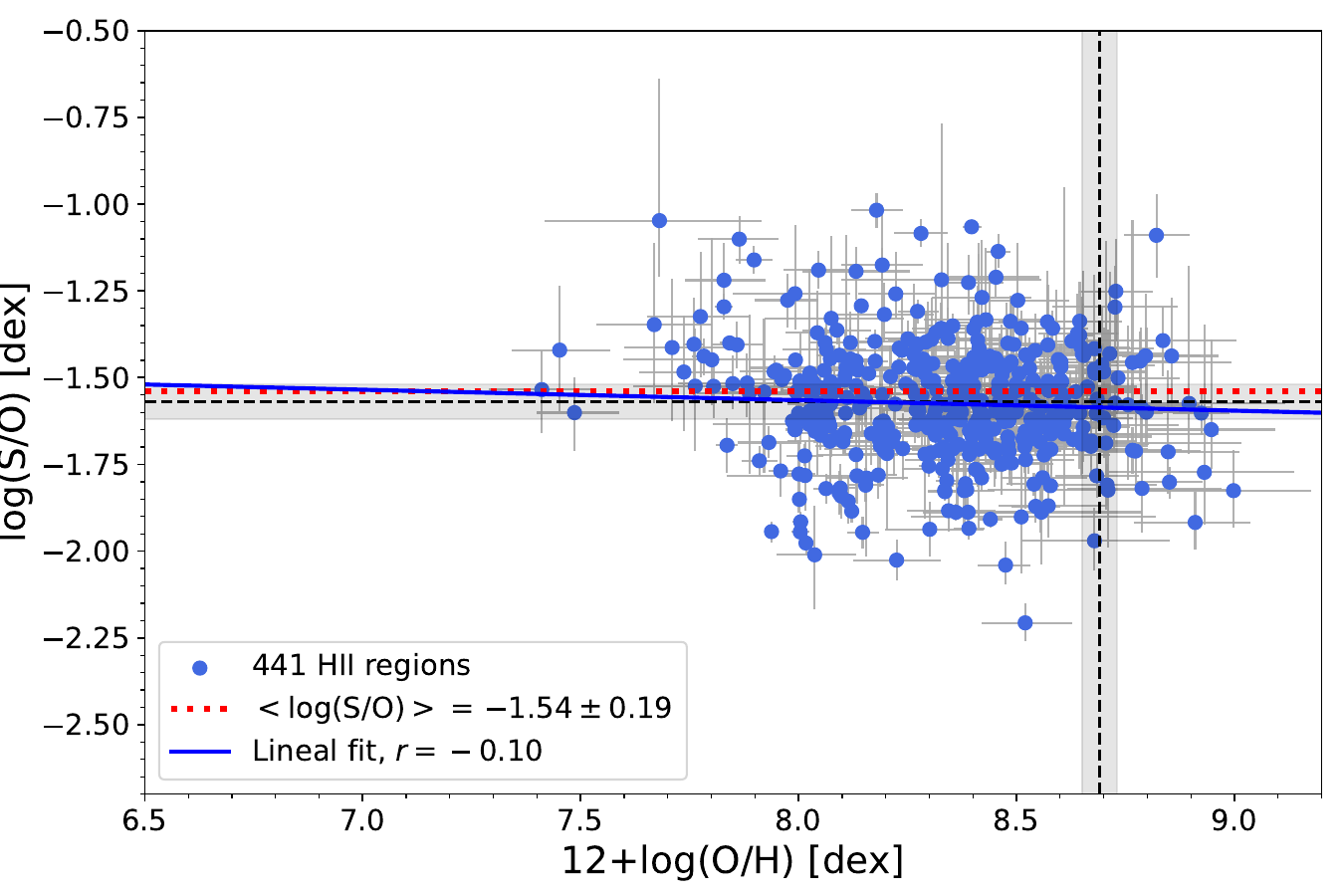}
\includegraphics[scale=0.38]{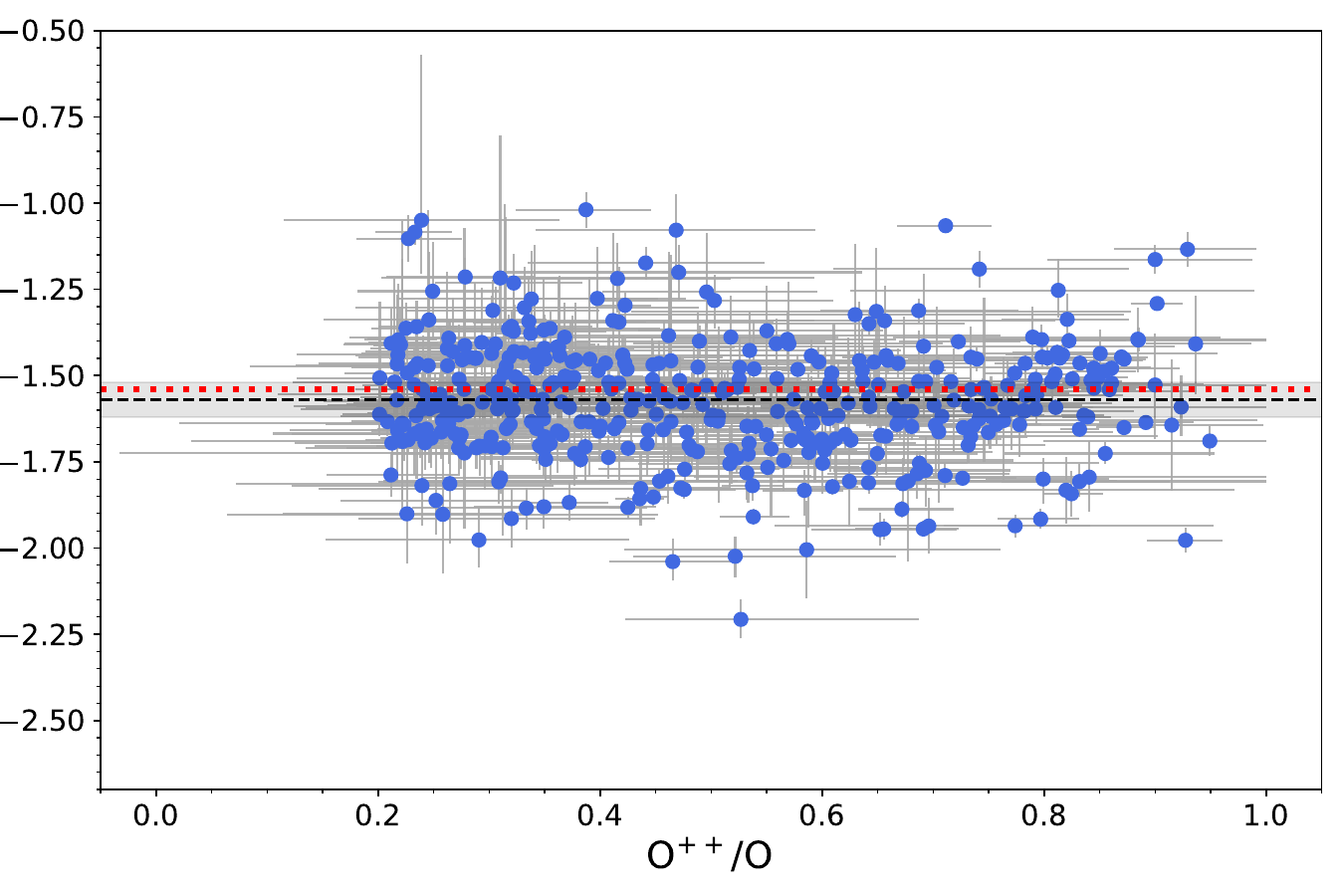}
\caption{log(S/O) as a function of 12+log(O/H) (left) and the ionisation degree, O$^{2+}$/O (right), for the DESIRED-E sample. The values of log(S/O) are calculated using the ICF(S) scheme by \citet{Izotov:06}. The top panels show the points corresponding to SFGs (black squares) and the bottom panels the corresponding to {\hii} regions (blue circles). The dotted red line represents the mean value of the log(S/O) obtained for each kind of object. The blue continuous lines represent linear fits to the data represented in each panel of the left column.  The black dashed lines and the grey bands show the solar 12+log(O/H) and log(S/O) and their associated uncertainties, respectively, from \citet{Asplund:21}.} 
\label{fig:SO_separated}
\end{figure*}

In Fig.~\ref{fig:NeO_stars} we represent the log(Ne/O) vs. 12+log(O/H) distribution of {\hii} regions and SFGs of the DESIRED-E sample, the curve defined by the CEM of \citet{Medina-Amayo:23} and abundances of O and B-type Galactic stars determined from quantitative spectroscopy by \citet{Nieva:12}, \citet{Wessmayer:2022} and \citet{Aschenbrenner:23}. The atmospheres of O and B-type stars represent the present-day chemical composition of the ISM and, in principle, should provide  abundance values similar to the ionised gas associated to star-forming regions. In Fig.~\ref{fig:NeO_stars}, we can see that the stellar O abundances are close to solar or oversolar --as expected because most of the stars are located in the solar neighborhood-- but quite separated from the locus of the bulk of the {\hii} regions. A systematic difference between nebular and stellar abundances of objects located in the same star-forming regions is a well-known fact.  Studies comparing the O abundances of Galactic {\hii} regions with that of their associated O and B-type stars \citep[e.g.][]{SimonDiaz:06, SimonDiaz:11, GarciaRojas:14}  find  stellar abundances $\sim$0.2 dex higher than nebular ones when they are calculated using CELs. However, such difference disappears when using faint recombination lines instead of CELs for deriving the nebular O abundance, this offset is the so-called abundance discrepancy problem. It may be related to the presence of temperature fluctuations inside the {\hii} regions \citep{Peimbert:67, GarciaRojas:07, MendezDelgado:23a}. On the other hand, Fig.~\ref{fig:NeO_stars} indicates that the O and B-type stars also show a quite high dispersion in their log(Ne/O) ratio. The mean and standard deviation of log(Ne/O) of the stars included in the figure is $-$0.56 $\pm$ 0.17, a fairly supersolar value. Although temperature fluctuations can alter the abundances calculated with CELs, the abundance ratios --as Ne/O, S/O or Ar/O-- obtained with this type of lines remain practically unchanged, so the observed log(Ne/O) offset between stellar and nebular objects is not expected to be related to this phenomenon \citep[e.g.][]{Esteban:20, MendezDelgado:24}.

\subsection{S/O}
\label{subsec:SO}

As in the case of the Ne/O ratio discussed in Sect.~\ref{subsec:NeO}, the ICF(S) by \citet{Izotov:06} is the one that provides a more featureless --flatter-- distribution between log(S/O) vs. 12+log(O/H) and vs. O$^{2+}$/O (Fig.~\ref{fig:ICFNe_analysis}), a tighter fit to the (S$^+$+S$^{2+}$)/O vs. O$^{2+}$/O distribution in the range of ionisation degrees in common with the others ICFs (Fig.~\ref{fig:ionic_ratios_NeS}) and a mean log(S/O) closer to the solar value of $-$1.57 $\pm$ 0.05 \citep{Asplund:21}. We will adopt the results obtained using this ICF(S) for the discussion we will develop in this section. The only drawback of the ICF(S) scheme by \citet{Izotov:06} is that its application is limited to objects with O$^{2+}$/O $\geq$ 0.2. In the case of the DESIRED-E sample this has a very limited impact. While no SFGs have to be removed applying this limit, the fraction of {\hii} regions that are  finally excluded is only 13.3\%. Considering the rough relation between ionisation degree and metallicity shown in Fig.~\ref{fig:Ovsion}, this implies that the removed points correspond mostly to high metallicity objects but --as we can see in the panels of the left column of Fig.~\ref{fig:ICFS_analysis}-- that area is not significantly depopulated when we compare the results using the different ICF(S) schemes. 

The results of most studies indicate that the S/O ratio remains basically constant with respect to O/H \citep[e.g.][]{Garnett:97, Kennicutt:03, Izotov:06, Guseva:11, Croxall:16, Berg:20, ArellanoCordova:20, ArellanoCordova:24}. Although there are others who claim that S/O decreases as O/H increases \citep[eg.][]{Vilchez:88b, Dors:16, Diaz:22, PerezDiaz:24, Brazzini:24}. In Fig.~\ref{fig:SO_separated}, we show the distribution of log(S/O) as a function of 12+log(O/H) and O$^{2+}$/O separately for SFGs and {\hii} regions. The S/O ratios have been calculated using the ICF(S) scheme by \citet{Izotov:06}. Unlike log(Ne/O) (see Fig.~\ref{fig:NeO_separated}), the behaviour of log(S/O) is qualitatively very similar in both types of objects. However, the mean value of log(S/O) and its standard deviation are somewhat different; in the case of SFGs we obtain $-$ 1.64 $\pm$ 0.13 (median $-$1.65) and  $-$1.54 $\pm$ 0.19 (median $-$1.58) for {\hii} regions (see Table~\ref{table:mean_ratios}), being closer to solar \citep[$-$1.57 $\pm$ 0.05;][] {Asplund:21} in this last type of objects. In Fig.~\ref{fig:SO_separated} we include the linear fits to log(Ne/O) vs. 12+log(O/H) for SFGs and {\hii} regions (their parameters are listed in Table~\ref{table:fits}), finding that the distribution of the data points is rather flat in both cases, specially for the objects classified as {\hii} regions, fact that is consistent with a constant and solar log(S/O). This is also confirmed by the large p-value of the fit for {\hii} regions or the combination of {\hii} regions plus SFGs, that are consistent with the null hypothesis.

\begin{figure}[ht!]
\centering    
\includegraphics[scale=0.38]{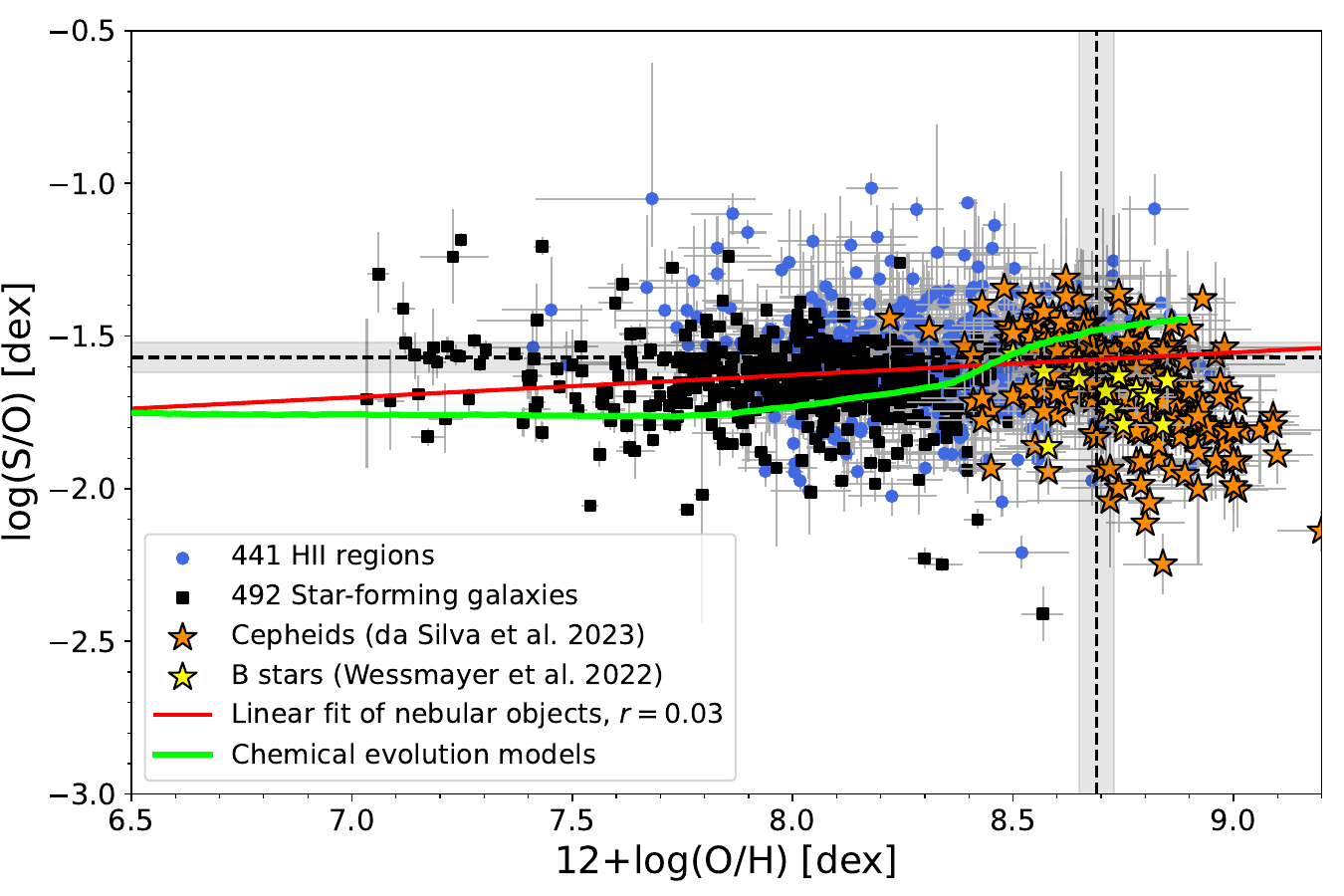}
\includegraphics[scale=0.38]{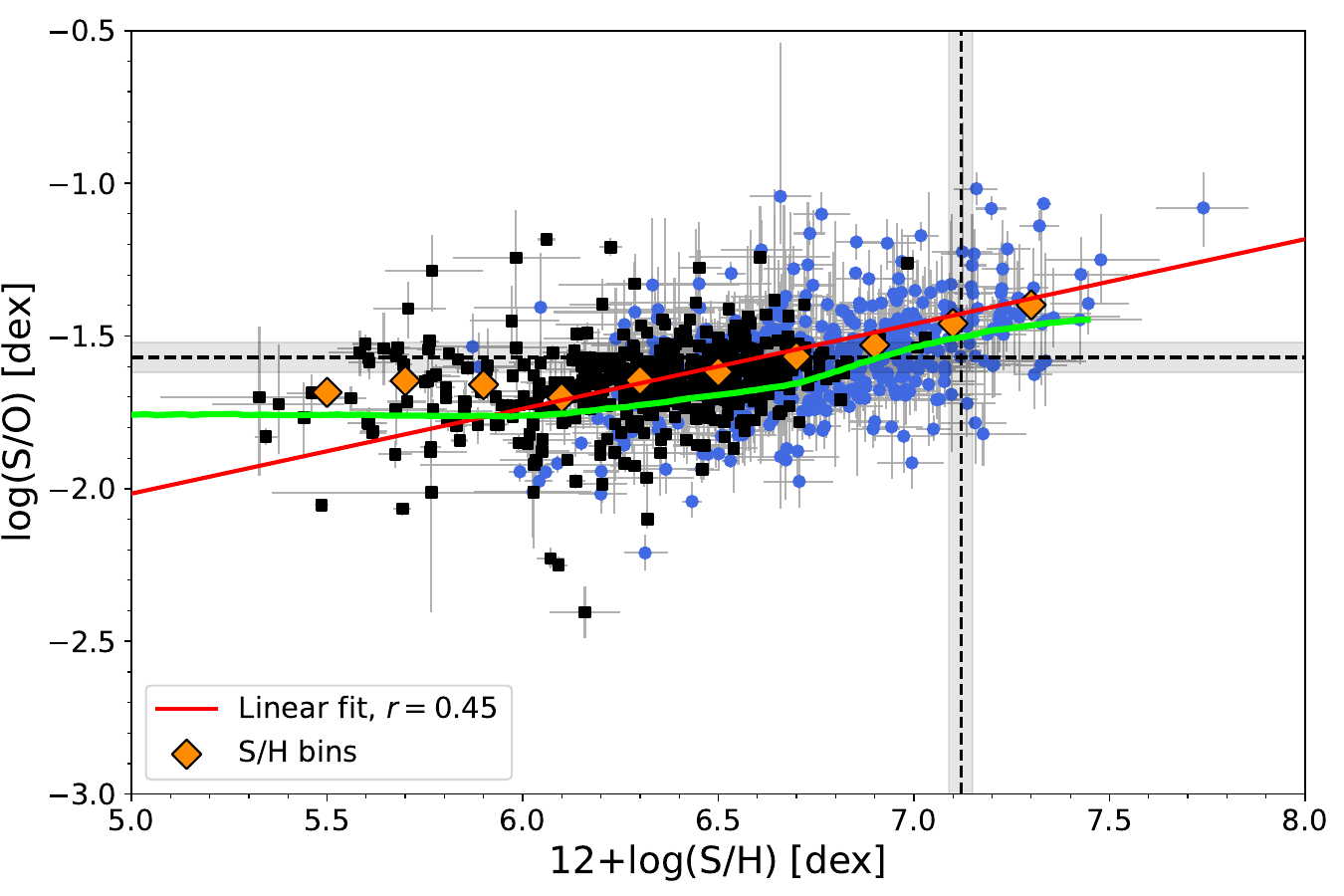}
\caption{log(S/O) as a function of 12+log(O/H) (top) and as a function of 12+log(S/H) (bottom). Black squares and blue circles represent SFGs and {\hii} regions of the DESIRED-E sample, which S/O ratios have been calculated using the ICF(S) scheme by \citet{Izotov:06}. The green continuous line shows the time evolution of S and O abundances of the ISM predicted by the CEM of the Milky Way of \citet{Kobayashi:20}. Yellow stars represent the abundances obtained from quantitative spectroscopic analysis of Galactic B stars by \citet{Wessmayer:2022}. Orange stars represent the abundances of Galactic classical Cepheids  determined by \citet{daSilva:23}. Orange diamonds indicate the mean log(S/O) values considering bins in 12+log(S/H). The red continuous lines represent linear fits to the data represented in each panel.  The black dashed lines and the grey bands show the solar 12+log(O/H), 12+log(S/H) and log(S/O) and their associated uncertainties, respectively, from \citet{Asplund:21}.} 
\label{fig:SO_stars}
\end{figure}

In Fig.~\ref{fig:SO_stars} we represent the log(S/O) vs. 12+log(O/H) (top) and vs. 12+log(S/H) (bottom) distributions of SFGs$+${\hii} regions of the DESIRED-E sample. In the top panel we also include abundances of Galactic B-type stars determined by \citet{Wessmayer:2022} and classical Cepheids observed and analysed by \citet{daSilva:23}. Classical or Population I Cepheids are variable stars with masses in the interval of 4 and 20 M$_\odot$ \citep{Turner:96} and ages $\lesssim$300 Myr \citep{Bono:05}, so --as B-type stars-- their atmospheric abundances should also be representative of the present-day chemical composition of the ISM. As in the case of Ne/O vs. O/H diagram shown in Fig.~\ref{fig:NeO_stars}, the O/H values in Galactic stars are close to solar or oversolar and are distributed in a separated locus from the bulk of the nebular objects, even the Galactic ones. This fact has already been discussed in Sect.~\ref{subsec:NeO}. On the other hand, the mean log(S/O) values of the B-type stars represented in Fig.~\ref{fig:SO_stars} is $-$1.68 $\pm$ 0.08 and that of the classical Cepheids $-$1.64 $\pm$ 0.15, both values below the solar log(S/O) of $-$1.57 $\pm$ 0.05. In Fig.~\ref{fig:SO_stars}, we can notice that the S/O ratio of classical Cepheids seem to show a clear decrease as O/H increases, a fact already noticed by \citet{daSilva:23}. 

The behaviour of the data points of the DESIRED-E sample shown in Figs.~\ref{fig:SO_separated} and \ref{fig:SO_stars} contrasts with the results obtained by some previous studies \citep{Dors:16,Diaz:22,Brazzini:24}. In particular, \citet{Diaz:22} find that S/O vs. O/H and vs. S/H relations are quite different for SFGs and {\hii} regions. \citet{Diaz:22} report that SFGs tend to show a S/O ratio lower than solar and that it increases as the S/H increases, but remains constant with O/H, results that are in agreement with ours. However, they find that their {\hii} regions show S/O ratios larger than solar and a clear tendency for lower S/O ratios as O or S abundances increase, a behaviour different to what we observe. In our case, for {\hii} regions, since log(S/O) remains basically constant with 12+log(O/H), it shows a clear increase as 12+log(S/H) increases (see Table~\ref{table:fits}).  
In trying to understand the origin of the discrepancy with the results by \citet{Diaz:22}, we realized that the O/H ratios of at least some of the {\hii} regions of the sample of \citet{Diaz:22} were lower than those determined by us for the same objects and clearly incorrect\footnote{For example, in their supplementary tables, \citet{Diaz:22} give a 12+log(O/H) = 7.92 for their object number 53, which corresponds to the slit position of the Orion Nebula observed by \citet{Esteban:04}, for which we obtain 8.50 (our object \#422). Another examples are the {\hii} regions of the Magellanic Clouds observed by \citet{Toribio:17} or \citet{DominguezGuzman:22}. For those objects, \citet{Diaz:22} find values of 12+log(O/H) between 7.10 and 8.52 (their objects number 232 to 239), while we find a narrower --and more reasonable-- range of values --between 8.02 and 8.43-- for them.}, producing spurious higher S/O ratios. It is beyond the scope of this paper to investigate the origin and quantify the errors in the abundances calculated in \citet{Diaz:22}, but we consider that part of their conclusions, at least those based on the chemical behaviour of the {\hii} regions --apparently no the SFGs-- should not be taken into account until the authors provide the  correct abundances. Unfortunately, the results obtained by \citet{Diaz:22} has led to studies exploring their origin and implications as, for example, \citet{Goswami:24}, who proposed that such high S/O values might be explained by the contribution of the ejecta of Pair Instability Supernovae (PISN) produced by very massive stars ($\geq$ 130 M$_\odot$) in combination of a bi-modal top-heavy IMF and an initial strong burst of star formation. Ultimately, it is not necessary to invoke such exotic scenarios to explain the correct abundance patterns. 

In Fig.~\ref{fig:SO_stars} we include the curves defined by the CEM of \citet{Kobayashi:20}\footnote{The model points represented have been obtained from figures 11 and 15 of Kobayashi et al (2020) assuming their same solar abundance ratios.} and a least-squares fit to all the data (SFGs+{\hii} regions) considering their errors (their parameters are listed in Table~\ref{table:fits}). The slopes of the linear fits are quite different and significantly higher for the log(S/O) vs. 12+log(S/H) distribution. The sample Pearson correlation coefficient of the log(S/O) vs. 12+log(O/H) fit is almost 0 and its p-value indicates that the distribution is indistinguishable of the null hypothesis. Therefore, we can conclude that the log(S/O) vs. 12+log(O/H) distribution of all the data is basically constant. This is not entirely consistent with the behaviour predicted by the CEM, where we would expect an increase of log(S/O) of the order of +0.30 dex in the range of metallicities covered by the objects. On the other hand, the fit to log(S/O) vs. 12+log(S/H) for SFGs+{\hii} regions shows a clearer trend with a more positive slope, a moderate sample Pearson correlation coefficient and extremely low $p$-values (see Fig.~\ref{fig:SO_stars} and Table~\ref{table:fits}). As we can see in Table~\ref{table:fits}, the same higher and positive slope of the fit when considering 12+log(S/H) instead of 12+log(O/H) remains when separating {\hii} regions and SFGs. \citet{Diaz:22} also find a greater dependence of the S/O ratio on S/H than on O/H in the case of SFGs, objects that apparently have correct O abundances in their paper. 
Fig.~\ref{fig:SO_stars} indicates that comparison with the CEM seems to be more consistent in this case. The model curve fits fairly well to the distribution of the different bins. In particular, the bins corresponding to 12+log(S/H) $\lesssim$ 6.0 show a flat trend, consistent with the behaviour and the average log(S/O) given by the model. At 12+log(S/H) $\gtrsim$ 6.0, a slight but continuous increase in S/O occurs, so the lockstep evolution of both elements is not longer achieved. Nucleosynthesis models for SNe Ia predict that these objects may produce a significant fraction of the S and Ar content in the solar neighbourhood \citep[e.g.][]{Iwamoto:99, Johnson:19, Kobayashi:20}. The Chandrasekhar-mass explosion models for SNe Ia by \citet{Kobayashi:20} predict that about 29\% of the total solar S originates from this process, although this fraction might be even higher if the contribution of sub-Chandrasekhar-mass SNe Ia is increased in their models.   

Finally, although the linear fit to log(S/O) vs. 12+log(O/H) presents a small --but rather uncertain-- positive slope that could be consistent with the hypothesis on O depletion put forward in Sect.~\ref{subsec:NeO} to explain the behaviour of the Ne/O ratio, it can not be verified in the case of S/O. Firstly because --as discussed in the previous paragraph-- the production of S by SNe Ia also produces an increase in S/O with O/H. Secondly, the fraction of S depleted onto dust grains in the ISM is a controversial issue because the spectral features used to study it are often highly saturated \citep[e.g][]{Jenkins:09}. In fact, it is often referred to as “the missing sulfur problem” and has raised a variety of questions over the past two decades. \citep[e.g][]{Keller:02,Slavicinska:24}.

\subsection{Ar/O}
\label{subsec:ArO}

\begin{figure*}[ht!]
\centering    
\includegraphics[scale=0.38]{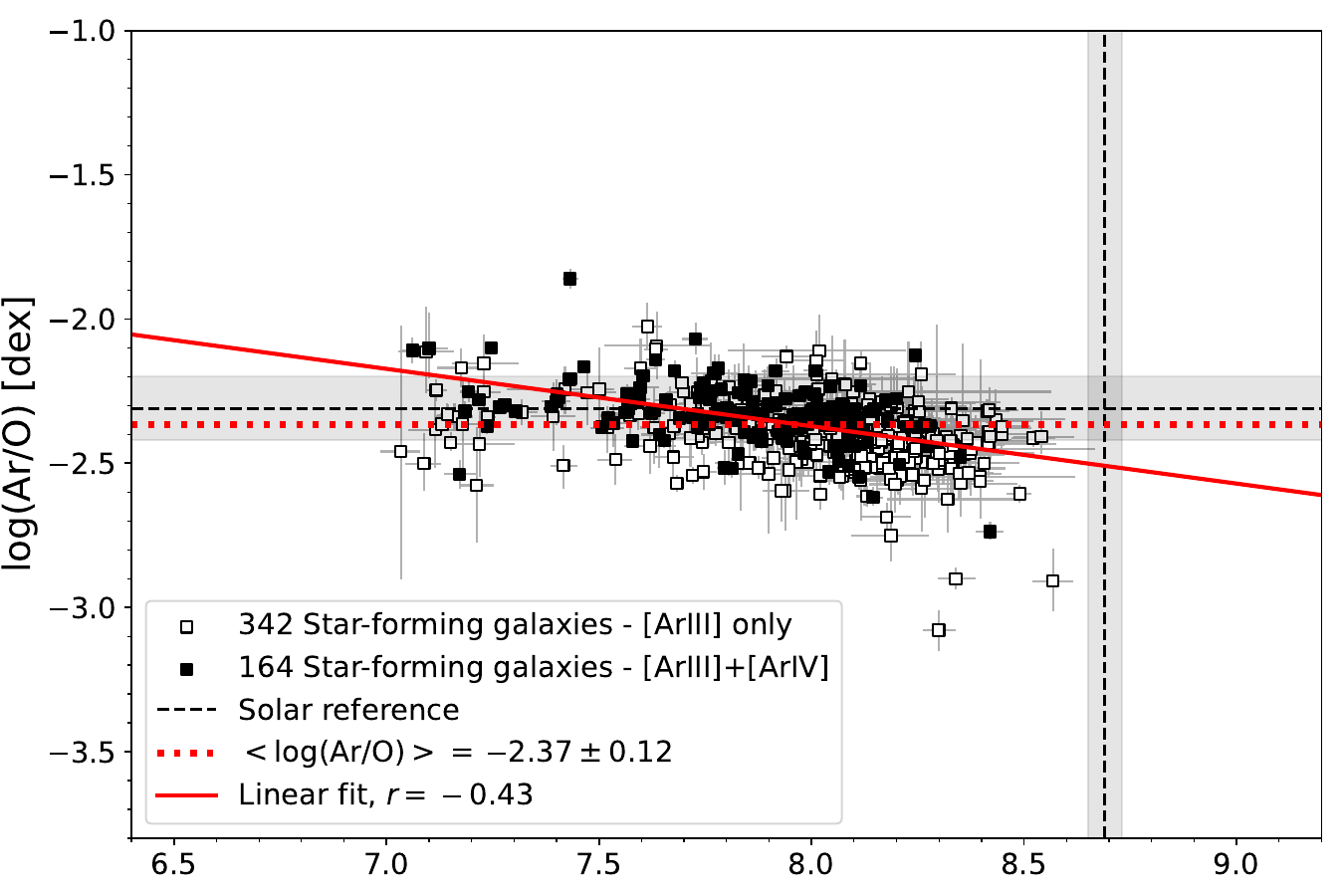}
\includegraphics[scale=0.38]{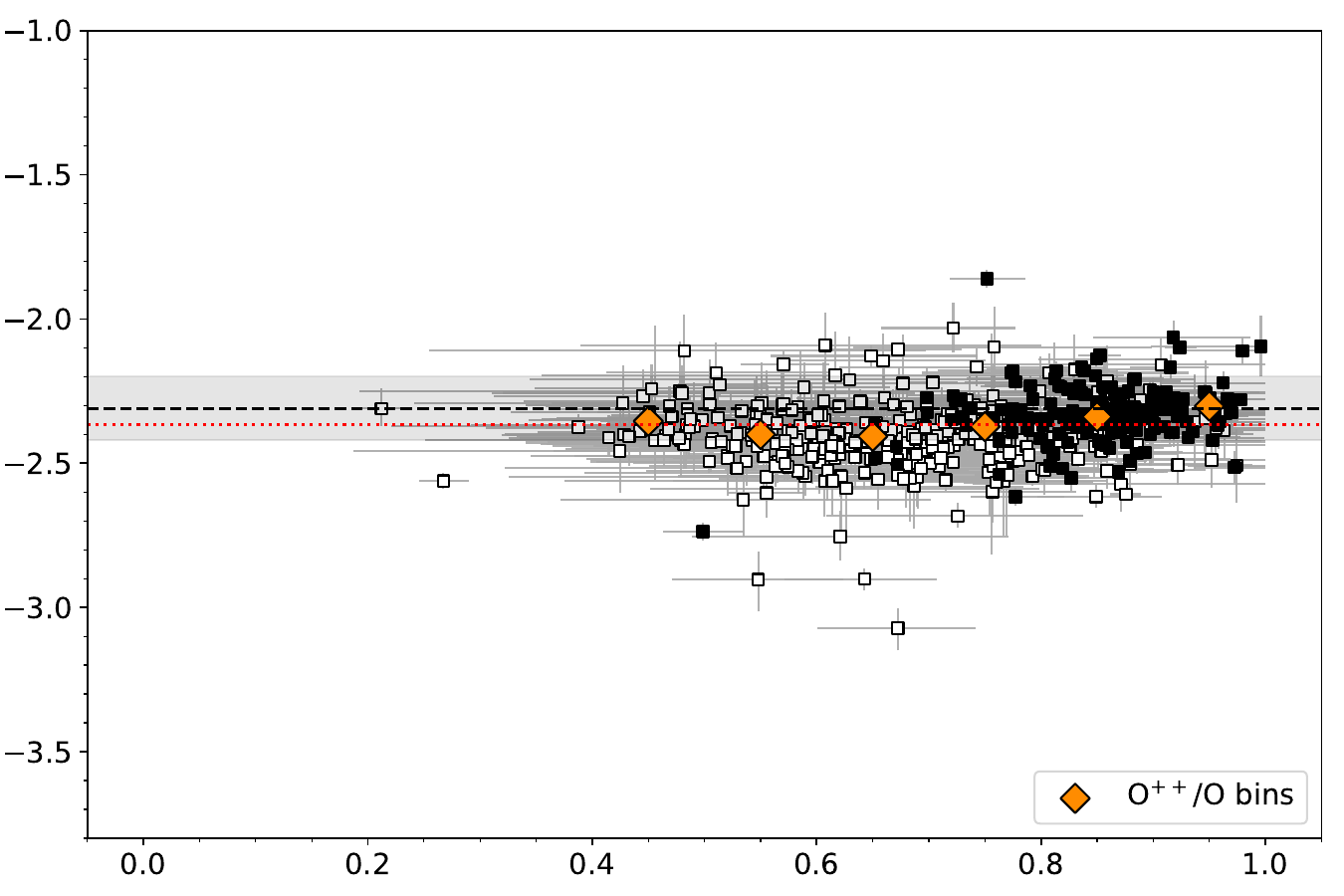}
\\
\includegraphics[scale=0.38]{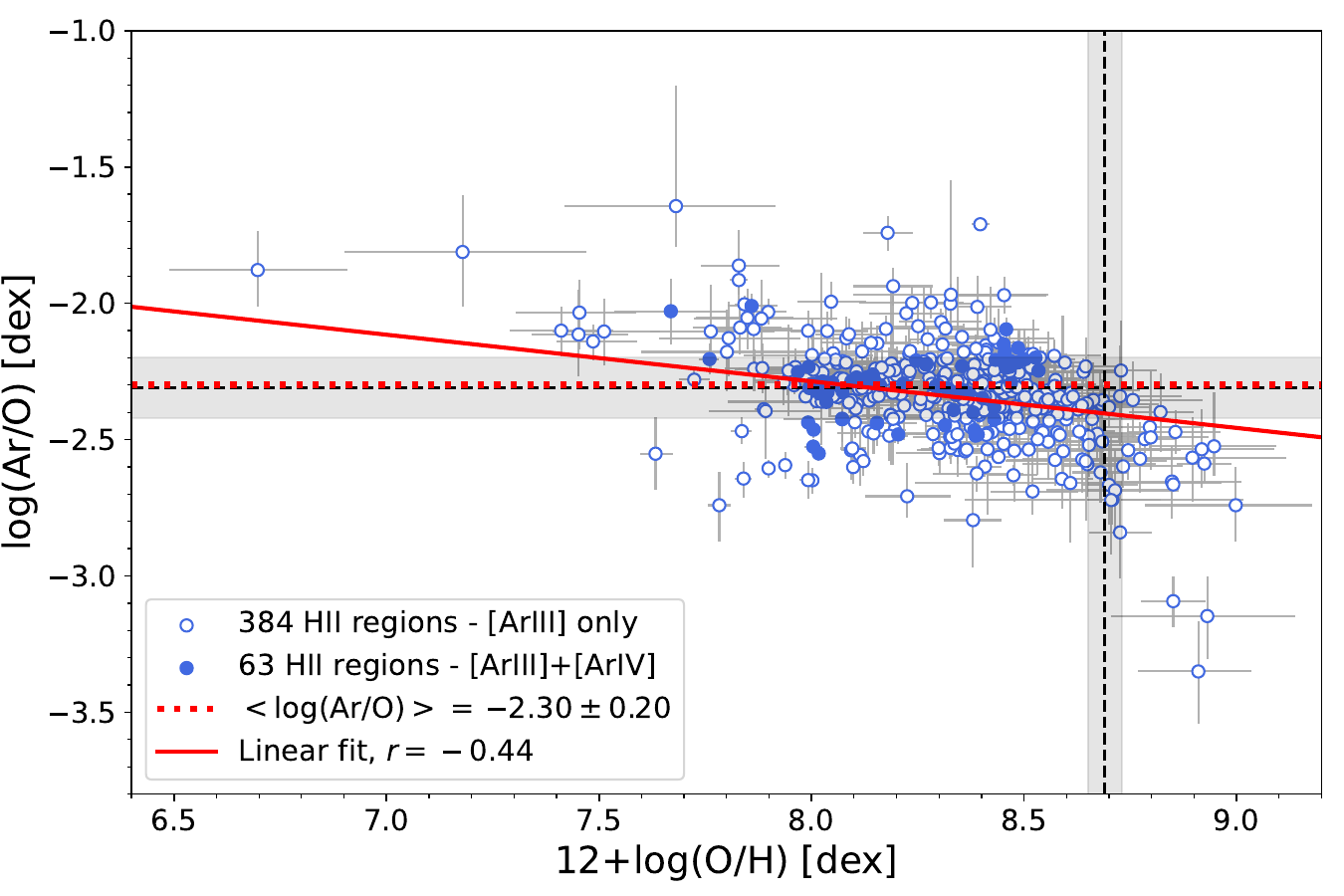}
\includegraphics[scale=0.38]{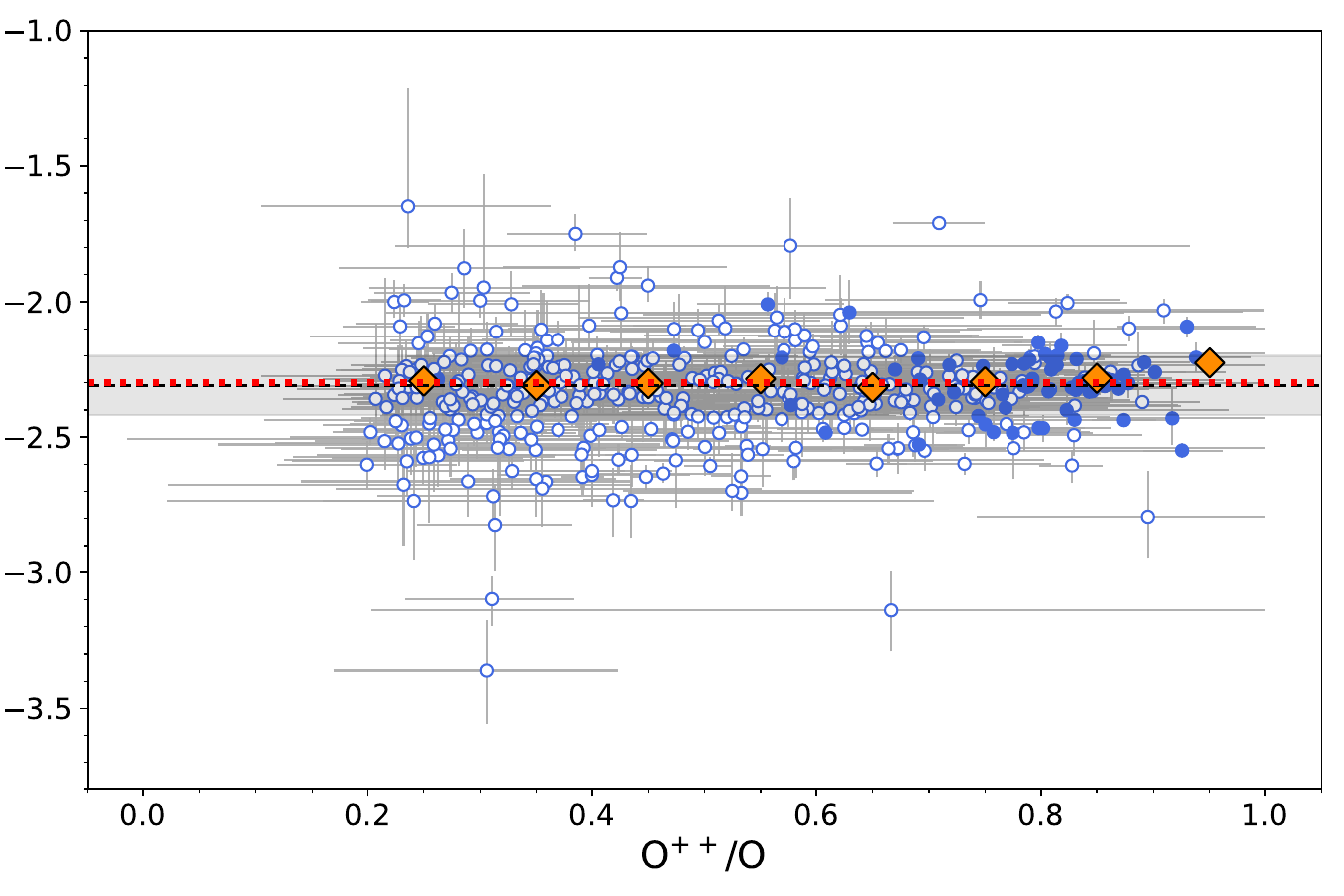}
\caption{log(Ar/O) as a function of 12+log(O/H) (left) and the ionisation degree, O$^{2+}$/O (right), for the DESIRED-E sample. The values of log(Ar/O) are calculated using the ICF(Ar) scheme by \citet{Izotov:06}. Empty symbols indicate Ar abundances determined solely from {\ariii} lines and full symbols those determined using {\ariii} and {\ariv} lines. The top panels show the points corresponding to SFGs (black squares) and the bottom panels the corresponding to {\hii} regions (blue circles).  Orange diamonds indicate the mean log(Ar/O) values considering bins in O$^{2+}$/O. The red continuous lines represent linear fits to the data represented in each panel. The dotted red line represents the mean value of the log(Ar/O) obtained for each kind of object.  The black dashed lines and the grey bands show the solar 12+log(O/H) and log(Ar/O) and their associated uncertainties, respectively, from \citet{Asplund:21}.} 

\label{fig:ArO_separated}
\end{figure*}

\begin{figure}[ht!]
\centering    
\includegraphics[scale=0.38]{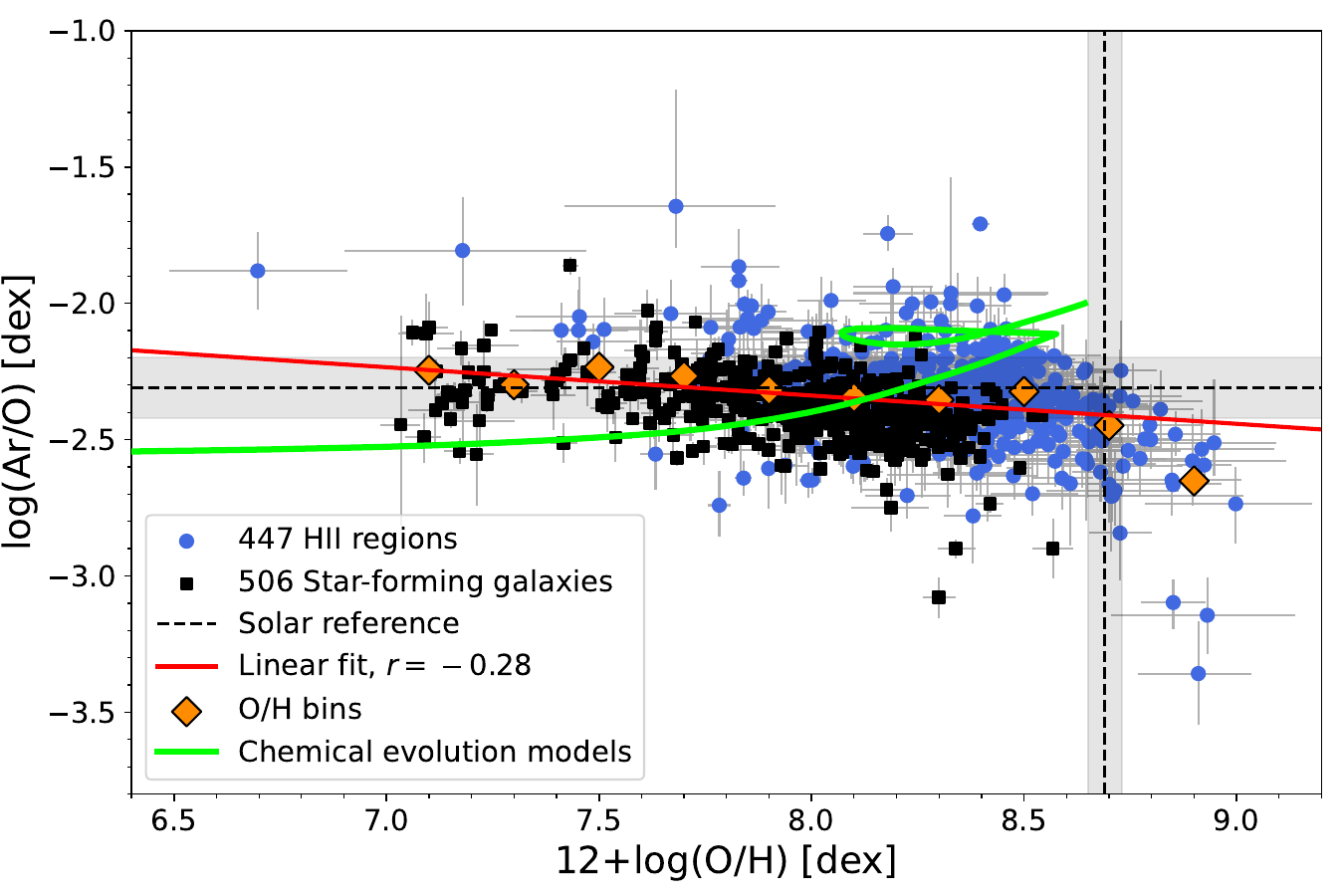}
\includegraphics[scale=0.38]{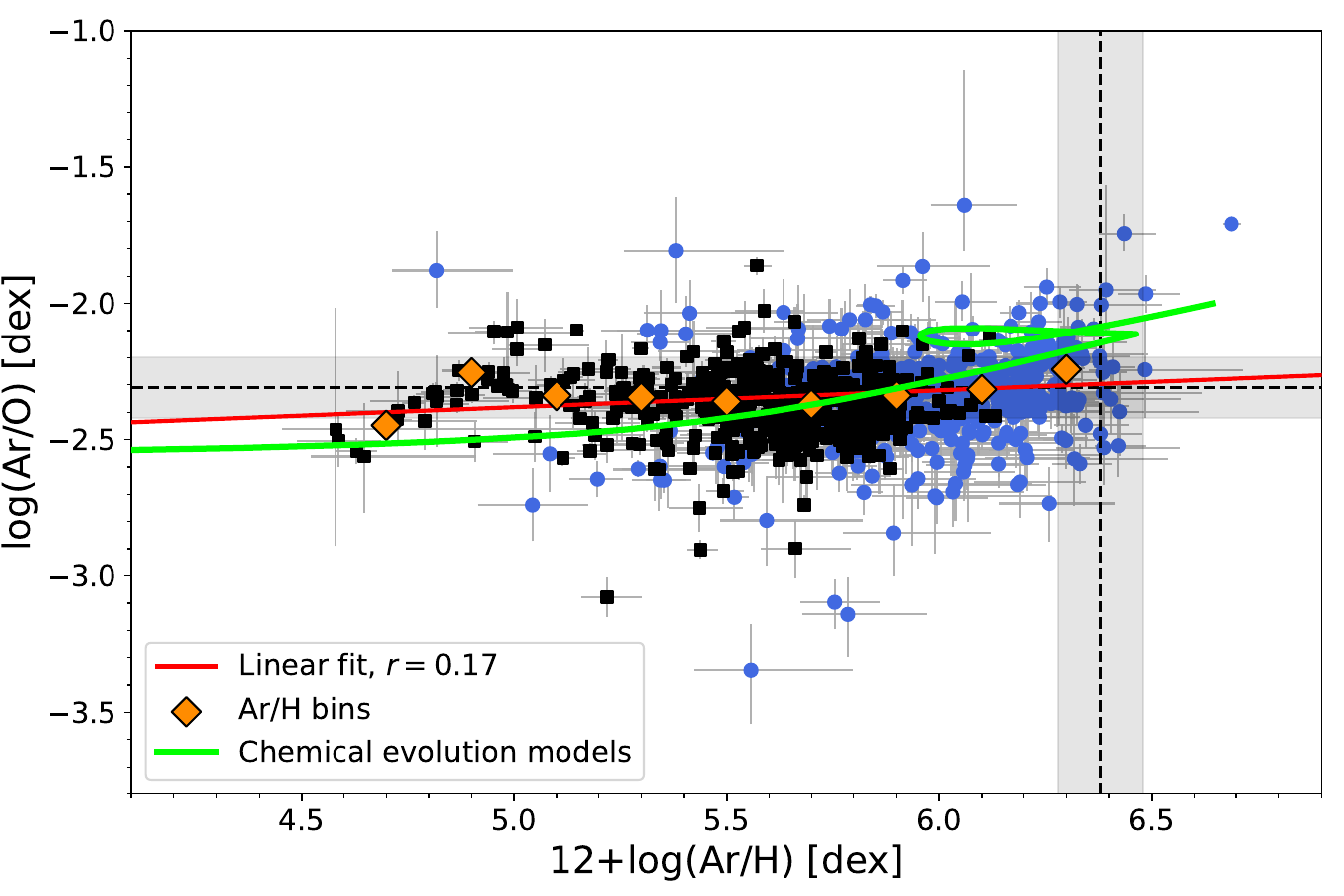}
\caption{log(Ar/O) as a function of 12+log(O/H) (top) and as a function of 12+log(Ar/H) (bottom). Black squares and blue circles represent star-forming galaxies and {\hii} regions of the DESIRED-E sample, which Ar/O ratios have been calculated using the ICF(Ar) scheme by \citet{Izotov:06}. The orange diamonds indicate the mean log(Ar/O) values considering bins in 12+log(O/H) or 12+log(Ar/H). The red continuous lines represent linear fits to the data represented in each panel. The green continuous line shows the time evolution of Ar and O abundances of the ISM predicted by the CEM of the Milky Way of \citet{Medina-Amayo:23}. The black dashed lines and the grey bands show the solar 12+log(O/H), 12+log(Ar/H) and log(Ar/O) and their associated uncertainties, respectively, from \citet{Asplund:21}.} 
\label{fig:ArO_Ar}
\end{figure}

In Sect.~\ref{subsec:argon}, we concluded that the different ICF(Ar) schemes considered show a very similar behaviour of the Ar/O ratio with respect to the O abundance and degree of ionisation, so the analysis of the Ar/O ratio seems to be independent of the final selection of the ICF(Ar) scheme. In fact, Fig~\ref{fig:comparion_ICFs} indicates that the differences in the Ar/H computed using any pair of ICF(Ar) schemes are lower than 0.04 dex when O$^{2+}$/O < 0.95 but not much larger than 0.10 for higher ionised objects. Due to that, we finally decided to use the ICF(Ar) by \citet{Izotov:06} in order to be consistent with the results obtained with the rest of alpha-elements and to remove the objects with lower ionisation degree (O$^{2+}$/O $<$ 0.2), where the contribution of unseen Ar$^+$ is larger. This allows the elimination of objects where the ICF(Ar) value is higher and minimizes the systematic uncertainty introduced by the use of an ICF.

Most works find an almost constant Ar/O vs. O/H distribution \citep[e.g.][]{Izotov:06,Hagele:08,Croxall:16,MirandaPerez:23}. However, several works find some correlation \citep[e.g.][]{PerezMontero:07, Arnaboldi:22,ArellanoCordova:24}. In Fig.~\ref{fig:ArO_separated}, we show the distribution of log(Ar/O) as a function of 12+log(O/H) and O$^{2+}$/O separately for {\hii} regions and SFGs. In all cases we have used the ICF(Ar) scheme by \citet{Izotov:06}. As in previous figures, we distinguish between those points whose Ar/H ratios have been derived solely using {\ariii} lines from those that have been determined using both {\ariii} and {\ariv} ones\footnote{There is a concentration of solid points in the lower left panel of Fig.~\ref{fig:ArO_separated} at 12+log(O/H) between 8.4 and 8.5 with oversolar Ar/O values. These points correspond to the nine spectra from different positions of the Orion Nebula where Ar$^{3+}$ abundances have been derived, so this apparent concentration is produced by a single object and has not statistical weight}. As can be noted in Fig.~\ref{fig:ArO_separated} and the fits parameters given in Table~\ref{table:fits}, the behaviour of log(Ar/O) vs. 12+log(O/H) and O$^{2+}$/O is very similar for {\hii} regions and SFGs. The log(Ar/O) vs. 12+log(O/H) distribution of both kinds of objects follow rather clear linear relationships with negative slopes. Their sample Pearson correlation coefficients are moderate but the corresponding $p$-values very small (see Table~\ref{table:fits}). The trend of decreasing Ar/O ratios as O/H increases is quite unexpected but it has been also observed in several previous works \citep{PerezMontero:07,Kojima:21,ArellanoCordova:24}. In the 
right panels of Fig.~\ref{fig:ArO_separated} we include also the mean log(Ar/O) in different O$^{2+}$/O bins of 0.1 units wide. In the case of {\hii} regions, the trend is basically constant with a maximum difference of log(Ar/O) of $\sim$0.05 dex. For SFGs the trend shows some undulation and log(Ar/O) increases slightly as O$^{2+}$/O increases, but the maximum difference is only 0.11 dex. The results shown in these two panels of Fig.~\ref{fig:ArO_separated} indicate that any trend we observe in the distribution of log(Ar/O) as a function of metallicity is basically independent of the ICF(Ar). The mean value of log(Ar/O) is $-$2.37 $\pm$ 0.12 in the case of SFGs and $-$2.30 $\pm$ 0.20 for {\hii} regions, both consistent with the solar value of $-$2.31 $\pm$ 0.11 \citep{Asplund:21}.

In Fig.~\ref{fig:ArO_Ar} we represent the log(Ar/O) vs. 12+log(O/H) (top) and vs. 12+log(Ar/H) (bottom) distributions of SFGs+{\hii} regions of the DESIRED-E sample. We have represented the data points with the same symbol regardless of whether {\ariii} or {\ariii}$+${\ariv} lines were used to determine the Ar abundance. As in some previous figures, in Fig.~\ref{fig:ArO_Ar} we also include a least-squares fit to the data represented in each panel. The two distributions show weak correlations with slopes of different signs. Although the sample Pearson correlation coefficients of the linear fits to log(Ar/O) vs. 12+log(O/H) and vs. 12+log(Ar/H) distributions are r = $-$0.28 and 0.17, respectively, their very low $p$-values indicate that both correlations can be considered real. As in the case of the fits separated by type of object, the fit to log(Ar/O) vs. 12+log(O/H) for SFGs+{\hii} regions indicates a decrease of log(Ar/O) of 0.24 dex in the whole interval of 12+log(O/H) represented in the top panel of Fig.~\ref{fig:ArO_Ar}. This behaviour is contradictory to what is expected from current nucleosynthesis and chemical evolution models. In fact, the curve predicted by the CEM of \citet{Medina-Amayo:23} included in Fig.~\ref{fig:ArO_Ar} would imply an increase of the Ar/O ratio with metallicity, in clear disagreement with the general behaviour shown by our data. This puzzling 
result deserves further study

In the case of the log(Ar/O) vs. 12+log(Ar/H) distribution shown in Fig.~\ref{fig:ArO_Ar}, we find a weak linear correlation with a small positive slope and a total increase of log(Ar/O) of 0.12 dex in the entire Ar/H range. Although the predictions of the CEM are now qualitatively consistent with the data, they are not quantitatively consistent. The CEM predicts an increase of 0.54 dex over the range of Ar abundances represented in Fig.~\ref{fig:ArO_Ar}, considerably higher than observed. On the other hand, unlike what happens in the cases of Ne and S, the CEM also does not reproduce the mean log(Ar/O) at low metallicities. The predicted value is about 0.10 dex lower than observed, but this cannot possibly be due to depletion of O onto dust grains because it is expected to be negligible at such low metallicities. 

Regarding the log(Ar/O) vs. 12+log(Ar/H) distribution obtained by other authors, the results are disparate. \citet{ArellanoCordova:24} and \citet{MirandaPerez:23} obtain a quite flat distribution in the log(Ar/O) vs. 12+log(Ar/H) relation at intermediate values of 12+log(Ar/H) but opposite trends in low and high values of the Ar abundance. In particular, \citet{MirandaPerez:23} find a rather strong increase of log(Ar/O) for objects with 12+log(Ar/H) $>$ 6.0, reaching values larger than in any SFG of our DESIRED-E sample. On the other hand, \citet{Arnaboldi:22} studied the behaviour of the Ar/O vs. Ar/H relation for a sample of planetary nebulae in the M31 galaxy. They find an increase of the Ar/O ratio as the Ar/H increases but in a rather narrow interval of 12+log(Ar/H), between 6.0 and 6.5 approximately, in agreement with the predictions of the chemical evolution models of the Milky Way thick disc and solar neighbourhood of \citet{Kobayashi:20}. \citet{Arnaboldi:22} interpret the increase in the log(Ar/O) vs. 12+log(Ar/H) plane they observe as due to the contribution of SNe Ia. In fact, the models by \citet{Kobayashi:20} predict that about 34\% of the total solar Ar is produced by SNe Ia. Fig.~\ref{fig:CEM} also indicates that in the CEM of \citet{Medina-Amayo:23} the increase of Ar/O with metallicity is produced basically by SNe Ia. We do not see an Ar/O increase as large as that reported by \citet{Arnaboldi:22}, and our results are broadly consistent with a constant log(Ar/O) distribution in most of the 12+log(Ar/H) range. However, the bins represented in the bottom panel of our Fig.~\ref{fig:ArO_Ar} suggest a slight increase --of only $\sim$0.1 dex-- for values of 12+log(Ar/H) $\gtrsim$ 6.0 which --as we demonstrate in Fig.~\ref{fig:ArO_separated}-- does not appear to be an artifact of the ICF(Ar). In principle, our linear fit to the log(Ar/O) vs. 12+log(Ar/H) distribution could be interpreted as an indication of a certain contribution by SNe Ia, although in a more limited proportion than that proposed by \citet{Arnaboldi:22}.

\section{Conclusions}

In this paper, we analyse the behaviour of Ne/O, S/O and Ar/O abundance ratios with respect to metallicity in star-forming regions of the local Universe making use of the spectroscopical data compiled in the DEep Spectra of ionised REgions Database (DESIRED) Extended project (DESIRED-E) \citep{MendezDelgado:23b, MendezDelgado:24}. DESIRED-E comprises almost two thousand high-quality spectra of {\hii} regions and star-forming galaxies (SFGs) from the literature having  direct determinations of {\tel}. For this study, we recalculate physical conditions and ionic and total abundances of O, Ne, S and Ar in a homogeneous manner for all the objects.  We have a total number of 1386 spectra for which we can determine the total abundance of one, two or three of the alpha-elements considered: Ne, S and Ar. Of that number, 50\% are classified as {\hii} regions and 50\% as SFGs. 

We compare the Ne/O, S/O and Ar/O abundance ratios obtained using three different ICF schemes for each element. Two of them: \citet{Amayo:21} and \citet{Izotov:06} are common to Ne, S and Ar, and the third ICF scheme is different for each element: \citet{Dors:13} for Ne, \citet{Dors:16} for S and \citet{PerezMontero:07} for Ar. After a careful analysis we conclude that the ICF scheme proposed by \citet{Izotov:06} seems to better reproduce the behaviour of the three abundance ratios. It produces a less dependent --flatter-- distribution of Ne/O, S/O and Ar/O with respect to O$^{2+}$/O and a smaller dispersion in the mean value of each ratio, this being consistent with the solar value. 

We derive Ne/O ratios for 1228 DESIRED-E spectra, of those 557 correspond to {\hii} regions and 671 to SFGs. The log(Ne/O) vs. 12+log(O/H) distribution of {\hii} regions shows a large dispersion and no evidence of correlation. Our analysis indicates that the position of those objects cannot be correctly reproduced by any of the ICF(Ne) schemes considered. Therefore, we do not recommend the use of these objects for the study of the cosmic evolution of Ne. On the other hand, the SFGs show much lower dispersion in their Ne/O ratios and rather similar log(Ne/O) vs. 12+log(O/H) and log(Ne/O) vs. 12+log(Ne/H) linear relations with positive but small slopes. These correlations may be due to several factors or a combination of them. Firstly, the effect of a metallicity-dependent O depletion fraction that increases throughout the metallicity range covered by the objects. Secondly, an effect of the ICF(Ne), which does not completely cancel out the Ne/O dependence on O$^{2+}$/O. A third possibility would be a certain unknown mechanism producing Ne, but chemical evolution models do not predict such scenario, at least in a significant amount. 

In order to achieve more confident determinations, we only consider objects for which their S/H ratio can be determined from the sum of S$^+$ and S$^{2+}$ abundances. Considering this, the log(S/O) vs. 12+log(O/H) relationship that we found separately for 441 and 492 spectra of {\hii} regions and SFG of our sample is compatible with a constant value, especially for {\hii} regions and when we combine all the sample objects ({\hii} regions + SFGs), in contrast with some previous works \citep{Dors:16, Diaz:22, Brazzini:24}. On this regard, we have found that at least part of the results of \citet{Diaz:22} are incorrect, so some of their conclusions should be revised.  We find a rather tight and positive linear fit for all the objects when we use the log(S/O) vs. 12+log(S/H) distribution. It is remarkable that the predictions of the CEM by \citet{Kobayashi:20} fit fairly well that distribution in the whole S/H range. We propose that the positive slope the log(S/O) vs. 12+log(S/H) relation and its change of slope at 12+log(S/H) $\sim$ 6.0 can be interpreted as an observational evidence of contribution of S produced by SNe Ia. 

The behaviour of log(Ar/O) vs. 12+log(O/H) and O$^{2+}$/O is very similar for our sample of 447 spectra of {\hii} regions and 506 of SFGs and seems to be independent of the O$^{2+}$/O ratio and the type of ICF(Ar) used, whether based on only the {\ariii} lines or on the sum of {\ariii} and {\ariv}. The fit to log(Ar/O) vs. 12+log(O/H) indicates a slight decrease of log(Ar/O) as 12+log(O/H) increases, an unexpected  behaviour. However, the log(Ar/O) vs. 12+log(Ar/H) relation shows a inverse trend, with a small positive slope that might be interpreted as an indication of a small contribution from SNe Ia, smaller than predicted by recent nucleosynthesis models.

The mean value of log(Ne/O) for our sample of SFGs is $-$0.64 $\pm$ 0.07, almost coincident with the solar value of $-$0.63 $\pm$ 0.06 recommended by \citet{Asplund:21}. Due to the low reliability of the ICF(Ne) for estimating the total Ne abundance of {\hii} regions, we cannot define a precise mean value of log(Ne/O) for these objects. In the case of log(S/O) the mean value for SFGs and {\hii} regions is $-$1.63 $\pm$ 0.13 and  $-$1.54 $\pm$ 0.19, respectively, the latter value being more consistent with the solar one of $-$1.57 $\pm$ 0.05. For log(Ar/O) our results are $-$2.37 $\pm$ 0.12 and $-$2.30 $\pm$ 0.20 for SFGs and {\hii} regions, respectively, again more consistent with the solar value of $-$2.31 $\pm$ 0.11 in the case of {\hii} regions. Considering that the mean 12+log(O/H) is 8.40 for our sample of {\hii} regions and 8.05 for SFGs, the fact that both the mean log(S/O) and log(Ar/O) are larger in {\hii} regions than in SFGs --about 0.11 and 0.07 dex, respectively-- indicates that both ratios increase with metallicity, consistent with a scenario of some additional contribution from S and Ar due to SNe Ia.
\begin{acknowledgements}
CE, JGR, ERR and MOG acknowledge financial support from the Agencia Estatal de Investigaci\'on of the Ministerio de Ciencia, Innovaci\'on y Universidades under grant ``The internal structure of ionised nebulae and its effects in the determination of the chemical composition of the interstellar medium and the Universe'' with reference PID2023-151648NB-I00 (DOI:10.13039/5011000110339). JGR also acknowledges financial support from the AEI-MCINN, under Severo Ochoa Centres of Excellence Programme 2020-2023 (CEX2019-000920-S), and from grant``Planetary nebulae as the key to understanding binary stellar evolution'' with reference PID-2022136653NA-I00 (DOI:10.13039/501100011033) funded by the Ministerio de Ciencia, Innovaci\'on y Universidades (MCIU/AEI) and by ERDF ``A way of making Europe'' of the European Union. JEMD  acknowledges support from project UNAM DGAPA-PAPIIT IG 101025, Mexico. LC acknowledges support from the Fundación Occident and the Instituto de Astrof\'\i sica de Canarias under the Visiting Researcher Programme 2022-2024 agreed between both institutions.
\end{acknowledgements}

\section*{Data Availability}

The complete tables of atomic data, references,
physical conditions and abundances for all the objects in our sample can be found at: \href{https://zenodo.org/records/15005362}{https://zenodo.org/records/15005362}.

\onecolumn
\begin{appendix} 

\section{Comparison between ICF schemes}
\label{sec:appendix_a}

\begin{figure*}[h]
\centering    
\includegraphics[scale=0.38]{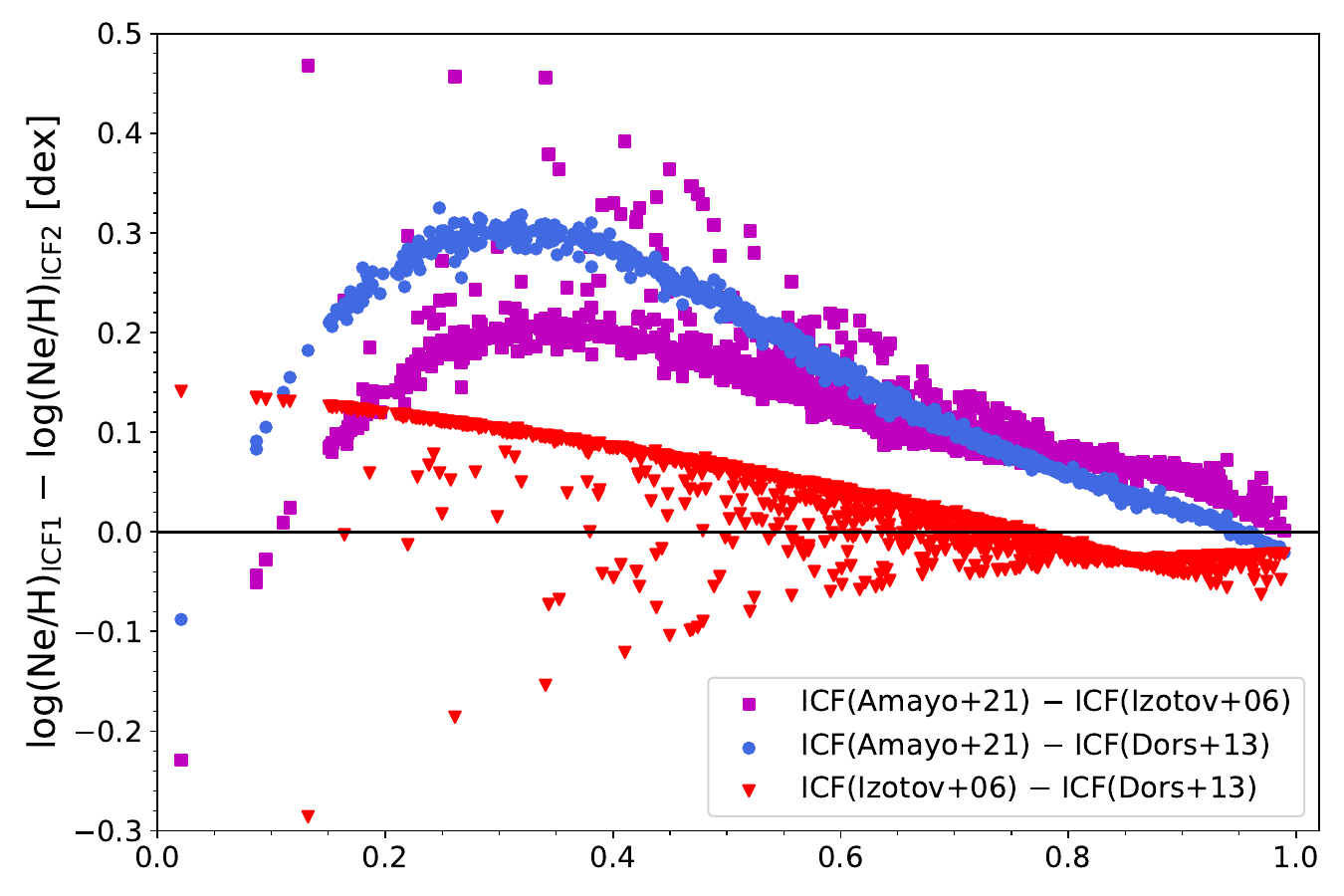}
\includegraphics[scale=0.38]{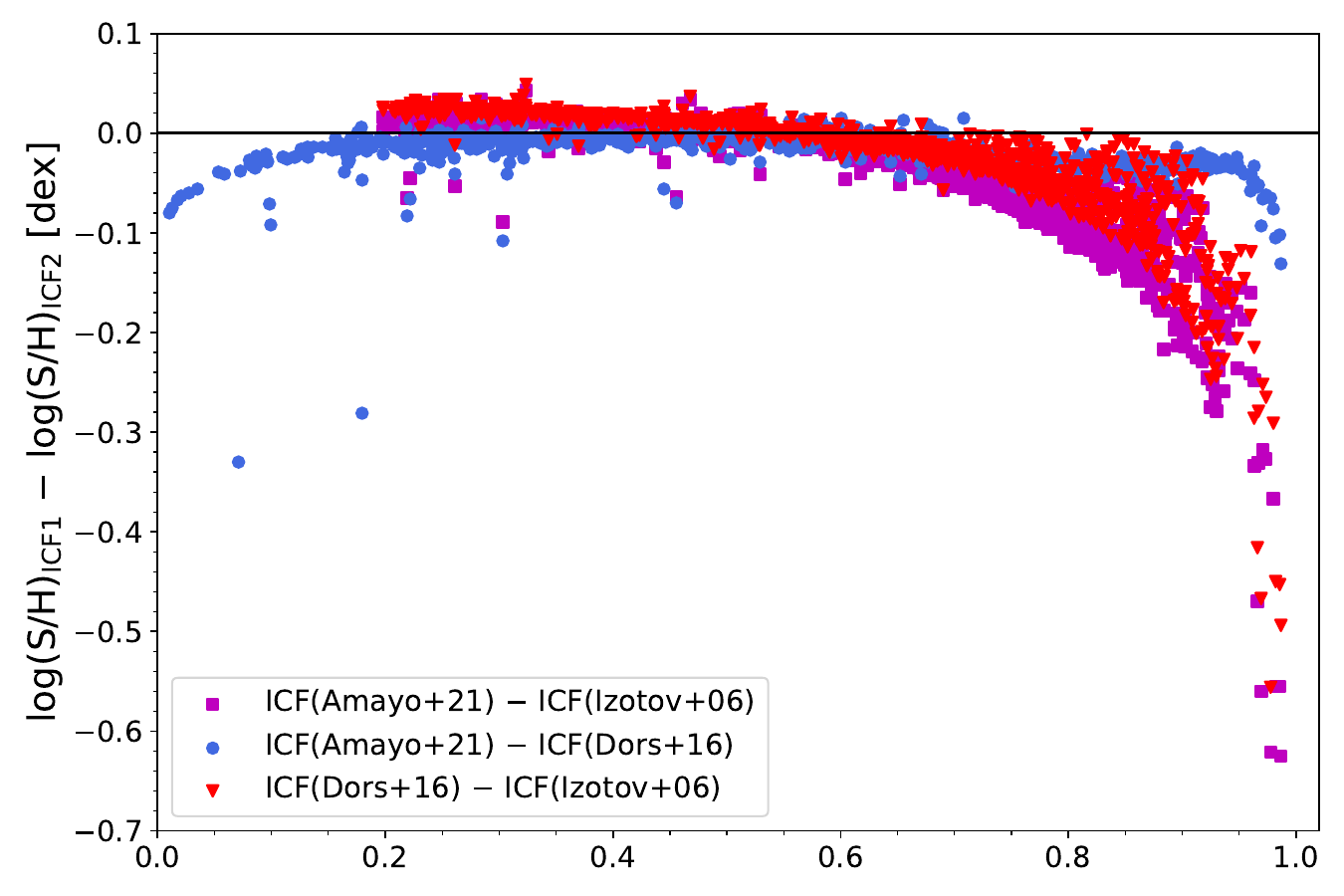}
\\
\includegraphics[scale=0.38]{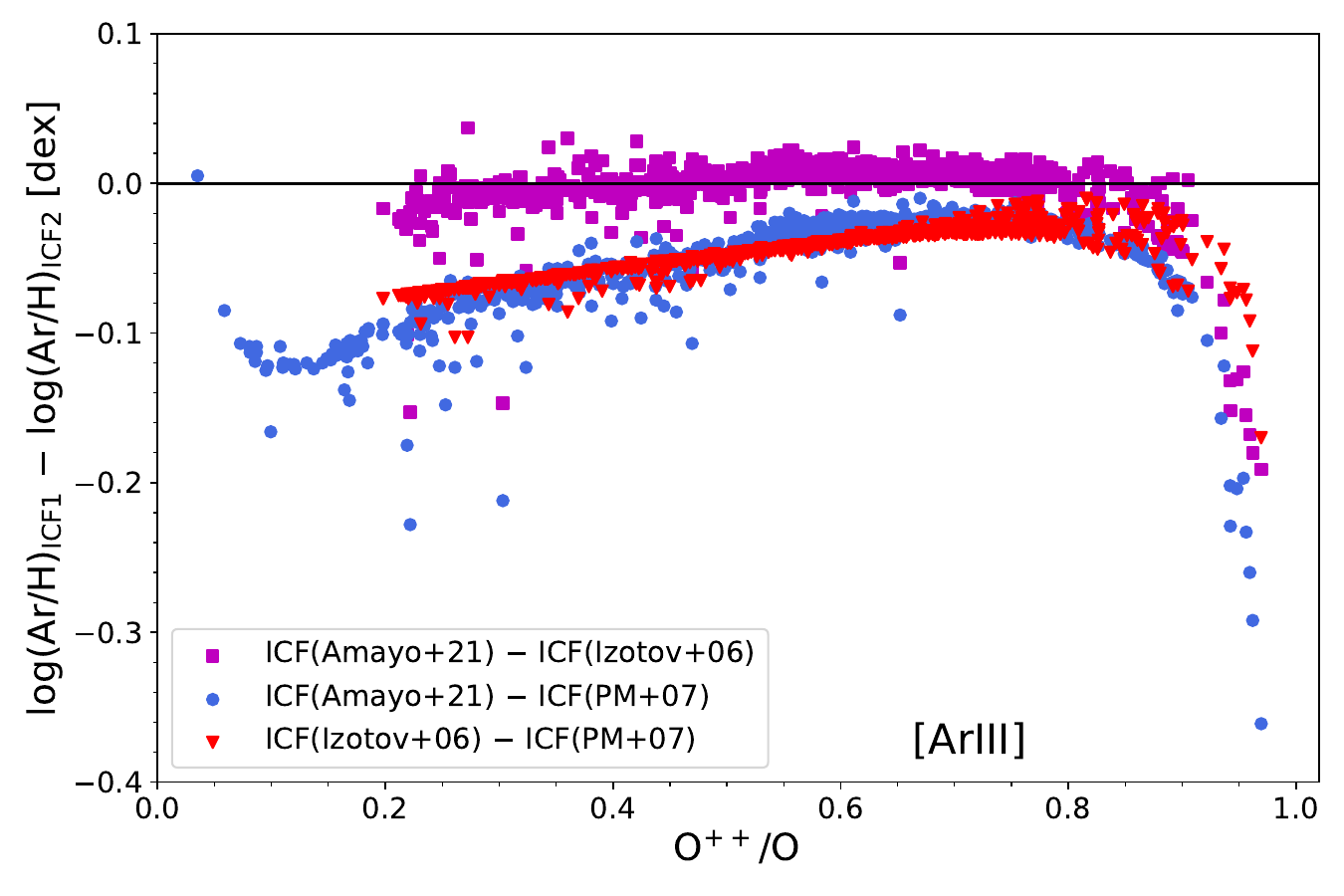}
\includegraphics[scale=0.38]{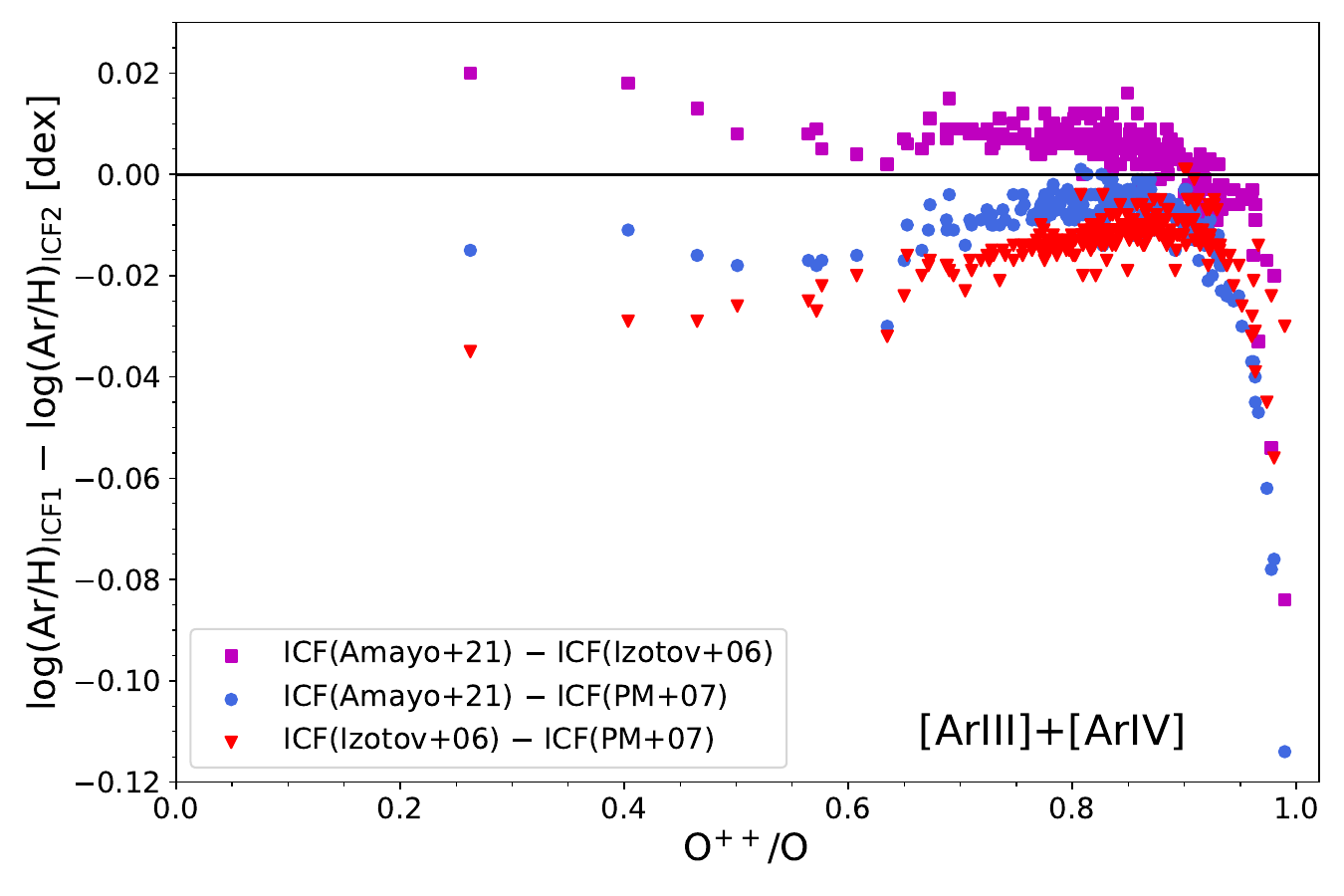}
\caption{Difference of the logarithmic Ne (top left), S (top right) and Ar (bottom panels) abundances calculated from the spectra of the DESIRED-E sample using various ICF schemes. In the bottom left panel we use ICF(Ar) using only {\ariii} lines and in the bottom right panel {\ariii} and {\ariv} lines. The different symbols represent the pairs of ICF used in each case. The horizontal black line indicates equal abundance. Amayo+21: \citet{Amayo:21}; Izotov+06: \citet{Izotov:06}; Dors+13: \citet{Dors:13}; Dors+16: \citet{Dors:16} and PM+07: \citet{PerezMontero:07}.} 
\label{fig:comparion_ICFs}
\end{figure*}

\section{The relation between the oxygen abundance and ionisation degree}
\label{sec:appendix_b}

\begin{figure*}[h]
\centering    
\includegraphics[scale=0.38]{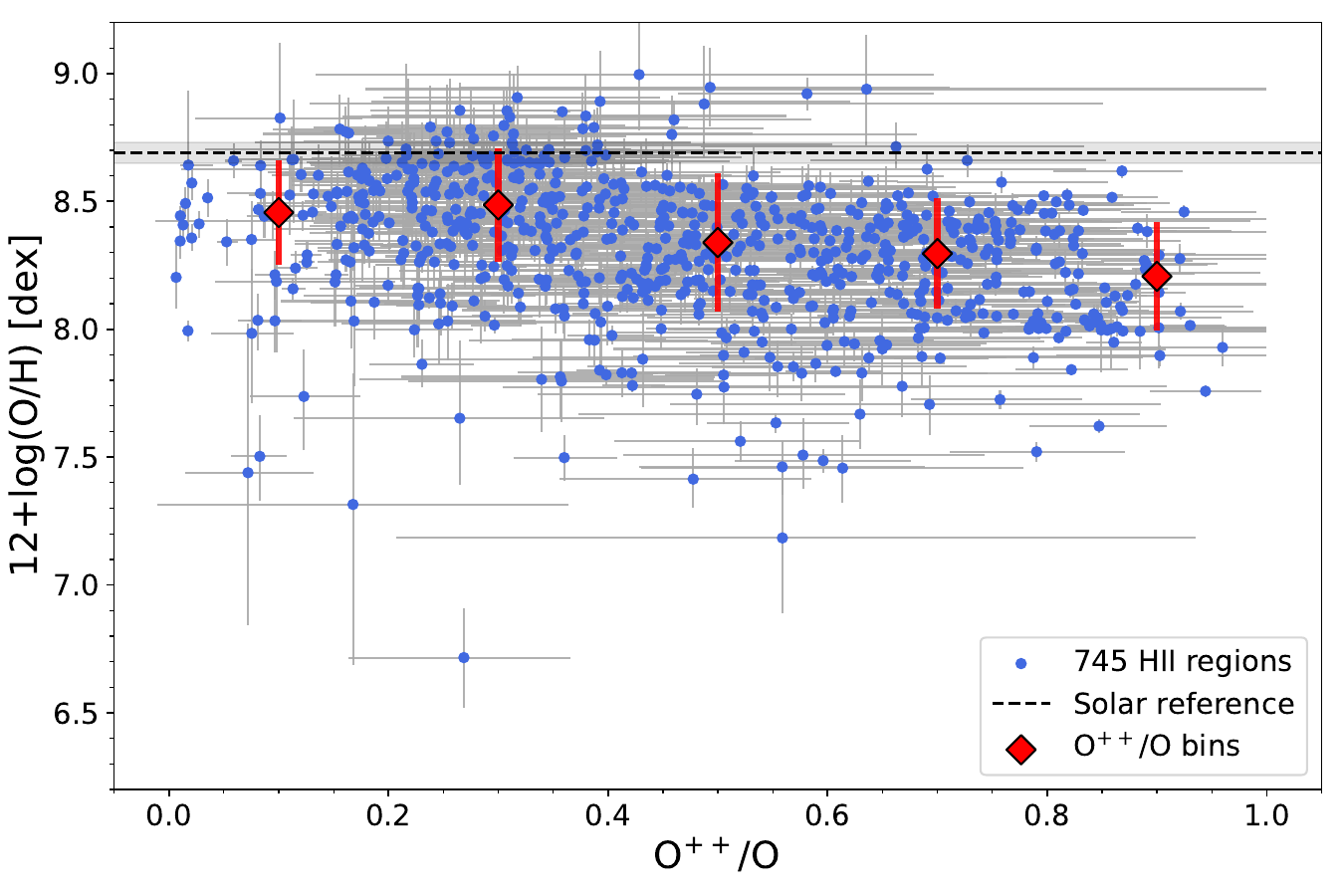}
\includegraphics[scale=0.38]{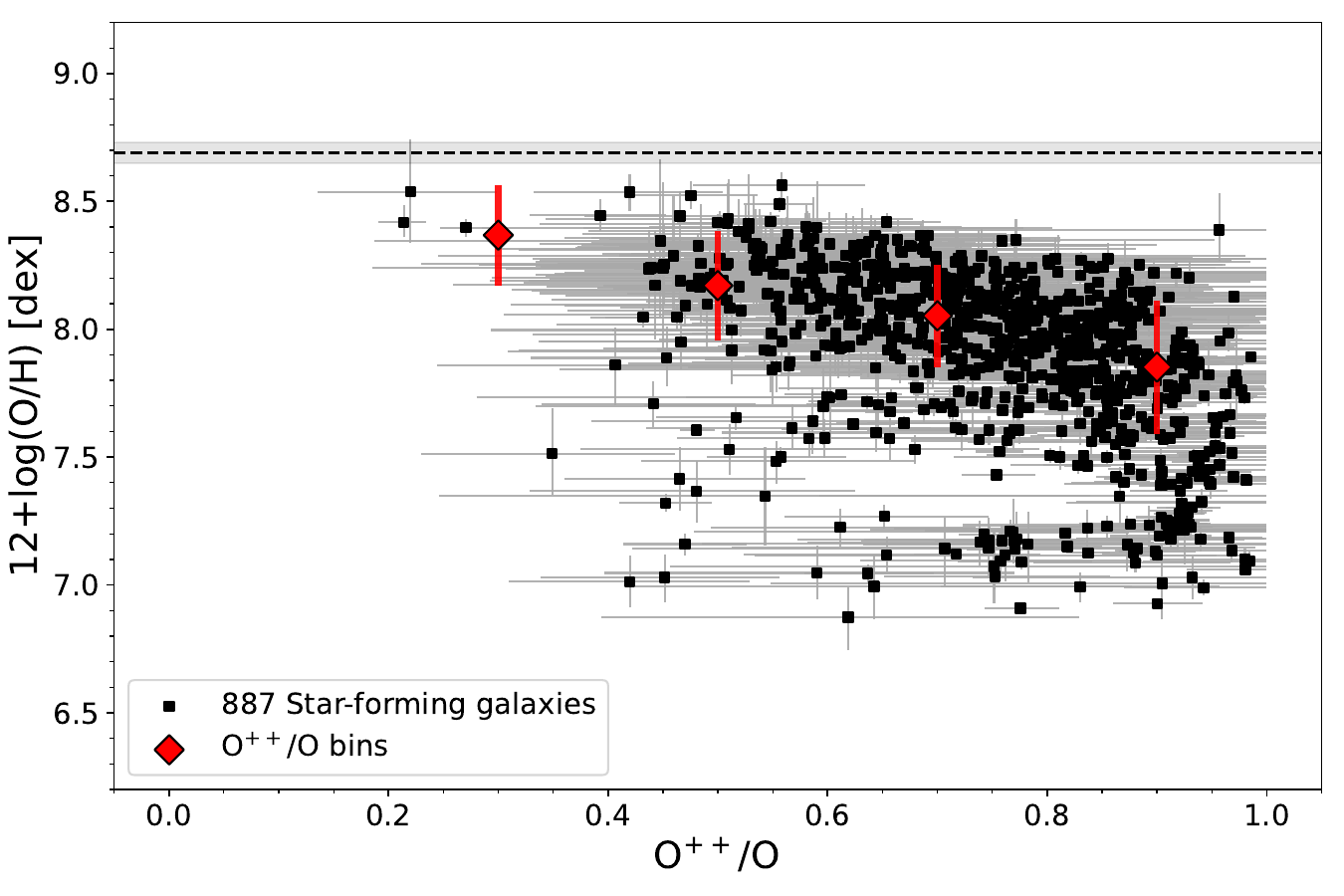}
\caption{12+log(O/H) as a function of the ionisation degree, O$^{2+}$/O, for DESIRED-E sample of Galactic and extragalactic \hii\ regions (blue circles, left panel) and star-forming galaxies (black squares, right panel). The red diamonds and lines indicate the mean 12+log(O/H) values and its standard deviation, considering bins of 0.2 in O$^{2+}$/O.  } 
\label{fig:Ovsion}
\end{figure*}

\section{Relative contribution to O, Ne, and Ar enrichment by 3 different types of stars considered in the chemical evolution model}
\label{sec:appendix_c}

\begin{figure*}[h]
\centering    
\includegraphics[scale=0.38]{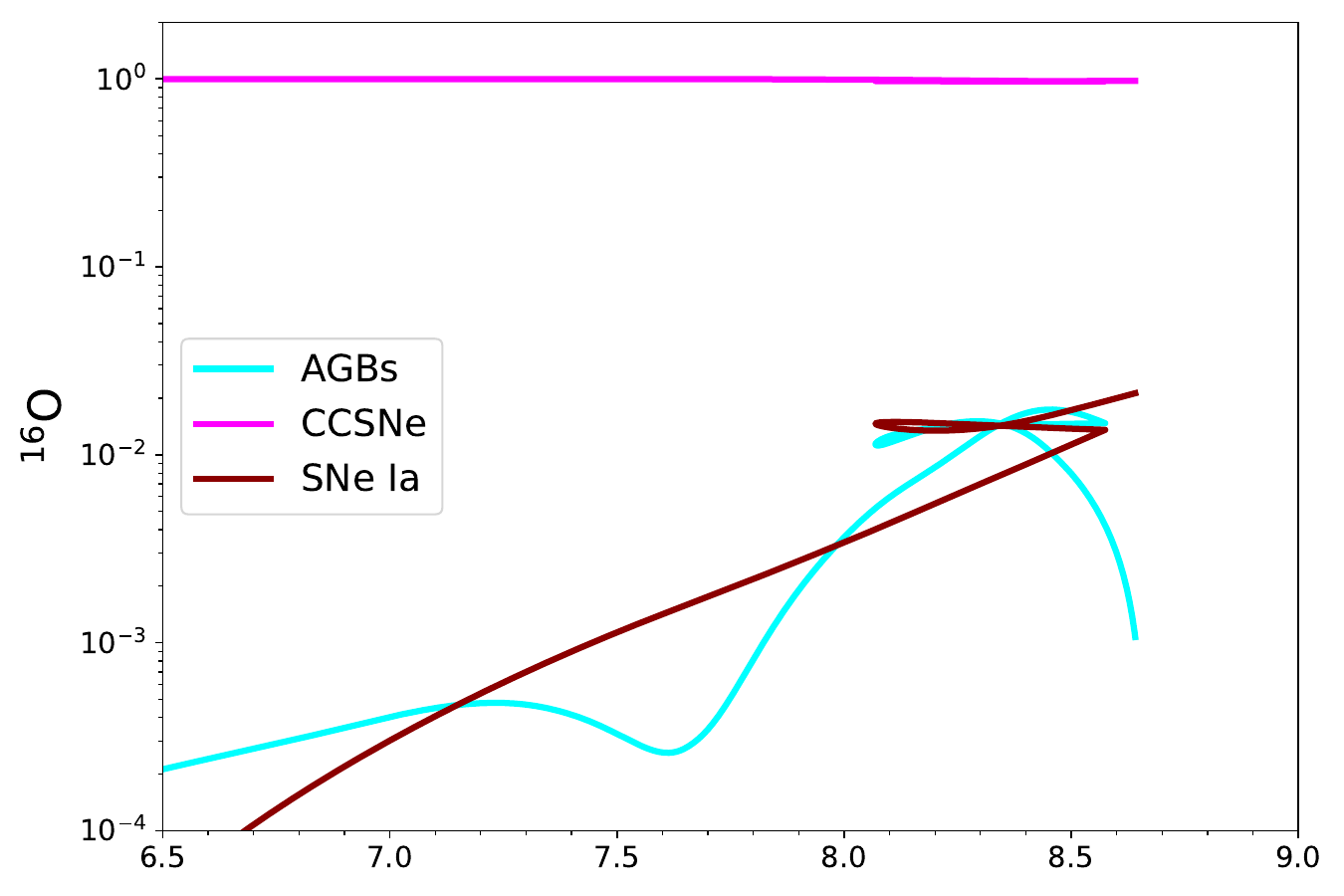}
\includegraphics[scale=0.38]{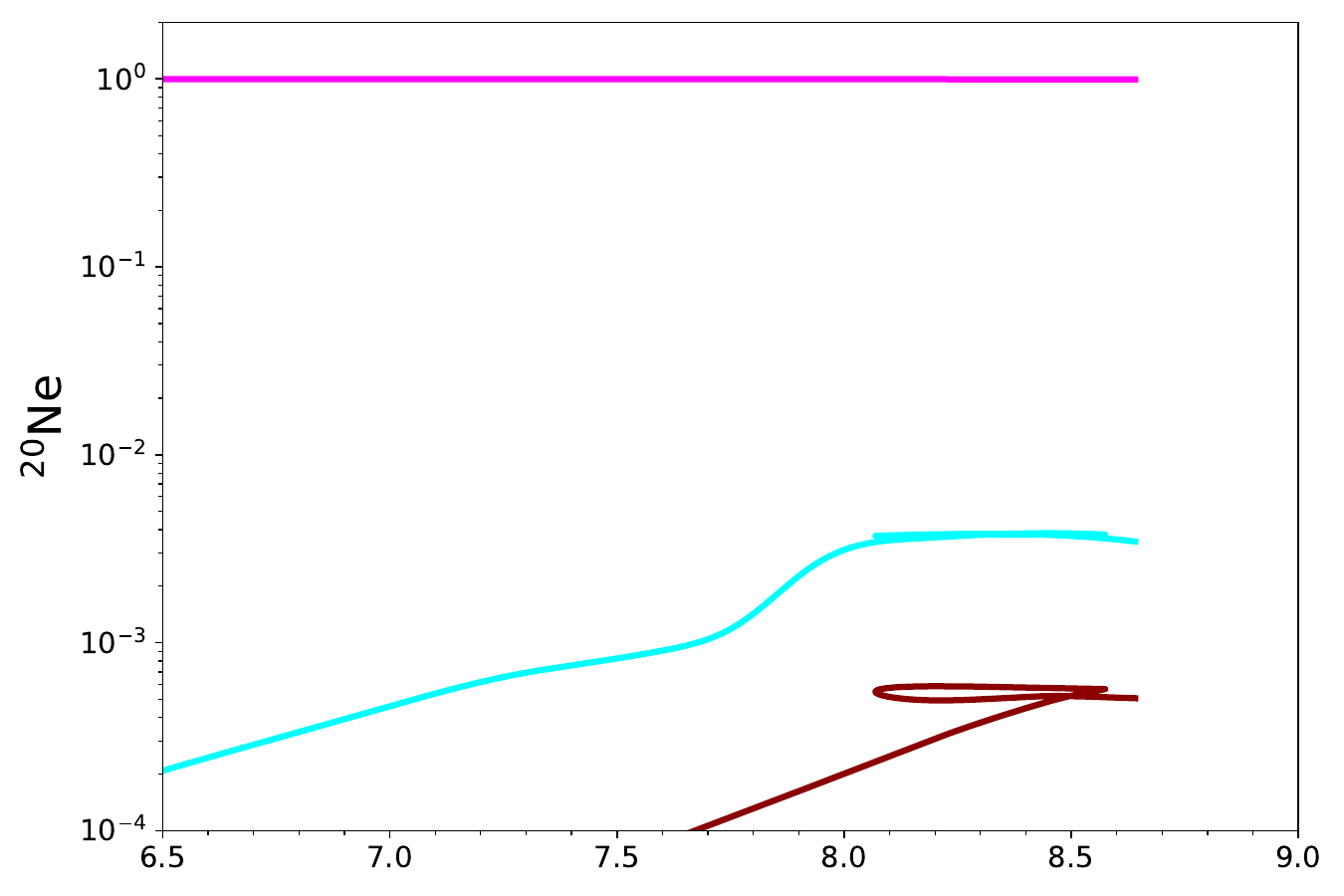}
\\
\includegraphics[scale=0.38]{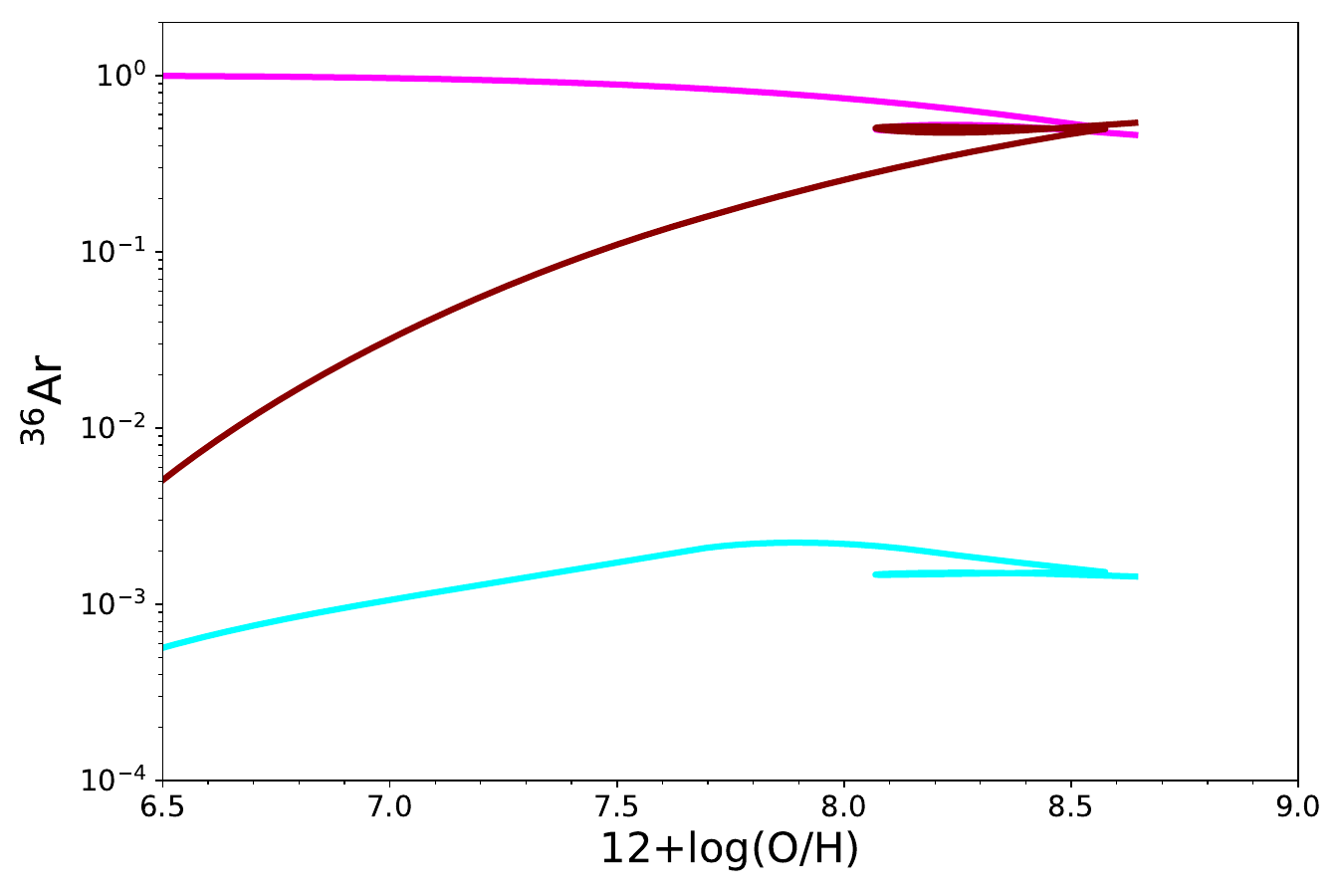}
\caption{Cumulative fraction of the contribution of different types of stars to $^{16}$O, $^{20}$Ne, and $^{36}$Ar (the most abundant stable isotopes of each element) enrichment as a function of 12+log(O/H) predicted by the chemical evolution model of \citet{Medina-Amayo:23}. Cyan line: asymptotic giant branch stars (AGBs); dark red line:  type Ia supernovae (SNe Ia); magenta line: core collapse supernovae (CCSNe).} 
\label{fig:CEM}
\end{figure*}

\section{Tables of atomic data, references, physical conditions and abundances}
\label{sec:appendix_d}

The complete reference tables associated with Appendix D can be found at: \href{https://zenodo.org/records/15005362}{https://zenodo.org/records/15005362}.

The references for the atomic data used per ion are: O$^{+}$: \citet{FroeseFischer:04}, \citet{Kisielius:09}; O$^{2+}$: \citet{Storey:00}, \citet{Storey:14}, \citet{Wiese:96}; Ne$^{2+}$: \citet{McLaughlin:11}; S$^{+}$: \citet{Irimia:05}, \citet{Tayal:10}; S$^{2+}$: \citet{FroeseFischer:06}, \citet{Grieve:14}; Cl$^{2+}$: \citet{Butler:89}, \citet{Fritzsche:99}; Ar$^{2+}$: \citet{Galavis:95},  \citet{Kaufman:86}, \citet{Mendoza:83b}; Ar$^{3+}$: \citet{Mendoza:82b}, \citet{Ramsbottom:97}; Fe$^{2+}$: \citet{Deb:09}, \citet{Mendoza:23}, \citet{Zhang:96}.

The references for the spectroscopic data are: \citet{ArellanoCordova:16, ArellanoCordova:21}; \citet{Aver:22}; \citet{Berg:13, Berg:15, Berg:16, Berg:20, Berg:21}; \citet{Bresolin:07a, Bresolin:11a, Bresolin:11b}; \citet{Bresolin:04, Bresolin:05, Bresolin:07b, Bresolin:09, Bresolin:09b, Bresolin:10}; \citet{Castellanos:02}; \citet{Crockett:06}; \citet{Croxall:15, Croxall:16}; \citet{DelgadoInglada:16}; \citet{Diaz:00}; \citet{DominguezGuzman:22}; \citet{Edmunds:84}; \citet{Egorova:21}; \citet{Esteban:18}; \citet{Esteban:04, Esteban:09, Esteban:13, Esteban:14, Esteban:17, Esteban:20}; \citet{Fernandez:18, Fernandez:22}; \citet{FernandezMartin:17}; \citet{GarciaRojas:04, GarciaRojas:05, GarciaRojas:06, GarciaRojas:07}; \citet{Garnett:94, Garnett:97}; \citet{Goddard:11}; \citet{GomezGonzalez:24}; \citet{Guseva:09, Guseva:11, Guseva:12, Guseva:13, Guseva:20, Guseva:24}; \citet{Hagele:06, Hagele:08, Hagele:11, Hagele:12}; \citet{Hsyu:17, Hsyu:18}; \citet{Isobe:22}; \citet{Izotov:98b, Izotov:04};  \citet{Izotov:97b, Izotov:99, Izotov:06, Izotov:09, Izotov:12, Izotov:16a, Izotov:17b, Izotov:18c, Izotov:19, Izotov:20a, Izotov:21a, Izotov:21b, Izotov:21c, Izotov:22, Izotov:24a, Izotov:24b}; \citet{Kehrig:11}; \citet{Kennicutt:03}; \citet{Kirsanova:23}; \citet{Kniazev:05, Kniazev:18}; \citet{Kojima:21}; \citet{Lee:05}; \citet{Lin:17}; \citet{LopezSanchez:07, LopezSanchez:15}; \citet{Luridiana:02}; \citet{Magrini:10}; \citet{MendezDelgado:21a, MendezDelgado:21b, MendezDelgado:22b}; \citet{MesaDelgado:09}; \citet{Nakajima:22}; \citet{Pagel:79, Pagel:80}; \citet{Patterson:12}; \citet{Peimbert:03, Peimbert:05, Peimbert:12}; \citet{Pena:07}; \citet{PenaGuerrero:12}; \citet{Pustilnik:05, Pustilnik:11, Pustilnik:16, Pustilnik:20}; \citet{Rogers:21, Rogers:22}; \citet{Rosolowsky:08}; \citet{Skillman:85}; \citet{Skillman:13}; \citet{Stanghellini:10}; \citet{Thuan:05, Thuan:22}; \citet{Toribio:16}; \citet{Valerdi:19, Valerdi:21}; \citet{Vilchez:88}; \citet{Watanabe:24}; \citet{Zurita:12}.

\end{appendix}

\end{document}